\documentclass[longauth]{aa} 

\usepackage{threeparttable}
\usepackage{longtable}
\usepackage{amssymb}
\usepackage[utf8]{inputenc}
\usepackage[T1]{fontenc}
\usepackage{xspace}
\usepackage{hyperref}
\usepackage{graphicx}
\usepackage{color}
\usepackage{lineno}
\usepackage{graphics}
\usepackage{subcaption}
\usepackage[section]{placeins}
\usepackage{longtable}
\usepackage[export]{adjustbox}
\usepackage{natbib}
\usepackage{txfonts}
\usepackage{tikz}
\usepackage{caption}
\usepackage{pdflscape}
\usepackage{geometry}
\usepackage{amsmath}
\usepackage{bm}
\usepackage{float}
\usepackage{rotating}
\usepackage{pdflscape}
\usepackage{afterpage}
\makeatletter
\newcommand\subsubsubsection{\@startsection{paragraph}{4}{\z@}%
  {-3.25ex \@plus -1ex \@minus -.2ex}%
  {1.5ex \@plus .2ex}%
  {\normalfont\normalsize}}
\makeatother

\newcommand{\tjdtdb}{\ensuremath{\rm {TJD_{TDB}}}}
\newcommand{\msun}{\ensuremath{\,M_\odot}}
\newcommand{\rsun}{\ensuremath{\,R_\odot}}
\newcommand{\lsun}{\ensuremath{\,L_\odot}}

\begin{document}

   \title{SOPHIE and TESS characterization of fourteen TESS Objects of Interest: nine planetary systems and five low-mass stellar companions}

   \author{N.~Heidari\inst{\ref{iap}},
          G.~H\'ebrard\inst{\ref{iap},\ref{OHP}}, 
          S. Sousa \inst{\ref{santos1},\ref{santos2}},
          T.~Forveille \inst{\ref{gronobl}},
          F. Destriez \inst{\ref{Lesia},\ref{iap}},
          J.~D.\ Eastman \inst{\ref{jason}},
          J. M.~Almenara\inst{\ref{geneva}},
          D.~ W. Latham \inst{\ref{jason}},
          B.~S.~Safonov\inst{\ref{stefano}},
          R. Brahm\inst{\ref{Jordan1},\ref{Jordan2}},
          K.~G.\ Stassun\inst{\ref{Stassun}},
          K. A. Collins\inst{\ref{jason}},
          D. Dragomir \inst{\ref{diana}},
          J. Jan\'{\i}k\inst{\ref{jan}}, 
          I. Boisse\inst{\ref{lam},\ref{OHP}},
          J.~G.~Barrientos\inst{\ref{Jonathan}},
          J.~Serrano Bell\inst{\ref{diaz}},
          F.~Kiefer \inst{\ref{Lesia}},
          A.~Bieryla\inst{\ref{jason}},
          X.~Bonfils\inst{\ref{gronobl}},
          Z.~Balk\'oov\'a\inst{\ref{Zuzana},\ref{Zuzana1}},
          G.~Conzo\inst{\ref{Conzo}},
          M.~Deleuil\inst{\ref{lam}},
          O. D. S. Demangeon \inst{\ref{santos1},\ref{santos2}},
          X.~Delfosse \inst{\ref{gronobl}},
          R.~ F. D\'iaz \inst{\ref{diaz}},
          Z. Essack\inst{\ref{diana}},
          J.~Eberhardt\inst{\ref{Heidelberg}},
          Y.~N.~E.~Eschen\inst{\ref{Warwick}},
          A.~C.~Frommer\inst{\ref{yale}},
          F.~R.~Frustagia\inst{\ref{Frustagia}},
          D.~ Feliz\inst{\ref{New_York}},
          A.~F.~Gupta\inst{\ref{arvind}},
          N.~Hara\inst{\ref{lam}},
          T.~Henning\inst{\ref{Heidelberg}},
          S.~B.~Howell\inst{\ref{nasa}},
          A.~Jord\'an\inst{\ref{Jordan1},\ref{Jordan3},\ref{Jordan4}},
          J.~K\"ohler\inst{\ref{Jana}},
          J.~ F.\ Kielkopf \inst{\ref{Louisville}},
          H.~A.~Knutson \inst{\ref{Jonathan}},
          J.~Lipt\'ak\inst{\ref{Zuzana},\ref{Zuzana1}},
          C.~R.~Mann\inst{\ref{mann}},
          E. ~Martioli\inst{\ref{eder},\ref{iap}},
          C.~Moutou \inst{\ref{Toulouse}},
          F.~Murgas\inst{\ref{Murgas1},\ref{Murgas2}},
          B.~Nicholson\inst{\ref{Australia}},
          P.~Pintr\inst{\ref{Plasma}},
          M.~Tala Pinto\inst{\ref{Jordan1},\ref{Jordan2}},
          S.~N.~Quinn\inst{\ref{jason}},
          H.~M.~Relles\inst{\ref{jason}},
          P.~Rowden\inst{\ref{Royal}},
          R.~P.~Schwarz\inst{\ref{jason}},
          R.~Sefako\inst{\ref{Africa}},
          A.~Shporer\inst{\ref{Massachusetts}},
          G.~Srdoc\inst{\ref{Croatia}},
          C.~Stockdale\inst{\ref{Hazelwood}},
          I.~A.~Strakhov\inst{\ref{stefano}},
          B.~Skinner\inst{\ref{diana}}, 
          B.~M.~Tofflemire\inst{\ref{Texas}},
          T.~Trifonov\inst{\ref{Heidelberg},\ref{Heidelberg2},\ref{Bulgaria}},
          S.~Ulmer-Moll\inst{\ref{inst20}},
          L.~Vanzi\inst{\ref{Vanzi},\ref{Vanzi1}},
          F.~P.~Wilkin\inst{\ref{Union}},
          C.~Ziegler\inst{\ref{Ziegler}}
          }

  \institute{Institut d'astrophysique de Paris, UMR 7095 CNRS université pierre et marie curie, 98 bis, boulevard Arago, 75014, Paris \label{iap}
   \and
   Observatoire de Haute-Provence, CNRS, Universit\'e d'Aix-Marseille, 04870 Saint-Michel-l'Observatoire, France \label{OHP},
   \and
     Instituto de Astrof\'isica e Ci\^encias do Espa\c{c}o, Universidade do Porto, CAUP, Rua das Estrelas, 4150-762 Porto, Portugal.\label{santos1},
   \and     
     Departamento de F\'isica e Astronomia, Faculdade de Ci\^encias, Universidade do Porto, Rua do Campo Alegre, 4169-007 Porto, Portugal.\label{santos2},
     \and
    Univ. Grenoble Alpes, CNRS, IPAG, 38000 Grenoble, France\label{gronobl},
    \and
    LIRA, Université Paris Cité, Observatoire de Paris, Université PSL, Sorbonne
Université, CY Cergy Paris Université, CNRS, F-92190 Meudon, France\label{Lesia}
   \and
    Center for Astrophysics \textbar \ Harvard \& Smithsonian, 60 Garden St, Cambridge, MA 02138, USA \label{jason},
    \and
     Observatoire de Gen\`eve,  Universit\'e de Gen\`eve, Chemin Pegasi, 51, 1290 Sauverny, Switzerland\label{geneva}
    \and
    Sternberg Astronomical Institute of Lomonosov Moscow State University, Moscow, 119234 Russia\label{stefano},
   \and
    Facultad de Ingenier\'{i}a y Ciencias, Universidad Adolfo Ib\'{a}\~{n}ez, Av. Diagonal las Torres 2640, 7941169 Pe\~{n}alol\'{e}n, Santiago, Chile \label{Jordan1}
    \and
   Departamento de Astronomía, Universidad de Chile, Casilla 36-D, Santiago, Chile \label{Jordan3}
   \and
   El Sauce Observatory --- Obstech, Coquimbo, Chile \label{Jordan4}
  \and
     Millennium Institute for Astrophysics, Chile \label{Jordan2}
     \and  
     Vanderbilt University, Department of Physics \& Astronomy, 6301 Stevenson Center Ln., Nashville, TN 37235, USA \label{Stassun}
    \and
     Department of Physics \& Astronomy, University of New Mexico, 1919 Lomas Boulevard NE, Albuquerque, NM 87131, USA \label{diana},
     \and
     Department of Theoretical Physics and Astrophysics, Faculty of Science, Masaryk University, Kotl\'{a}\v{r}sk\'{a} 2, CZ-611 37, Czech Republic \label{jan},
     \and
     Aix Marseille Univ, CNRS, CNES, LAM, Marseille, France.\label{lam},
         \and
    Division of Geological and Planetary Sciences, California, Institute of Technology, Pasadena, CA 91125, USA\label{Jonathan},
      \and
    International Center for Advanced Studies (ICAS) and ICIFI (CONICET), ECyT-UNSAM, Campus Miguelete, 25 de Mayo y Francia, (1650) Buenos Aires, Argentina.\label{diaz}
     \and
     Astronomical Institute, Czech Academy of Sciences, Fri\v{c}ova 298, 25165
Ond\v{r}ejov, Czech Republic\label{Zuzana},
     \and
     Astronomical Institute of Charles University, V Hole\v{s}ovi\v{c}k\'{a}ch 2, CZ-180
00 Prague, Czech Republic\label{Zuzana1},
    \and
    Gruppo Astrofili Palidoro, Fiumicino, Italy\label{Conzo},
    \and
    Max-Planck-Institut für Astronomie, Königstuhl 17, D-69117, Heidelberg, Germany\label{Heidelberg},
    \and
    Department of Physics, University of Warwick, Gibbet Hill Road, Coventry CV4 7AL, UK\label{Warwick},
    \and
    Department of Astronomy, Yale University, New Haven, CT 06510, USA\label{yale}
    \and
    Frustaglia Private Observatory, Spain\label{Frustagia},
    \and
    Flatiron Institute Center for Computational Astrophysics, New York, NY 10010, USA\label{New_York}
    \and
   U.S. National Science Foundation National Optical-Infrared Astronomy Research Laboratory, 950 N.\ Cherry Ave., Tucson, AZ 85719, USA\label{arvind}
    \and
    NASA Ames Research Center, Moffett Field, CA 94035, USA\label{nasa} 
    \and
    Th\"uringer Landessternwarte, D-07778 Tautenburg, Germany\label{Jana}
    \and
    Department of Physics and Astronomy, University of Louisville, Louisville, KY 40292, USA\label{Louisville} 
    \and
    National Research Council Canada, Herzberg Astronomy \& Astrophysics Research Centre, 5071 West Saanich Road, Victoria, BC V9E 2E7, Canada\label{mann},
    \and
   Laboratório Nacional de Astrofísica, Rua Estados Unidos 154, 37504-364 Itajubá, MG, Brazil \label{eder}
   \and
    Univ. de Toulouse, CNRS, IRAP, 14 avenue Belin, 31400 Toulouse, France \label{Toulouse}
   \and
    Instituto de Astrof\'isica de Canarias (IAC), E-38205 La Laguna, Tenerife, Spain\label{Murgas1}
   \and
    Departamento de Astrof\'isica, Universidad de La Laguna (ULL), E-38206 La Laguna, Tenerife, Spain\label{Murgas2},
    \and
    Centre for Astrophysics, University of Southern Queensland, West St., Toowoomba, Queensland, 4350, Australia\label{Australia},
    \and
    Institute of Plasma Physics ASCR, v. v. i. U Slovanky 2525/1a 182 00 Prague 8, Liben\label{Plasma},
    \and
    Royal Astronomical Society, Burlington House, Piccadilly,London W1J 0BQ, UK\label{Royal},
    \and
  South African Astronomical Observatory, P.O. Box 9, Observatory, Cape Town 7935, South Africa\label{Africa},
     \and
     Department of Physics and Kavli Institute for Astrophysics and Space Research, Massachusetts Institute of Technology, Cambridge, MA 02139, USA\label{Massachusetts},
     \and
     Kotizarovci Observatory, Sarsoni 90, 51216 Viskovo, Croatia\label{Croatia},
     \and
     Hazelwood Observatory, Australia\label{Hazelwood},
     \and
     Sternberg Astronomical Institute Lomonosov Moscow State University\label{IVA},
     \and
      Department of Astronomy, Sofia University 'St Kliment Ohridski', 5 James Bourchier Blvd, BG-1164 Sofia, Bulgaria\label{Bulgaria}
    \and
     Landessternwarte, Zentrum f\"ur Astronomie der Universit\"at Heidelberg, K\"onigstuhl 12, D-69117 Heidelberg, Germany\label{Heidelberg2}
     \and
     SETI Institute, Mountain View, CA 94043, USA/NASA Ames Research Center, Moffett Field, CA 90345 USA\label{Texas}
     \and
     Leiden Observatory, Leiden University, P.O. Box 9513, 2300 RA Leiden, The Netherlands\label{inst20}
     \and
     Department of Electrical Engineering, Pontificia Universidad Católica de Chile, Av.Vicuña Mackenna 4860, 782-043 Santiago, Chile\label{Vanzi}
     \and
      Instituto de Astrof\'isica, Pontificia Universidad Cat\'olica de Chile, Av. Vicu\~na
Mackenna 4860, 7820436 Macul, Santiago, Chile\label{Vanzi1}
     \and
     Department of Physics and Astronomy, Union College, 807 Union St., Schenectady, NY 12308, USA\label{Union}
     \and
     Department of Physics, Engineering and Astronomy, Stephen F. Austin State University, 1936 North Street, Nacogdoches, TX 75962, USA\label{Ziegler}
   }

   \date{Received hopefully soon; accepted hopefully soon}

 
  \abstract
   {We present a study of fourteen \textit{TESS} Objects of Interest (TOIs) observed as part of the SOPHIE–\textit{TESS} follow-up program. Using \textit{TESS} photometry, SOPHIE radial velocities (RVs), ground-based photometry, and high-spatial resolution imaging, we characterized all targets and established their true nature. Among them, nine systems—TOI-6457, TOI-6883A, TOI-6649, TOI-6842A, TOI-5217A, TOI-6893A, TOI-7025, TOI-7176, and TOI-5893—are shown to host planets. We find that TOI-6649~b, TOI-6842A~b, and TOI-5217A~b are transiting hot Jupiters with orbital periods of approximately 4.4–6.0 d, radii of 1.3–1.6~$R_{\rm J}$, and masses of 0.3–1.3~$M_{\rm J}$, corresponding to bulk densities of 0.1–0.6~g~cm$^{-3}$. Our RV data further reveal that TOI-5217A hosts an outer companion, TOI-5217A~c, with a period of $\sim$1761~d, a minimum mass of $\sim$ 13~$M_{\rm J}$, and a significantly eccentric orbit ($e = 0.53 \pm 0.02$). We also characterize TOI-5893~b, TOI-6893A~b, TOI-6457~b, TOI-7176~b, and TOI-7025~b as transiting warm Jupiters, a relatively scarce population. Their orbital periods range from 40.4 to 67.0~days, with radii of 0.8–1.2~$R_{\rm J}$ and masses of 0.5–1.0~$M_{\rm J}$. In addition, TOI-6883A~b—a recently reported transiting warm Jupiter—is further characterized using new ground- and space-based photometry together with SOPHIE and TRES RVs. The remaining five systems—TOI-5621AB, TOI-5321AB, TOI-3678AB, TOI-5906AB and TOI-5308AB—exhibit large RV semi-amplitudes indicative of eclipsing binaries. The companions have estimated masses of 0.1–0.6~$M_{\odot}$ and radii of 0.1–0.6~$ R_\odot$. Among our sample, TOI-6842A~b stands out as an extremely low-density giant planet ($\rho_{p} = 0.08 \pm 0.01\,\mathrm{g\,cm^{-3}}$), placing it in the super-puff category and ranking it among the nine lowest-density planets known to date. With a transmission spectroscopy metric of $\sim$250, it represents an excellent target for atmospheric characterization. Additionally, transiting planets TOI-6883A b, TOI-6893A b, TOI-6457 b, TOI-5893 b, TOI-7176 b, and TOI-7025 b—each with equilibrium temperatures between 377 and 846 K—provide valuable low-irradiation benchmarks for control models of hot-Jupiter inflation. Furthermore, TOI-7025 b stands out due to its orbital eccentricity of 0.82$\pm$0.02, one of the largest planetary eccentricities discovered so far for a planet orbiting a single star. Finally, using the \textit{Gaia} astrometric excess noise, we detect TOI-5906~B, marking the first discovery of an object identified using all three techniques: transit, RV, and astrometry.}

   \keywords{planets and satellites: detection – techniques: photometric, radial velocities – stars: TOI-6457, TOI-6649, TOI-6842A, TOI-7025, TOI-6893A, TOI-5621, TOI-5321, TOI-3678, TOI-5217A, TOI-7176, TOI-6883A, TOI-5893, TOI-5906, TOI-5308}
\titlerunning{Detection and characterization of 14 TOIs}
\authorrunning{N. Heidari \& G.~H\'ebrard et al.}
\maketitle

%

\section{Introduction}

Hot Jupiters were among the first exoplanets identified through radial-velocity (RV) and transit surveys, and they remain some of the most extensively studied planetary systems to date. Yet several of their physical properties continue to challenge current theoretical models. One of the most notable open questions concerns their origin: whether these planets formed in situ or instead migrated inward from beyond the ice line \citep[e.g.,][]{dawson2018origins,fortney2021hot}. Another long-standing puzzle is the phenomenon of radius inflation. Hot Jupiters with equilibrium temperatures above roughly 1000 K display radii larger than those predicted by standard irradiative and structural evolution models. The proposed explanations include tidal mechanisms \citep[e.g.,][]{leconte2010tidal}, delayed cooling and contraction \citep[e.g.,][]{burrows2007possible}, and additional internal heating processes \citep[e.g.,][]{chabrier2007heat,sarkis2021evidence,batygin2011evolution}.

Warm Jupiters, which orbit at larger distances and therefore experience weaker tidal forces, reduced atmospheric erosion, and lower equilibrium temperatures, provide an essential comparison sample for assessing these mechanisms. As such, warm Jupiters offer crucial diagnostics for distinguishing between different evolutionary scenarios and for advancing our understanding of giant-planet formation and migration.

A radius-inflation problem also arises at the lower end of the main sequence, though it is driven by different physical mechanisms than those invoked for hot Jupiters. Low-mass stars exhibit radii that exceed the predictions of stellar evolution models by approximately 5–10\% \citep[e.g.,][]{kraus2011mass,birkby2012discovery,dittmann2017discovery}. The source of this discrepancy remains under debate, with enhanced magnetic activity \citep[e.g.,][]{chabrier2007evolution,kraus2011mass} and non-solar metallicity effects \citep[e.g.,][]{berger2006first,lopez2007correlation} proposed as possible explanations. Understanding this problem is particularly important because low-mass stars constitute the most common stellar population in the Galaxy \citep{kroupa2001variation}. Moreover, improving the accuracy of their fundamental parameters directly enhances the precision with which the properties of their orbiting planets are inferred. Systems containing such stars in eclipsing binaries offer an especially valuable framework for investigating these effects, since their masses, radii, and luminosities can be determined with high precision.

In this work, we present the discovery and characterization of fourteen \textit{TESS} Objects of Interest (TOIs) observed as part of the SOPHIE–\textit{TESS} follow-up program, including both hot and warm Jupiter systems as well as giant-planet candidates revealed to be low-mass stellar companions. Section~\ref{observation} describes the observational datasets, including \textit{TESS} photometry, RV follow-up observations, high-angular-resolution imaging, and complementary ground-based photometry. Sections~\ref{stellar} and~\ref{ident} present, respectively, the characterization of the host stars and the photometric and spectroscopic analyses of the systems. In Section~\ref{Astrometry}, we combine RV measurements with Hipparcos/\textit{Gaia} astrometry to investigate the presence of additional companions within these systems. Finally, Section~\ref{Discussion} discusses our findings, and Section~\ref{summary} summarizes our conclusions.

\section{Observational datasets}
\label{observation}
\subsection{\textit{TESS} photometry}
\label{data_tess}

All the stars analyzed in this study were observed by NASA’s Transiting Exoplanet Survey Satellite \citep[TESS;][]{ricker2015transiting}. Our analysis includes \textit{TESS} observations collected up to April 2025, spanning Sectors 1 through 90. Some of the targets are expected to be re-observed in future \textit{TESS} sectors. To ensure the reliability of the light curves, data points with NaN values or nonzero data-quality flags were masked. We used both short-cadence data with 20- and 120-second exposures, as well as full-frame image (FFI) data with exposure times ranging from 200 to 1800 seconds. The light curves were retrieved using the \texttt{lightkurve} package \citep{collaboration2018lightkurve} from the Mikulski Archive for Space Telescopes (MAST\footnote{\url{https://mast.stsci.edu/portal/Mashup/Clients/Mast/Portal.html}}
). Details of the \textit{TESS} observations are summarized in Table~\ref{tess_obs}.

The data products analyzed in this work were primarily processed using the Presearch Data Conditioning–Simple Aperture Photometry (PDC-SAP) pipeline \citep{Stumpe2012,smith2012kepler,stumpe2014multiscale,2020RNAAS...4..201C}, provided by the Science Processing Operations Center \citep[SPOC;][]{jenkins2016tess} at NASA Ames Research Center. These data have been corrected for long-term systematic trends, including instrumental artifacts and flux dilution. When the SPOC pipeline PDC-SAP light curves were not available, we used SAP light
curves from the MIT Quick Look Pipeline \citep[QLP;][]{huang2020photometry,2020RNAAS...4..206H} which is corrected from flux contamination. This was the case for TOI-5217 in Sectors 40, 41, and 54, for TOI-7176 in Sector 83, and for TOI-6883 in Sector 81. Both the SPOC and QLP pipelines search their extracted light curves for transit-like signals, which are then vetted by the \textit{TESS} Science Office. Objects that pass all vetting procedures are assigned a \textit{TESS} TOI designation. All objects analyzed in this study have been assigned TOI numbers.

For all targets studied here, we examined neighboring stars within the photometric aperture, considering sources with magnitude contrasts of up to $\sim$6 mag. All detected nearby stars have TIC identifiers; therefore, their flux contamination is already accounted for in the SPOC and QLP light-curve processing. No additional sources were detected in our high-resolution imaging (see Sect.~\ref{image}), so we did not apply any further dilution factor in our analysis. Among the nearby stars, the only close companion located within both the \textit{TESS} and SOPHIE apertures is \textit{Gaia}~DR3~2788767177766645120 (TIC~611378244), situated 2.5~arcsec from TOI-5321 and 4.7 magnitudes fainter. This star is sufficiently faint that it is not expected to significantly contaminate the SOPHIE RVs.

We additionally compared our list of planet-candidate hosts with the catalog of binary stars from \textit{Gaia}~DR3 presented by \citet{eeles2025high}. Three targets—TOI-5217, TOI-6842, and TOI-6893—have nearby stars with similar parallaxes and proper motions, consistent with bound companions. Specifically, \textit{Gaia}~DR3~2031448238910493312 (TIC~215013974) is located 5.5~arcsec from TOI-5217 and 0.6 magnitudes fainter; \textit{Gaia}~DR3~2589776773097516928 (TIC~58858007) is 5.0~arcsec from TOI-6842A and 3.1 magnitudes fainter; and \textit{Gaia}~DR3~2741377638198866304 (TIC~611078492) is 4.12~arcsec from TOI-6893 and 6.7 magnitudes fainter. In addition, \citet{conzo2025characterization} reported that TOI-6883 is part of a bound binary system. Its companion, \textit{Gaia}~DR3~4233021429372256512 (TIC~393818340), is 1.5~mag fainter and separated by approximately 6.5~arcsec. As mentioned above, all these contaminations have already been corrected by the SPOC and QLP light-curve processing and are therefore not expected to have any significant impact on our analysis.

\subsection{SOPHIE spectroscopy}
\label{sophie_observation}

We observed the stars analyzed in this work using the SOPHIE high-resolution spectrograph on the 1.93-m telescope at the Observatoire de Haute-Provence \citep{perruchot2008sophie, bouchy2013sophie+}. SOPHIE offers a resolving power of $R \approx 75{,}000$ in high-resolution mode and $R \approx 40{,}000$ in high-efficiency mode. All targets were observed in simultaneous sky mode to monitor potential moonlight contamination. Depending on the stellar magnitude and the weather conditions, the average exposure time ranged from 542 to 1934 s, yielding an average signal-to-noise ratio (S/N) per pixel at 550 nm between 22 and 49. Details of the individual star observations are provided in Table~\ref{rv_ob}.

All spectra were extracted using the SOPHIE Data Reduction System \citep[DRS;][]{bouchy2009sophie}, along with the optimized procedure described in \cite{heidari2022overcoming} and  \cite{heidari2024sophie}.We used a G2 numerical mask to derive the RVs \citep{baranne1996elodie, pepe2002coralie}.  Finally, we removed suspicious measurements from our dataset. Specifically, we excluded data points that did not meet the following criteria: 1. the uncertainty on the RV measurement must be less than three times its mean value for that target; 2. the S/N  per pixel at 550\,nm must exceed half its mean value for that target. With applying these conditions, we removed 11 data points from all the stars observations. The final RVs are provided in the supplementary material.

\subsection{FEROS spectroscopy}
\label{FEROS_observation}

In addition to SOPHIE, we also observed TOI-5893, TOI-5321 and TOI-6893 with the FEROS spectrograph \citep{feros} in the context of the Warm gIaNts with tEss collaboration
\citep[WINE,][]{wine1,wine2,wine3}. FEROS is fiber-fed to the MPG 2.2m
telescope at the ESO La Silla Observatory, has a resolving power of 48,000,
and covers the whole optical spectral range. All observations with FEROS were done using the simultaneous
wavelength calibration technique. FEROS data was processed from raw images
with the automated \texttt{ceres} pipeline \citep{ceres}, which also
computes precision RVs using the cross-correlation technique
using binary masks as templates. A summary of the observations is presented in Table~\ref{rv_ob}, and final RVs are provided in the supplementary material.

\subsection{TRES spectroscopy}
\label{tres_observation}
TOI-5217A, TOI-5308, TOI-5321, TOI-6457, TOI-6883A, TOI-6893, TOI-7025, and
TOI-7176 were also observed with the Tillinghast Reflector Echelle
Spectrograph \citep[TRES;][]{gaborthesis} located at the Fred Lawrence Whipple
Observatory (FLWO) in Arizona, USA. TRES is mounted on the 1.5m Tillinghast
Reflector telescope. TRES has a resolving power of about 44,000 and operates in
the wavelength range 390-910nm. The spectra were extracted as described in \cite{buchhave2010}. Average exposure times ranged from 207-1029 s
resulting in average S/N per resolution element of 27-43
for the TOIs . A multi-order spectral analysis was performed by cross
correlating each spectrum, order by order, against a co-added spectrum used
as a template to determine the relative velocities. The specific details
for each target are listed in Table~\ref{rv_ob}.

\subsection{PLATOSpec spectroscopy}
\label{PS_observation}
TOI-6893 was observed with PLATOSpec \citep{platospec}, which is mounted on the 1.52m telescope at the ESO La Silla Observatory, in Chile. PLATOSpec is a high-resolution fiber-fed echelle spectrograph designed to support space missions such as TESS, PLATO and Ariel, covering the spectral region between 380 and 680 nm with a resolving power of R= 70,000 \citep[e.g.][]{ps1,ps2,ps3}. PLATOSpec data are processed by a fully automated dedicated pipeline based on the \texttt{ceres} package \citep{ceres}. We adopted an exposure time of 900 s resulting in spectra with S/N per resolution element ranging from 80 to 150 depending on the observing conditions.

\subsection{High angular resolution observations}
\label{image}
Except TOI-7176 and TOI-6457, we observed all other targets using speckle interferometry with the Speckle Polarimeter (SPP), a facility instrument on the 2.5 m telescope at the Caucasian Observatory of the Sternberg Astronomical Institute (SAI) of Lomonosov Moscow State University. We used an Electron Multiplying CCD Andor iXon 897 \citep{safonov2017speckle} as a detector for the observations of TOI-3678 and a CMOS Hamamatsu ORCA-quest \citep{strakhov2023speckle} for the rest of the objects. The pixel scale in both cases was 20.6 mas/pixel. Dates of observations and filters used for each object are listed in Table \ref{high_image}. The power spectrum was estimated from 4000 frames per star with 30 ms exposure time. The atmospheric dispersion compensator was employed. We did not detect a stellar companion for any of the objects, the resulting sensitivity limits are presented in Table \ref{high_image} and the resulting sensitivity curves are shown in Fig. \ref{image_high}.

Additionally, we observed TOI-7176 and TOI-6457 with the ‘Alopeke speckle imager mounted on the Gemini North 8.1-m telescope \citep{2021FrASS...8..138S}. ‘Alopeke provides simultaneous speckle imaging in two bands (562 nm and 832 nm), producing reconstructed images and robust contrast limits for the detection of close companions. For TOI-7176, we obtained seven sets of 1000 exposures with an individual integration time of 0.06\,s, while five such sets were collected for TOI-6457. The data were processed using the standard reduction pipeline described by \citet{2011AJ....142...19H}. Figure~\ref{image_high} shows the final 5$\sigma$ contrast curves alongside the reconstructed speckle images at 562 nm and 832 nm, and the resulting sensitivity limits are summarized in Table~\ref{high_image}.

We therefore conclude that no close stellar companions are detected for any of the targets.

\subsection{Ground-based lightcurve Photometry\label{sec:groundtransits}}

The \textit{TESS} pixel scale is $\sim 21\arcsec$ pixel$^{-1}$ and
photometric apertures typically extend out to roughly 1 arcminute,
generally causing multiple stars to blend in the \textit{TESS} photometric
aperture. To determine the true source of the \textit{TESS} detections, we
acquired ground-based time-series follow-up photometry of the fields around
TOI-5217A, TOI-6649, TOI-6842A, TOI-6883A as part of the \textit{TESS}
Follow-up Observing Program \citep[TFOP;][]{collins:2019}\footnote{
https://tess.mit.edu/followup}. A summary of the observations is provided
in Table \ref{table:groundobservatories}. These observations correspond to the planetary candidates with well-constrained orbital period presented in this work (see Sect.~\ref{detection}). TOI-6457 is not included in this sample of follow-up observations.

Eight of the observations were conducted with the Las Cumbres Observatory
Global Telescope \citep[LCOGT;][]{Brown:2013} 0.4\,m and 1.0\,m network
nodes at McDonald Observatory near Fort Davis, Texas, United States (McD),
Cerro Tololo Inter-American Observatory in Chile (CTIO), Teide Observatory
on the island of Tenerife (TEID), and Cerro Tololo Inter-American
Observatory in Chile (CTIO). The 0.4\,m Planewave Delta Rho 350 telescopes
are equipped with a $9576\times6388$ QHY600 CMOS camera having an image
scale of $0\farcs73$ per pixel, resulting in a $114\arcmin\times72\arcmin$
full field of view. We used the optional $30\arcmin\times30\arcmin$ sub
field of view for a faster detector read-out. The 1-m telescopes are
equipped with $4096\times4096$ SINISTRO cameras having an image scale of
$0\farcs389$ per pixel, resulting in a $26\arcmin\times26\arcmin$ field of
view. The images were calibrated by the standard LCOGT {\tt BANZAI}
pipeline \citep{McCully:2018}. Another transit was observed from the
Frustaglia Private Observatory 0.3\,m telescope located near Fregenal de la
Sierra, Spain (RFAC). The telescope is equipped with a $6248\times4176$ QHY
268M camera having an image scale of $0\farcs32$ per pixel, resulting in a
$33\arcmin\times22\arcmin$ field of view. We also observed an intended full
transit of TOI-6649 from the University of Louisville Manner Telescope
0.61\,m located at the Steward observatory near Tucson, AZ (ULMT). However,
based on the revised ephemeris from this work, the data are all
out-of-transit, and were not included in the analysis of the system.

\begin{table*}
\caption{Stellar properties of the stars analysed in this work.}
\centering
\resizebox{1.5\columnwidth}{!}{%
\begin{tabular}{lllll}
\hline
Parameter & TOI-3678 & TOI-5621 & TOI-5893 & TOI-5217A \\
\hline
ID (TIC) & 34551972 & 219467837 & 321790238 & 215018906 \\
ID (\textit{Gaia} DR3) & 116922311314227840 & 1712229330056985216 & 1743373271792149760 & 2031448238910494208 \\
$\alpha$ (h m s)$^{1}$ & 02:56:55.99 & 14:00:15.35 & 21:11:42.35 & 19:39:41.37 \\
$\delta$ (d m s)$^{1}$ & +28:46:17.31 & +75:05:03.91 & +08:58:50.42 & +28:34:20.02 \\
Parallax (mas)$^{1}$ & 2.000$\pm$0.021 & 1.400$\pm$0.010 & 3.037$\pm$0.014 & 1.887$\pm$0.014 \\
T (mag)$^{2}$ & 10.5873$\pm$0.0096 & 12.64$\pm$0.0073 & 11.4879$\pm$0.0093 & 10.9992$\pm$0.0061 \\
G (mag)$^{1}$ & 10.970 & 13.04 & 11.92 & 11.43 \\
J (mag)$^{3}$ & 10.063$\pm$0.021 & 12.051$\pm$0.020 & 10.880$\pm$0.026 & 10.373$\pm$0.041 \\
H (mag)$^{3}$ & 9.868$\pm$0.016 & 11.778$\pm$0.019 & 10.608$\pm$0.022 & 10.105$\pm$0.040 \\
K (mag)$^{3}$ & 9.810$\pm$0.019 & 11.711$\pm$0.019 & 10.532$\pm$0.021 & 10.049$\pm$0.035 \\
W1 (mag)$^{4}$ & 9.751$\pm$0.022 & 11.682$\pm$0.023 & 10.492$\pm$0.022 & 9.585$\pm$0.019 \\
W2 (mag)$^{4}$ & 9.765$\pm$0.021 & 11.715$\pm$0.021 & 10.533$\pm$0.020 & 9.662$\pm$0.018 \\
W3 (mag)$^{4}$ & 9.775$\pm$0.055 & 11.827$\pm$0.167 & 10.318$\pm$0.077 & 9.802$\pm$0.042 \\
W4 (mag)$^{4}$ & 8.182 & 9.303 & 8.862 & 8.611$\pm$0.404 \\
$T_{\mathrm{eff}}$ (K)$^{5}$ & $6777\pm205$ & $5983\pm82$ & $5963\pm68$ & $6142\pm75$ \\

$T_{\mathrm{eff}}$ (K)$^{6}$ &
--- &
--- &
5895$\pm$50 &
6175$\pm$50 \\

{[Fe/H]}$^{5}$ & $-0.03\pm0.22$ & $+0.014\pm0.059$ & $+0.149\pm0.047$ & $+0.299\pm0.053$ \\
{[Fe/H]}$^{6}$ &
--- &
--- &
0.15$\pm$0.08 &
0.39$\pm$0.08 \\

$\log g$ (dex)$^{5}$ & $3.90\pm0.4$ & $4.20\pm0.06$ & $4.38\pm0.03$ & $3.98\pm0.03$ \\
$v\sin i^{5} (km/s)$ & $>~20$ & --- & $3.9\pm1.0$ & $6.6\pm1.0$ \\
$v\sin i^{6} (km/s)$ & --- & --- & $4.3\pm0.5$ & $8.5\pm0.5$ \\

$\log (R'_{HK})$ &
--- &
---&
$-4.4\pm0.1$ &
$-5.2\pm0.1$\\

$P_{\rm rot} (d)^{7}$ &
--- &
--- &
5 $\pm$ 3 &
15 $\pm$ 2 \\

$M_\star (M_\odot)^{5}$ & $1.60\pm0.10$ & $1.198\pm0.072$ & $1.17\pm0.07$ & $1.51\pm0.09$ \\
$R_\star (R_\odot)^{5}$ & $2.35\pm0.14$ & $1.469\pm0.042$ & $1.192\pm0.030$ & $2.309\pm0.072$ \\
\hline
\end{tabular}}
\par\medskip
\resizebox{1.5\columnwidth}{!}{%
\begin{tabular}{llllll}
\hline
Parameter & TOI-5906 & TOI-7025 & TOI-5308 & TOI-6842A & TOI-5321 \\
\hline
ID (TIC) & 276376289 & 219803922 & 422653398 & 58858008 & 243153196 \\
ID (\textit{Gaia} DR3) & 2065995581392009472 & 1439496329781024000 & 2507775019804197760 & 2589776768804421760 & 2788767173470883584 \\
$\alpha$ (h m s)$^{1}$ & 20:53:57.50 & 17:44:29.33 & 01:56:00.30 & 01:22:32.87 & 01:02:44.52 \\
$\delta$ (d m s)$^{1}$ & +42:44:36.70 & +62:33:18.42 & +00:10:47.22 & +13:21:08.02 & +20:03:01.52 \\
Parallax (mas)$^{1}$ & 3.409$\pm$0.014 & 2.309$\pm$0.011 & 2.908$\pm$0.056 & 2.512$\pm$0.022 & 2.172$\pm$0.018 \\
T (mag)$^{2}$ & 11.2922$\pm$0.0061 & 11.368$\pm$0.006 & 11.0208$\pm$0.0074 & 11.5955$\pm$0.0088 & 10.7304$\pm$0.0072 \\
G (mag)$^{1}$ & 11.73 & 11.76 & 11.50 & 12.04 & 11.06 \\
J (mag)$^{3}$ & 10.635$\pm$0.026 & 10.80$\pm$0.02 & 10.377$\pm$0.026 & 10.946$\pm$0.026 & 10.263$\pm$0.021 \\
H (mag)$^{3}$ & 10.312$\pm$0.033 & 10.55$\pm$0.02 & 10.035$\pm$0.022 & 10.653$\pm$0.029 & 10.062$\pm$0.022 \\
K (mag)$^{3}$ & 10.226$\pm$0.018 & 10.494$\pm$0.019 & 9.936$\pm$0.019 & 10.578$\pm$0.026 & 9.992$\pm$0.017 \\
W1 (mag)$^{4}$ & 10.152$\pm$0.023 & 10.459$\pm$0.023 & 9.885$\pm$0.022 & 10.377$\pm$0.023 & 9.973$\pm$0.022 \\
W2 (mag)$^{4}$ & 10.198$\pm$0.020 & 10.496$\pm$0.020 & 9.930$\pm$0.020 & 10.387$\pm$0.021 & 10.002$\pm$0.019 \\
W3 (mag)$^{4}$ & 9.835$\pm$0.062 & 10.403$\pm$0.044 & 9.884$\pm$0.050 & 10.365$\pm$0.091 & 9.892$\pm$0.047 \\
W4 (mag)$^{4}$ & 8.747 & 9.305 & 8.601 & 8.873 & 8.656$\pm$0.320 \\

$T_{\mathrm{eff}}$ (K)$^{5}$ & $5959\pm68$ & $6164\pm67$ & $5820\pm97$ & $5786\pm64$ & $6530\pm140$ \\

$T_{\mathrm{eff}}$ (K)$^{6}$ &
5940$\pm$50 &
6020$\pm$50 &
5650$\pm$50 &
5720$\pm$50 &
6590$\pm$50 \\

{[Fe/H]}$^{5}$ & $-0.185\pm0.048$ & $0.08\pm0.05$ & $+0.188\pm0.070$ & $-0.007\pm0.045$ & $-0.22\pm0.13$ \\

{[Fe/H]}$^{6}$ &
-0.17$\pm$0.08 &
-0.03$\pm$0.08 &
0.26$\pm$0.08 &
-0.02$\pm$0.08 &
0.08$\pm$0.08 \\

$\log g$ (dex)$^{5}$ & $4.38\pm0.03$ & $4.21\pm0.03$ & $4.20\pm0.04$ & $4.23\pm0.04$ & $3.99\pm0.16$ \\
$v\sin i^{5} (km/s)$ & $4.0\pm1.0$ & $5.3\pm1.0$ & $6.5\pm1.0$ & $3.3\pm1.0$ & $15.1\pm1.0$ \\
$v\sin i^{6} (km/s)$ & $4.2\pm0.5$ & $6.8\pm0.5$ & $8.0\pm0.5$ & $3.4\pm0.5$ & $17.3\pm0.5$ \\

$\log\left(R'_{\mathrm{HK}}\right)$ & $-4.6 \pm 0.1$ & $-4.6 \pm 0.1$ & $-5.1 \pm 0.1$ & $-4.5 \pm 0.1$ & $-5.4 \pm 0.1$\\

$P_{\rm rot} (d)^{7}$ &
12 $\pm$ 4 &
11 $\pm$ 3 &
57 $\pm$ 7 &
7 $\pm$ 3 &
29 $\pm$ 3 \\

$M_\star (M_\odot)^{5}$ & $1.00\pm0.06$ & $1.27 \pm 0.08$& $1.200\pm0.072$ & $1.135\pm0.068$ & $1.39\pm0.08$ \\
$R_\star (R_\odot)^{5}$ & $1.239\pm0.038$ & $1.534 \pm 0.035$& $1.648\pm0.066$ & $1.420\pm0.035$ & $1.972\pm0.095$ \\
\hline
\end{tabular}}
\par\medskip
\resizebox{1.5\columnwidth}{!}{%
\begin{tabular}{llllll}
\hline
Parameter & TOI-6649 & TOI-6883A & TOI-6893A & TOI-6457 & TOI-7176 \\
\hline
ID (TIC) & 258940847 & 393818343 & 272125414 & 328502522 & 44161614\\
ID (\textit{Gaia} DR3) & 2806600569174401920 & 4233021433662981760 & 2741377642494388864 & 890269449705929344& 1890072101738525696\\
$\alpha$ (h m s)$^{1}$ & 00:36:49.97 & 20:41:10.01 & 00:11:34.25 & 07:14:14.81 & 23:01:23.96\\
$\delta$ (d m s)$^{1}$ & +24:16:04.05 & +03:38:17.82 & +04:46:01.96 &+33:04:13.3& +32:56:58.89 \\
Parallax (mas)$^{1}$ & 4.950$\pm$0.016 & 10.612$\pm$0.015 & 6.0084$\pm$0.0194 &3.781$\pm$0.022 & 14.471 $\pm$0.021 \\
T (mag)$^{2}$ & 9.7664$\pm$0.0073 & 8.9797$\pm$0.006 & 8.189$\pm$0.006 & 10.4475 $\pm$0.0061 & 10.4361$\pm$ 0.0061\\
G (mag)$^{1}$ & 10.13 & 9.44 & 8.59 & 10.84 & 11.20\\
J (mag)$^{3}$ & 9.266$\pm$0.021 & 8.401$\pm$0.019 & 7.659$\pm$0.026 &9.925 $\pm$0.027 &9.371$\pm$0.024\\
H (mag)$^{3}$ & 9.097$\pm$0.030 & 8.094$\pm$0.040 & 7.515$\pm$0.046 & 9.650$\pm$0.032 & 8.780$\pm$0.018 \\
K (mag)$^{3}$ & 9.000$\pm$0.019 & 8.052$\pm$0.033 & 7.349$\pm$0.020 & 9.635 $\pm$0.02 &8.646$\pm$0.013 \\
W1 (mag)$^{4}$ & 8.984$\pm$0.023 & 7.933$\pm$0.033 & 7.214$\pm$0.035 & 9.592$\pm$0.030& 8.566$\pm$0.038\\
W2 (mag)$^{4}$ & 9.016$\pm$0.019 & 8.003$\pm$0.021 & 7.345$\pm$0.021 & 9.638$\pm$0.030 & 8.643$\pm$0.037\\
W3 (mag)$^{4}$ & 9.049$\pm$0.032 & 8.026$\pm$0.027 & 7.346$\pm$0.018 & 9.559$\pm$0.051 & 8.560$\pm$0.039 \\
W4 (mag)$^{4}$ & 9.100$\pm$0.518 & 7.648 & 7.103$\pm$0.112 & 8.249 & 8.481$\pm$0.376 \\

$T_{\mathrm{eff}}$ (K)$^{5}$ & 6397$\pm$69 & 5831$\pm$63 & 6239$\pm$69 &6230$\pm$67  &4484$\pm$171 \\

$T_{\mathrm{eff}}$ (K)$^{6}$ &
6280$\pm$50 &
5730$\pm$50 &
--- &
6060$\pm$50 &
4400$\pm$100 \\

$[Fe/H]^{5}$ & $+0.181\pm0.047$ & $+0.275\pm0.043$ & $0.25\pm0.05$ & $0.23\pm0.05$ & $-0.08\pm0.09$ \\

{[Fe/H]}$^{6}$ &
0.12$\pm$0.08 &
0.19$\pm$0.08 &
--- &
0.19$\pm$0.08 &
-0.12$\pm$0.08 \\

$\log g$ (dex)$^{5}$ & 4.37$\pm$0.03 & 4.42$\pm$0.03 & 3.88$\pm$0.03 & 4.31$\pm$0.03 &4.52$\pm$ 0.12 \\

$v\sin i^{5} (km/s)$ & 6.5$\pm$1.0 & 3.8$\pm$1.0 & 8.5$\pm$1.0 &4.9$\pm$1.0 & --- \\
$v\sin i^{6} (km/s)$ & $9.1\pm0.5$ & $3.6\pm0.5$ & --- & $7.3\pm0.5$ & $2.4\pm0.5$ \\

$\log\left(R'_{\mathrm{HK}}\right)$ & $-5.2\pm0.1$ & $-5.2\pm0.1$ & $-5.3\pm0.1$ & $-4.9\pm0.1$ & --- \\

$P_{\rm rot} (d)^{7}$ (d) &
11 $\pm$ 1 &
52 $\pm$ 6 &
28 $\pm$ 3 &
8 $\pm$ 1 &
--- \\

$M_\star (M_\odot)^{5}$ & 1.32$\pm$0.08 & 1.14$\pm$0.07 &1.39 $\pm$ 0.08 & 1.30 $\pm$ 0.08&  0.70$\pm$0.04\\
$R_\star (R_\odot)^{5}$ & 1.374$\pm$0.031 & 1.068$\pm$0.024 & 2.524 $\pm$ 0.073  & 1.384 $\pm$ 0.035 & 0.700$\pm$0.054 \\
\hline
\end{tabular}}
\tablefoot{$^{1}$ \textit{Gaia} DR3, $^{2}$ \textit{TESS} Input Catalog,
$^{3}$ 2MASS, $^{4}$ WISE, $^{5}$ this work, using SOPHIE spectroscopic
parameters (see Sect.~\ref{stellar}), $^{6}$ this work, using TRES spectra,
$^{7}$ rotation periods inferred from the measured $\log R'_{\rm HK}$ values.}
\label{star3}
\end{table*}

For all observations, differential photometric data were extracted using
{\tt AstroImageJ} \citep{Collins:2017}. We applied parametric detrending to
each lightcurve by considering the improvement to the transit model fit
after iterating over airmass (AM), time (TM), full-width-half-maximum
(FWHM), sky background (SKY), target x-position (X), target y-position (Y),
and total comparison star ensemble counts (TOTC) as a proxy for common mode
systematics. The best zero, one, or two parametric detrend vectors were
retained if joint linear fits to the vectors plus a transit model fit
decreased the Bayesian Information Criterion (BIC) of the fit by at least
two per detrend parameter.

Additionally, we observed a transit of TOI-7025.01 in J-band with the Wide-field Infrared Camera
(WIRC; \citealt{Wilson2003}) on the 5.1 m Hale Telescope at Palomar Observatory,
California, USA. To improve our observing efficiency and photometric precision, we
used a beam-shaping diffuser with a FWHM of 3\arcsec
\citep{Stefansson2017, Vissapragada2020}. We reduced the observations following \citet{Vissapragada2020} and detrended the target-star light curve as described in \citet{GomezBarrientos2025}.

We observed one egress of TOI‑5893.01 on the night of 28 May 2026 with the Exoplanets in Transits and their Atmospheres facility \citep[ExTrA;][]{Bonfils2015}, located at La Silla Observatory in Chile. ExTrA is a low-resolution near-infrared (0.85 to 1.55~$\mu$m) multi-object spectrograph fed by three 60~cm telescopes. We observed with the three telescopes and used 8$\arcsec$ aperture fibers and the lowest-resolution mode ($R$$\sim$20) of the spectrograph with an exposure time of 60~seconds. The ExTrA data were reduced and the photometry extracted using custom software.

The detrended lightcurves are included in the
joint modeling described in Section \ref{ident}. These ground-based
observations confirm that the transit signals originate from the target
stars TOI-6842A, TOI-5217A, TOI-6649, TOI-6883A, and TOI-7025, and that contamination from nearby sources within the \textit{TESS} apertures is not significant.

\section{Stellar properties}
\label{stellar}

We began our stellar parameter analysis by estimating the projected rotational velocity ($v \sin i$) from the FWHM of the SOPHIE cross-correlation profile using the \citet{boisse2010sophie} calibration. This calibration is valid only for stars with a $B-V$ colour index in the range 0.4--1.3 and for $v \sin i$ below 20\,km\,s$^{-1}$. Therefore, $v\sin i$ could not be estimated for TOI-7176 and TOI-5621 because their $B-V$ colours lie outside the validity range of the calibration. For TOI-3678, whose projected rotational velocity exceeds the calibrated range, we report only a lower limit of $v\sin i > 20$~km\,s$^{-1}$ rather than a precise measurement. To derive the effective temperature ($T_{\rm eff}$), metallicity ([Fe/H]), and surface gravity ($\log g$), we applied the methodologies described in \citet{sousa2011spectroscopic}, \citet{santos2013sweet}, \citet{SWEET-Cat}, and \citet{sousa2018sweet} to the combined SOPHIE spectra of each star. However, the spectroscopic determination of $T_{\rm eff}$, [Fe/H], and $\log g$ could not be reliably performed for TOI-3678 and TOI-5321 because their high projected rotational velocities ($v \sin i >~20$ km/s and $15.1 \pm 1.0$ km/s, respectively) produce substantially broadened spectral lines. The resulting spectroscopic parameters, together with the other derived stellar properties, are reported in Tables~\ref{star3}.

Additionally, we independently derived $v\sin i$, [Fe/H], and $T_{\rm eff}$ from the TRES spectra using the Stellar Parameter Classification (SPC; \citealt{Buchhave2012}) tool. For the targets with measurements from both instruments, the stellar parameters derived from SOPHIE and TRES are in good agreement, with most targets consistent within $1\sigma$ and all consistent within approximately $2\sigma$.

To improve the S/N in the Ca\,{\sc ii} H\&K region, we co-added all available SOPHIE spectra for each target prior to measuring the chromospheric activity index. The $\log(R'_{\rm HK})$ values were then derived following the methodologies of \citet{noyes1984rotation} and \citet{boisse2010sophie}. The calibration is formally valid for main-sequence stars with $0.43\lesssim B-V \lesssim 1.2$ \citep{boisse2010sophie}, corresponding approximately to effective temperatures of about $4200 \lesssim T_{\rm eff} \lesssim 6550$\,K \citep{Pecaut2013}. The resulting $\log(R'_{\rm HK})$ values are listed in Table~\ref{star3}, except for TOI-3678, whose effective temperatures lie outside the validity range of the calibration, and TOI-7176 and TOI-5621, whose $B-V$ colours ($1.7 \pm 0.4$ and $0.3 \pm 0.1$, respectively) fall outside the calibrated range. The measured $\log(R'_{\rm HK})$ values span a wide range of chromospheric activity levels. TOI-5217A, TOI-5308, TOI-6649, TOI-6883A, TOI-5321, TOI-6893A, and TOI-6457 are chromospherically inactive, with $\log(R'_{\rm HK}) < -4.75$ \citep{henry1996survey}. TOI-5906, TOI-7025, TOI-6842A, and TOI-5893 are chromospherically active, with $\log(R'_{\rm HK})$ values between $-4.4$ and $-4.6$. The potential impact of stellar activity on the derived planetary parameters for these active systems is discussed in Sect.~\ref{detection}.

In addition, we estimated stellar rotation periods from the measured $\log(R'_{\rm HK})$ values using the empirical calibrations of \citet{mamajek2008improved}. The resulting rotation periods are listed in Table~\ref{star3}. For TOI-5321, TOI-5308, TOI-6883A, and TOI-6893A, these activity-based estimates are not fully consistent with the rotation period dervied using reported stellar radii and $v\sin i$. The rotation periods of these stars should therefore be regarded as uncertain.

To obtain empirical measurements of the stellar radii, we followed the procedure of \citet{Stassun:2016,Stassun:2017,Stassun:2018} and analyzed the broadband spectral energy distribution (SED) of the stars together with {\it Gaia} DR3 parallaxes \citep[with no systematic offset applied; see, e.g.,][]{StassunTorres:2021}. We pulled the available broadband photometry, which for most of the stars included the $JHK_S$ magnitudes from {\it2MASS} \citep{skrutskie2006two}, the $G_{\rm BP} G_{\rm RP}$ magnitudes from {\it Gaia} \citep{gaia2023gaia}, and the W1--W3 magnitudes from {\it WISE} \citep{wright2010wide}. We also utilized the absolute flux calibrated spectrophotometry from {\it Gaia}, when available. We note that for TOI-5217A, the {\it WISE} photometry was strongly contaminated by the nearby star \textit{Gaia} DR3 2031448238910493312, located 5.5~arcsec away from TOI-5217A with a magnitude difference of~0.6. Consequently, these data points were removed.

We fitted each star’s SED using PHOENIX stellar atmosphere models \citep{Husser:2013}, adopting the spectroscopic values of $T_{\rm eff}$ and [Fe/H] as primary parameters when available. If these values were not available, we considered them as free parameters. The extinction, $A_V$, was fitted and limited to the line-of-sight maximum from the dust maps of \citet{Schlegel:1998}. The stellar radius, $R_\star$, was derived from the Stefan-Boltzmann relation. In addition, we estimated the stellar mass using the empirical relations of \citet{Torres:2010}. The resulting parameters are listed in Tables~\ref{star3}.

As an additional check, we compared these values with results obtained using an alternative modeling method in \texttt{EXOFASTv2}, performed simultaneously with the planetary fit (see Sect.~\ref{exofast_results}), except for TOI-5217A, TOI-5321, TOI-6893A, TOI-7025, and TOI-3678 (see Sect. \ref{exofast_results} for more information). Overall, the derived stellar masses and radii agree within about $\sim$1$\sigma$. A notable exception is the mass of TOI-5906AB, a double-lined spectroscopic binary (SB2), which is modeled in Sect.~\ref{exofast_results} using a simultaneous two-star fit.

Finally, we investigated the chemical and kinematic properties of the host stars to assess their membership in the thin disk, thick disk, or halo of the Milky Way. Because Galactic environment can influence planet formation \citep{Nielsen_Planet_Formation_2023} and planet occurrence rates \citep{Johnson_Giant_Occurrence_Metallicity_2010,Hallatt_thickdisk_planets_2025}, identifying Galactic membership enables demographic studies and exploration of trends between stellar and planetary properties \citep[e.g.,][]{Thorngren_mass_metallicity_giants_2016}.

Following \citet{Johnson_Soderblom_galactic_space_velocity_1987} and \citet{Bensby_lsr_2003}, we computed the Galactic space velocities $(U, V, W)$ for all 15 targets using \textit{Gaia} DR3 astrometry and radial velocities \citep{Gaia_DR3_Release_Summary_2023}. These velocities were corrected to the Local Standard of Rest using the solar motion from \citet{Almeida_Fernandes_Stellar_Ages_LSR_2018}. Membership probabilities were then calculated following \citet{Bensby_lsr_2003} and \citet{Reddy_Thick_Disk_2006}, adopting the velocity dispersions and normalization fractions from \citet{Chen_PAST_I_2021}. TOI-6457 has a 71\% probability of belonging to the thin disk, while TOI-5621 is most likely a thick-disk star (88\%). The remaining 13 targets are consistent with thin-disk membership, each with probabilities exceeding 90\%.


\section{Detection and characterization of the systems}
\label{ident}

\subsection{First evaluation of the planetary candidates}
\label{detection}

\begin{figure*}[h!]
\centering
\includegraphics[width=1.5\columnwidth]{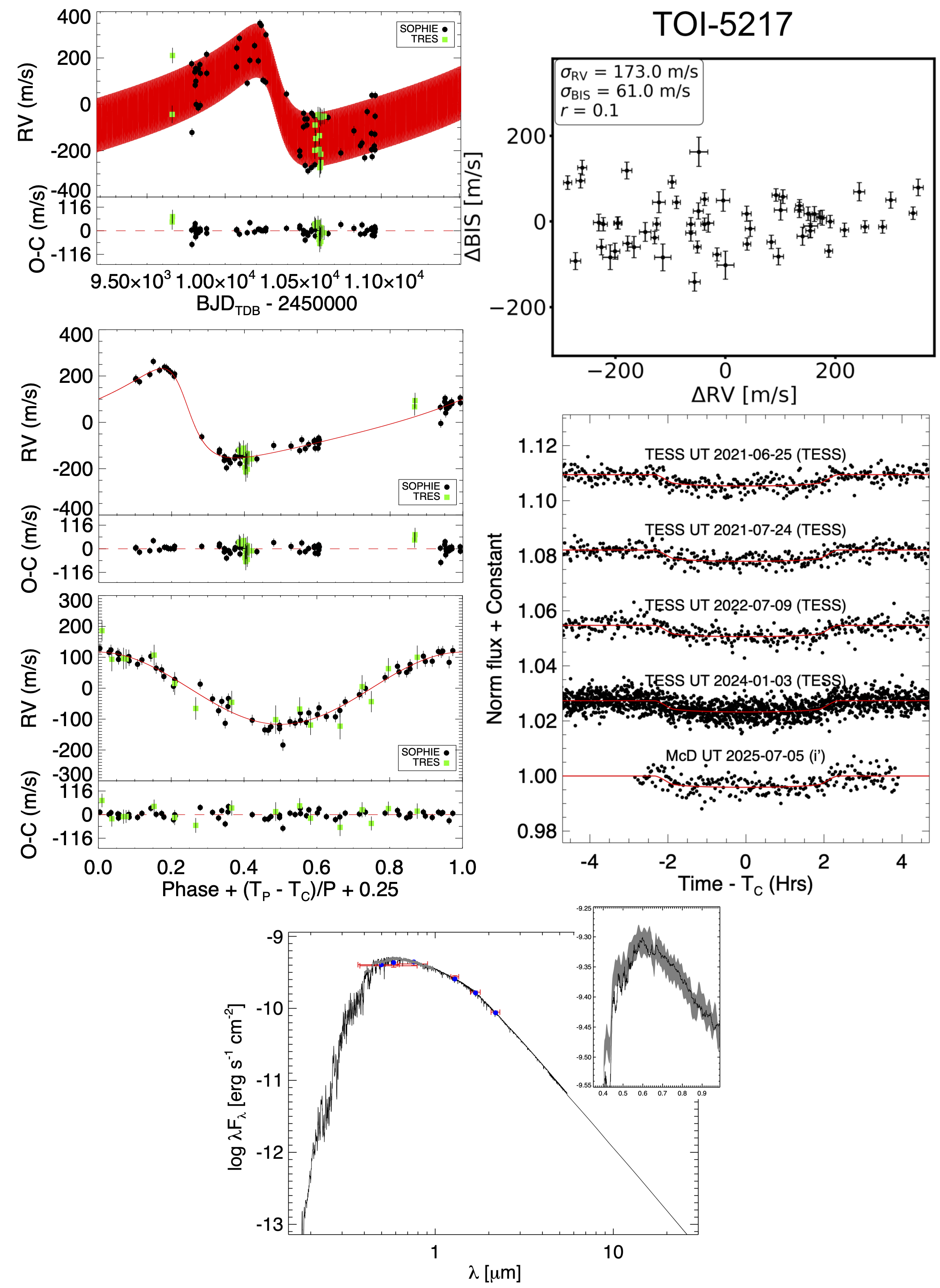}
\caption{\footnotesize{Summary of the spectroscopic, photometric, and SED analyses for TOI-5217A. \emph{Top-left}: RV time series with the best-fit Keplerian model overplotted in red; residuals are shown below. \emph{Second-left}: Phase-folded RV curve for TOI-5217Ac with the best-fit model (red line) and residuals. \emph{Third-left}: Phase-folded RV curve for the TOI-5217Ab with the best-fit model (red line) and residuals. \emph{Top-right:} Bisector span versus RV. \emph{Second-right:} \textit{TESS} and ground-based light curves, phase-folded on the best-fit period and time of conjunction, with their respective transit models overplotted in red. \emph{Bottomn:} SED analyses. Red symbols represent the observed photometric measurements, where the horizontal bars represent the effective width of the broadband photometry. Blue symbols are the model fluxes from the best-fit PHOENIX atmosphere model (black). The absolute flux-calibrated {\it Gaia \/} spectrophotometry is shown as the gray swathe. TOI-5217A\,b has an orbital period of $P = 4.3712411 \pm 0.0000082$~d, a radius of $R_{\rm p} = 1.385^{+0.056}_{-0.055}\,R_{\rm J}$, and a mass of $M_{\rm p} = 1.282\pm0.060\,M_{\rm J}$. The outer companion, TOI-5217A\,c, has a minimum mass of approximately $13\,M_{\rm J}$ on an eccentric orbit ($e = 0.529\pm0.017$), while its orbital period remains poorly constrained.}}
\label{image_5217}
\end{figure*}

We began our analysis by performing an RV-only fit, adopting wide uniform priors for all parameters. For all targets except TOI-5217A, a single-Keplerian model provided an adequate description of the data. In the case of TOI-5217A (see Fig.~\ref{image_5217}), the RVs clearly show both long- and short-period variations, and we therefore adopted a two-Keplerian model. The purpose of this initial analysis was to compare the time of conjunction ($T_c$) and orbital period derived from the RV-only fit with the transit epoch and period measured by SPOC and QLP, as a diagnostic of whether the RV and photometric signals originate from the same astrophysical source.

Except for five cases (TOI-5906.01, TOI-5893.01, TOI-6893.01, and TOI-7176.01), the periods and $T_c$ values derived from the RV-only models are consistent with the \textit{TESS} measurements, indicating that the RV and photometric signals originate from the same source. These five systems are discussed in more detail below.

For TOI-5906.01, the derived period is consistent with the SPOC values, but the RV-derived $T_c$ coincides with the expected timing of a secondary eclipse rather than the primary transit. This timing, combined with the large RV semi-amplitude ($K \simeq 36{,}440$~m~s$^{-1}$) and high orbital eccentricity ($e = 0.5$), strongly suggests that TOI-5906.01 is an eccentric binary system in which the primary transit is undetected \citep[see discussion in][]{santerne2013contribution}. The event observed in the \textit{TESS} data therefore corresponds to the secondary eclipse of TOI-5906~B. This configuration is similar to those of KOI-419AB, KOI-698AB, and TOI-4168AB, as discussed by \citet{santerne2012sophie} and \citet{heidari2025characterization}. We also note that the system is a SB2, but the secondary cross-correlation function (CCF) is significantly weaker than the primary. The contamination of the primary CCF by the secondary is therefore negligible, and we did not attempt to separate the two components.

For three systems—TOI-5893.01, TOI-6893.01, and TOI-7176.01—\textit{TESS} detected a single high-S/N transit, and no orbital period was reported by the SPOC pipeline. Our RV-only analysis reveals a well-defined periodic signal for each target: TOI-5893.01 exhibits a period of $40.1 \pm 0.2$~d with a semi-amplitude of $K = 62 \pm 5$~m~s$^{-1}$; TOI-6893.01 shows a period of $56.3 \pm 0.1$~d and $K = 27.2 \pm 2.4$~m~s$^{-1}$; and TOI-7176.01 has a period of $47.8 \pm 0.2$~d with $K = 43 \pm 5$~m~s$^{-1}$. In all three cases, the transit midpoints inferred from the RV models are consistent with the \textit{TESS} transit timings, indicating that the photometric and RV signals originate from the same object. For TOI-5893.01, we further validated this interpretation by obtaining ground-based photometric observations at the transit times predicted from the RV solution, successfully detecting a second transit and significantly improving the precision of the orbital period. Finally, joint RV–photometric modeling (see Sect.~\ref{join-model}) confirms that the inferred masses and radii fall within the planetary regime, and no additional significant periodicities are detected in the RV residuals.

For TOI-5621.01, TOI-5321.01, TOI-5308.01, and TOI-3678.01, this initial evaluation showed that the measured RV semi-amplitudes are approximately 6.8, 9.6, 9.2, and 13.7~km~s$^{-1}$, respectively. These large amplitudes are incompatible with planetary-mass companions, pointing instead to low-mass stellar companions. We therefore conclude that these four systems host transiting low-mass stars rather than exoplanets (see Sect.~\ref{join-model} for a full analysis), as TOI-5906 discussed above.

In contrast, the RV semi-amplitudes of TOI-6649.01, TOI-6842.01, TOI-5217.01, TOI-7025.01, TOI-6457.01, and TOI-6883.01 are $67 \pm 4$, $37 \pm 3$, $117 \pm 5$, $77\pm 15$, $24 \pm 6$, and $410 \pm 2$~m~s$^{-1}$, respectively. These candidates fall within the planetary regime and show RV signals whose times of conjunction and orbital periods that are consistent with the \textit{TESS} transit ephemerides, motivating a detailed investigation of potential false-positive scenarios. We present that analysis below, for those 6 objects as well as for the 3 single-transit planets reported above.

To evaluate whether the observed planetary transiting candidates could be produced by blended or background eclipsing binaries located within the \textit{TESS} photometric aperture, we employed Equation~4 from \citet{2019ApJ...881L..19V}. This equation estimates the maximum brightness contrast of a contaminating source capable of reproducing the measured transit depths. At the $3\sigma$ level, a blended eclipsing star would need to satisfy the following constraints: $\Delta m \leq 7$~mag for TOI-6649.01, $\Delta m \leq 4$~mag for TOI-7025.01, $\Delta m \leq 3$~mag for TOI-5893.01, $\Delta m \leq 2$~mag for TOI-6893.01 and TOI-7176.01, and $\Delta m \leq 1$~mag for TOI-6457.01, TOI-6842.01, TOI-5217.01, and TOI-6883.01. No nearby stars meeting these criteria are present within the SOPHIE aperture according to \textit{Gaia} DR3 and high-resolution imaging (Sect.~\ref{image}). We therefore confidently rule out background eclipsing binaries as the source of the observed photometric variations.

To rule out unresolved eclipsing binaries in these nine cases, we analyzed bisector span variations and their correlation with the RVs, as significant variability or correlation would indicate blending. The measured SOPHIE RV dispersions are 50.9, 173.0, 193.6, 33.2, 22.9, 46.3, 23.3, 33.9, and 36.0~m~s$^{-1}$ for TOI-6649, TOI-5217A, TOI-6883A, TOI-6842A, TOI-6893A, TOI-5893, TOI-6457, TOI-7176, and TOI-7025, respectively, with corresponding bisector dispersions of 36.4, 61.0, 11.7, 12.2, 20.8, 21.5, 22.3, 13.8, and 16.9~m~s$^{-1}$ (see Fig.~\ref{image_5217} and all related plots in Appendix~\ref{all_plot}). In all cases, the bisector dispersions are smaller than the RV dispersions.

To search for correlations between RV and bisector span, we used the Pearson correlation coefficient ($r$), which measures linear correlation, and the Spearman rank coefficient ($\rho$), which measures monotonic correlation. The statistical significance was assessed using two-sided p-values under the null hypothesis of no correlation. None of the targets show statistically significant correlations (all $p > 0.05$), further supporting the planetary interpretation. We note that TOI-6842A ($r = -0.40$, $\rho = -0.40$) and TOI-6457 ($r \simeq 0.40$, $\rho \simeq 0.41$) exhibit moderate correlations, but these remain non-significant. This lack of significance could be due to the limited number of measurements. Nonetheless, a stellar companion would need to be brighter than $\Delta m \leq 1$~mag (corresponding to $G \leq 11$~mag for TOI-6842 and $G \leq 10$~mag for TOI-6457) to produce a transit depth comparable to that of TOI-6842.01 and TOI-6457.01. Given the absence of such bright companions in our high-resolution imaging, the small bisector variations, and the lack of any secondary peak in the RV CCF residuals after removing the primary's CCF, we conclude that even if the correlations were statistically significant, they would most likely be driven by stellar activity rather than a blended eclipsing binary.

Considering (i) the absence of bisector variations or significant correlations with RV, (ii) the absence of sufficiently bright nearby sources in the SOPHIE apertures, (iii) RV semi-amplitudes consistent with planetary masses, and (iv) the agreement between the RV-derived $T_c$ and the \textit{TESS} transit epochs for all candidates, together with consistent orbital periods when available, we conclude that TOI-6649.01, TOI-6842.01, TOI-6457.01, TOI-7176.01, TOI-5217.01, TOI-6883.01, TOI-7025.01, TOI-6893.01, and TOI-5893.01 are bona fide planetary systems. We hereafter refer to these transiting planets as TOI-6649~b, TOI-6842A~b, TOI-6457~b, TOI-7176~b, TOI-5217A~b, TOI-6883A~b, TOI-7025~b, TOI-6893A~b, and TOI-5893~b.

As discussed in Sect.~\ref{stellar}, among the planetary host stars studied here, the stars TOI-7025, TOI-6842A, and TOI-5893 are the most magnetically active, with chromospheric activity indices of $\log(R'_{\rm HK}) = -4.6 \pm 0.1$, $-4.5 \pm 0.1$, and $-4.4 \pm 0.1$, respectively. Magnetic activity can induce apparent RV variations through the presence of spots, faculae, and plages on the stellar surface. These activity-induced signals are expected to occur primarily at the stellar rotation period and its harmonics and may therefore bias the measured RV semi-amplitude when they overlap with the planetary orbital period \citep{boisse2011disentangling}.

To assess the possible impact of stellar activity on the derived RV semi-amplitude, we compared the stellar rotation periods, the expected activity-induced RV jitter estimated from the empirical relation of \citet{hojjatpanah2020correlation}, and the correlations between the RVs and the bisector span for each system.

For TOI-7025, the stellar rotation period ($P_{\rm rot}=11\pm3$\,d) is well separated from the planetary orbital period ($P_{\rm orb}=67$\,d). Similarly, for TOI-5893, the stellar rotation period ($P_{\rm rot}=5\pm3$\,d) differs substantially from the planetary orbital period ($P_{\rm orb}=40$\,d). Consequently, significant contamination of the planetary RV signals by rotationally modulated stellar activity is unlikely. This conclusion is supported by the weak RV--bisector correlations ($r=-0.3$ and $r=0.1$ for TOI-7025 and TOI-5893, respectively). Furthermore, the expected activity-induced RV jitter is approximately 8 and 16\,m\,s$^{-1}$ for TOI-7025 and TOI-5893, respectively. These values are comparable to the mean RV uncertainty of $\sim14$\,m\,s$^{-1}$ for both systems and remain smaller than the measured planetary RV semi-amplitudes ($K=77\pm15$ and $57.6\pm3.5$\,m\,s$^{-1}$, respectively). In addition, the RV observations span multiple stellar rotation cycles, which should partially help to mitigate the impact of stellar activity by averaging over epochs with varying activity levels. We therefore conclude that stellar activity is unlikely to dominate the measured RV signals or significantly bias the inferred planetary masses of TOI-7025\,b and TOI-5893\,b, although it may contribute to the quoted uncertainties.

TOI-6842A represents the least favorable case. Its estimated stellar rotation period ($P_{\rm rot}=7\pm3$\,d) is closer to the planetary orbital period ($P_{\rm orb}=5.4$\,d), and the RV--bisector correlation is moderately negative ($r=-0.5$), indicating that stellar activity could contribute to the observed RV variability. However, the expected activity-induced RV jitter ($\sim11$\,m\,s$^{-1}$) remains comparable to the median RV uncertainty ($\sim12$\,m\,s$^{-1}$) and smaller than the measured planetary semi-amplitude ($K=35\pm4$\,m\,s$^{-1}$).

To investigate whether correlated stellar activity affects the inferred planetary mass, we explored two Gaussian-process (GP) kernels, namely the exponential \citep{ambikasaran2015fast} and Matern kernels \citep{foreman2017fast}, using \texttt{juliet}\citep{Espinoza2019}. Bayesian model comparison revealed no significant preference for either GP model over the simpler Keplerian model ($\Delta\ln Z < 1$). Furthermore, the GP hyperparameters remained poorly constrained, likely because of the limited number (17) and uneven temporal sampling of the RV observations. We therefore adopt the simpler Keplerian model with an additional RV jitter term for the final analysis presented in Sect.~\ref{exofast_results}. Nevertheless, we note that residual stellar activity cannot be completely excluded and may still introduce a modest systematic uncertainty into the inferred planetary mass.

\begin{table*}[t!]
\centering
\caption{Stellar and planetary parameters for the nine planetary systems.}
\resizebox{2.0\columnwidth}{!}{%
\begin{tabular}{l l l l l l l}
\hline
Parameter & Definition & TOI-6883A & TOI-6649 & TOI-6842A & \multicolumn{2}{c}{TOI-5217A} \\
\hline
$M_*$ & Mass ($M_\odot$) & $1.055\pm0.052$ & $1.259^{+0.059}_{-0.069}$ & $1.002^{+0.057}_{-0.047}$ & \multicolumn{2}{c}{$1.537\pm0.088$} \\
$R_*$ & Radius ($R_\odot$) & $1.068\pm0.019$ & $1.366^{+0.041}_{-0.039}$ & $1.393^{+0.029}_{-0.028}$ & \multicolumn{2}{c}{$2.275\pm0.071$} \\
$L_*$ & Luminosity ($L_\odot$) & $1.172^{+0.085}_{-0.071}$ & $2.652^{+0.099}_{-0.097}$ & $1.880^{+0.10}_{-0.095}$ & \multicolumn{2}{c}{$6.63^{+0.78}_{-0.72}$} \\
$\rho_*$ & Density (cgs) & $1.228^{+0.033}_{-0.051}$ & $0.696^{+0.072}_{-0.071}$ & $0.524^{+0.030}_{-0.028}$ & \multicolumn{2}{c}{$0.184^{+0.021}_{-0.019}$} \\
$\log g$ & Surface gravity (cgs) & $4.405^{+0.012}_{-0.015}$ & $4.267^{+0.032}_{-0.036}$ & $4.152^{+0.020}_{-0.019}$ & \multicolumn{2}{c}{$3.911^{+0.035}_{-0.036}$} \\
$T_{\rm eff}$ & Effective temperature (K) & $5814^{+85}_{-79}$ & $6302^{+90}_{-89}$ & $5725^{+76}_{-72}$ & \multicolumn{2}{c}{$6140^{+150}_{-140}$} \\
$[{\rm Fe/H}]$ & Metallicity (dex) & $0.254\pm0.076$ & $0.184\pm0.091$ & $-0.020^{+0.096}_{-0.085}$ & \multicolumn{2}{c}{$0.299^{+0.080}_{-0.081}$} \\
Age & Age (Gyr) & $4.6^{+2.4}_{-2.0}$ & $2.4^{+1.5}_{-1.1}$ & $9.6 \pm 2.1$ & \multicolumn{2}{c}{---} \\
$EEP$ & Equal Evolutionary Phase$^{3}$ & $364^{+27}_{-21}$ & $351^{+41}_{-19}$ & $441.3^{+5.7}_{-7.5}$ & \multicolumn{2}{c}{---} \\
$A_V$ & V-band extinction (mag) & $0.137^{+0.093}_{-0.083}$ & $0.055^{+0.030}_{-0.035}$ & $0.094^{+0.039}_{-0.055}$ & \multicolumn{2}{c}{---} \\
$d$ & Distance (pc) & $93.95\pm0.16$ & $200.79^{+0.77}_{-0.76}$ & $392.7 \pm 3.7$ & \multicolumn{2}{c}{---}\\
\hline
\\
&&TOI-6883Ab&TOI-6649b&TOI-6842Ab&TOI-5217Ab&TOI-5217Ac\\
\\
$P$ & Period (days) & $16.2499265\pm0.0000077$ & $5.9877995^{+0.000010}_{-0.0000098}$ & $5.3917436 \pm 0.0000025$ & $4.3712411\pm0.0000082$ & $\sim 1761$ \\
$R_P$ & Radius ($R_J$) & $1.073^{+0.021}_{-0.020}$ & $1.25^{+0.74}_{-0.45}$ & $1.647 \pm 0.039$ & $1.385^{+0.056}_{-0.055}$ & ---\\
$M_P$ & Mass ($M_J$) & $4.26\pm0.14$ & $0.645^{+0.032}_{-0.033}$ & $0.302 \pm 0.035$ & $1.282\pm0.060$ & --- \\
$T_C$ & Time of conjunction (BJD$_{TDB}$) & $2459811.24052\pm0.00025$ & $2460587.9255\pm0.0015$ & $2460209.33933^{+0.00028}_{-0.00027}$ & $2460315.3526\pm0.0010$ & $\sim 242461150$ \\
$a$ & Semi-major axis (AU) & $0.1280\pm0.0021$ & $0.0697^{+0.0011}_{-0.0013}$ & $0.06023^{+0.0011}_{-0.00097}$ & $0.0604^{+0.0011}_{-0.0012}$ & $\sim3$ \\
$i$ & Inclination (Degrees) & $89.60^{+0.28}_{-0.34}$ & $84.66^{+0.34}_{-0.40}$ & $86.67^{+0.22}_{-0.21}$ & $82.98\pm0.53$ & undefined \\
$e$ & Eccentricity &$0.6037\pm0.0018$ & $0.039^{+0.039}_{-0.026}$ &0 (fixed) & 0 (fixed) & $0.529\pm0.017$ \\
$\omega_*$ & Arg of periastron (Degrees) & $1.73\pm0.43$& $-59^{+64}_{-27}$ & 90 (fixed)& 90 (fixed) & $68.5^{+2.8}_{-2.7}$ \\
$T_{\rm eq}$ & Equilibrium temp (K) & $809^{+12}_{-11}$ & $1345^{+16}_{-15}$ & $1327^{+19}_{-18}$ & $1817^{+55}_{-53}$ & $\sim246$ \\
$K$ & RV semi-amplitude (m/s) & $412.5\pm2.4$ & $61.8\pm2.3$ & $34.8^{+3.8}_{-3.9}$ & $118.7\pm3.3$ & $195.6\pm3.7$ \\
$R_P/R_*$ & Radius ratio&
$0.10331^{+0.00050}_{-0.00046}$&
$0.094^{+0.054}_{-0.033}$&
$0.12147^{+0.00067}_{-0.00070}$&
$0.06257^{+0.0010}_{-0.0010}$&
--- \\
$a/R_*$ & Semi-major axis in stellar radii & $25.82^{+0.23}_{-0.36}$ & $10.97^{+0.37}_{-0.39}$ & $9.30 \pm 0.17$ & $5.71^{+0.21}_{-0.20}$ & --- \\
$\delta_{\rm TESS}$ & Transit depth in \textit{TESS} (frac) & $0.01270^{+0.00017}_{-0.00016}$ & $0.001155^{+0.000097}_{-0.000093}$ & $0.01634 \pm 0.00018$ & $0.00407\pm0.00012$ & ---\\
$\tau$ & In/egress transit duration (days) & $0.01629^{+0.00056}_{-0.00021}$ & $0.0283^{+0.0015}_{-0.0014}$ & $0.0269 \pm 0.0012$ & $0.0217^{+0.0019}_{-0.0017}$ & --- \\
$T_{14}$ & Total transit duration (days) & $0.17203^{+0.00065}_{-0.00060}$ & $0.0567^{+0.0031}_{-0.0028}$ & $0.1819 \pm 0.0011$ & $0.1975\pm0.0024$ & --- \\
$b$ & Transit impact parameter & $0.114^{+0.093}_{-0.078}$ & $1.048^{+0.060}_{-0.041}$ & $0.541^{+0.024}_{-0.027}$ & $0.697^{+0.027}_{-0.030}$ &---\\
$\rho_P$ & Density (cgs) & $4.29^{+0.14}_{-0.21}$ & $0.41^{+1.2}_{-0.31}$ & $0.084^{+0.011}_{-0.010}$ & $0.599^{+0.083}_{-0.073}$ &---\\
$M_P\sin i$&Minimum mass ($M_J$)&$4.26\pm0.14$&$0.642^{+0.032}_{-0.033}$&$0.302\pm0.035$&$1.272^{+0.061}_{-0.060}$&$\sim 13$\\
$T_P$ &Time of Periastron (BJD$_{TDB}$)&$2459810.136\pm0.013$&$2460585.49^{+1.1}_{-0.50}$&$2460209.33898^{+0.00028}_{-0.00027}$&$2460315.3523\pm0.0010$&$\sim242461110$\\
\hline
\end{tabular}%
}
\label{tab:GP_params_complete1}
\end{table*}
\begin{table*}[h!]
\centering
\resizebox{2\columnwidth}{!}{%
\begin{tabular}{l l c c c c c}
\hline
Parameter & Definition & TOI-6457 & TOI-7176 & TOI-6893A & TOI-7025 & TOI-5893\\
\hline
$M_*$ & Mass ($M_\odot$) &
$1.237^{+0.061}_{-0.068}$ &
$0.700^{+0.026}_{-0.024}$ &
{$1.375^{+0.079}_{-0.080}$} &
{$1.258^{+0.080}_{-0.079}$} &
{$1.058^{+0.066}_{-0.065}$} \\

$R_*$ & Radius ($R_\odot$) &
$1.359^{+0.041}_{-0.035}$ &
$0.688\pm0.019$ &
{$2.543\pm0.069$} &
{$1.537\pm0.030$} &
{$1.169^{+0.035}_{-0.034}$} \\

$L_*$ & Luminosity ($L_\odot$) &
$2.42\pm0.12$ &
$0.177^{+0.012}_{-0.014}$ &
{$8.74^{+0.95}_{-0.87}$} &
{$3.07^{+0.18}_{-0.17}$} &
{$1.52\pm0.12$} \\

$\rho_*$ & Density (cgs) &
$0.694^{+0.068}_{-0.072}$ &
$3.03^{+0.25}_{-0.23}$ &
{$0.118^{+0.012}_{-0.011}$} &
{$0.488^{+0.043}_{-0.040}$} &
{$0.934^{+0.099}_{-0.089}$} \\

$\log g$ & Surface gravity (cgs) &
$4.264^{+0.030}_{-0.037}$ &
$4.608^{+0.024}_{-0.023}$ &
{$3.766^{+0.034}_{-0.035}$} &
{$4.164^{+0.031}_{-0.032}$} &
{$4.327\pm0.034$} \\

$T_{\rm eff}$ & Effective temperature (K) &
$6171^{+86}_{-88}$ &
$4507^{+82}_{-87}$ &
{$6220\pm140$} &
{$6164^{+66}_{-67}$} &
{$5930\pm100$} \\

$[{\rm Fe/H}]$ & Metallicity (dex) &
$0.230\pm0.048$ &
$-0.014^{+0.075}_{-0.046}$ &
{$0.255^{+0.099}_{-0.100}$} &
{$0.080\pm0.050$} &
{$0.136^{+0.090}_{-0.087}$} \\

Age & Age (Gyr) &
$3.0^{+1.8}_{-1.3}$ &
$9.7^{+2.8}_{-4.0}$ &
--- &
--- &
{$5.9^{+3.1}_{-2.5}$} \\

$EEP$ & Equal Evolutionary Phase &
$368^{+38}_{-28}$ &
$338.7^{+6.9}_{-14}$ &
--- &
--- &
{$396^{+22}_{-38}$} \\

$A_V$ & V-band extinction (mag) &
$0.101^{+0.057}_{-0.060}$ &
$0.30^{+0.13}_{-0.17}$ &
--- &
--- &
{$0.219^{+0.095}_{-0.11}$} \\

$d$ & Distance (pc) &
$262.3\pm1.6$ &
$68.96\pm0.11$ &
--- &
--- &
{$325.0\pm1.8$} \\
\hline
\\
&& TOI-6457\,b & TOI-7176\,b & TOI-6893A\,b & TOI-7025\,b &
TOI-5893\,b \\
\\
$P$ & Period (days) &
$41.418325^{+0.000099}_{-0.00010}$ &
$47.880^{+0.059}_{-0.065}$ &
{$56.222^{+0.087}_{-0.10}$} &
{$66.995436\pm0.000034$} &
{$40.369376^{+0.000100}_{-0.00010}$} \\

$R_P$ & Radius ($R_J$) &
$0.915^{+0.034}_{-0.029}$ &
$0.772\pm0.026$ &
{$1.036^{+0.038}_{-0.035}$} &
{$1.152\pm0.029$} &
{$0.973^{+0.036}_{-0.035}$} \\

$M_P$ & Mass ($M_J$) &
$0.49\pm0.12$ &
$0.632\pm0.063$ &
{$0.627^{+0.078}_{-0.076}$} &
{$0.873^{+0.099}_{-0.095}$} &
{$1.012^{+0.073}_{-0.072}$} \\

$T_C$ & Time of conjunction (BJD$_{\rm TDB}$) &
$2460280.04018\pm0.00094$ &
$2459835.66800\pm0.00046$ &
{$2460231.5480\pm0.0015$} &
{$2460340.02854\pm0.00063$} &
{$2459817.1737\pm0.0020$} \\

$a$ & Semi-major axis (AU) &
$0.2515^{+0.0041}_{-0.0047}$ &
$0.2292^{+0.0028}_{-0.0027}$ &
{$0.3194^{+0.0060}_{-0.0063}$} &
{$0.3485^{+0.0072}_{-0.0074}$} &
{$0.2348^{+0.0048}_{-0.0049}$} \\

$i$ & Inclination (Degrees) &
$89.58^{+0.21}_{-0.16}$ &
$89.564^{+0.085}_{-0.055}$ &
{$89.00^{+0.57}_{-0.28}$} &
{$84.66^{+0.55}_{-0.66}$} &
{$89.105^{+0.067}_{-0.066}$} \\

$e$ & Eccentricity &
0 (fixed) &
$0.189^{+0.076}_{-0.071}$ &
{$0.167^{+0.065}_{-0.089}$} &
{$0.815^{+0.022}_{-0.021}$} &
0 (fixed) \\

$\omega_*$ & Arg of periastron (Degrees) &
90 (fixed) &
$40^{+28}_{-21}$ &
{$123^{+36}_{-29}$} &
{$77.3^{+10.0}_{-9.8}$} &
90 (fixed) \\

$T_{\rm eq}$ & Equilibrium temperature (K) &
$692.6^{+9.1}_{-8.8}$ &
$376.9^{+6.1}_{-7.0}$ &
{$846^{+24}_{-23}$} &
{$624\pm11$} &
{$638\pm12$} \\

$K$ & RV semi-amplitude (m/s) &
$25.1^{+6.1}_{-6.0}$ &
$45.8\pm4.3$ &
{$27.4^{+3.1}_{-3.2}$} &
{$64.5^{+8.1}_{-7.3}$} &
{$57.8^{+3.3}_{-3.5}$} \\

$R_P/R_*$ & Radius ratio &
$0.06919^{+0.00071}_{-0.00068}$ &
$0.1152^{+0.0020}_{-0.0021}$ &
{$0.04184^{+0.00093}_{-0.00078}$} &
{$0.0770^{+0.0011}_{-0.0012}$} &
{$0.0856\pm0.0012$} \\

$a/R_*$ & Scaled semi-major axis &
$39.8^{+1.3}_{-1.4}$ &
$71.6^{+1.9}_{-1.8}$ &
{$27.00^{+0.91}_{-0.88}$} &
{$48.7\pm1.4$} &
{$43.2^{+1.5}_{-1.4}$} \\

$\delta_{\rm TESS}$ & Transit depth in TESS (frac) &
$0.005427^{+0.00010}_{-0.000099}$ &
$0.01633^{+0.00035}_{-0.00033}$ &
{$0.001951\pm0.000055$} &
{$0.00557\pm0.00012$} &
{$0.00773\pm0.00021$} \\

$\tau$ & In/egress transit duration (days) &
$0.0239^{+0.0020}_{-0.0016}$ &
$0.0246^{+0.0031}_{-0.0030}$ &
{$0.0266^{+0.0070}_{-0.0042}$} &
{$0.0216\pm0.0027$} &
{$0.0347^{+0.0028}_{-0.0026}$} \\

$T_{14}$ & Total duration (days) &
$0.3413^{+0.0023}_{-0.0021}$ &
$0.1896^{+0.0028}_{-0.0026}$ &
{$0.5559^{+0.0067}_{-0.0053}$} &
{$0.0941^{+0.0022}_{-0.0024}$} &
{$0.2532^{+0.0049}_{-0.0048}$} \\

$b$ & Impact parameter &
$0.289^{+0.098}_{-0.14}$ &
$0.474^{+0.076}_{-0.11}$ &
{$0.41^{+0.16}_{-0.25}$} &
{$0.851^{+0.015}_{-0.020}$} &
{$0.675^{+0.028}_{-0.032}$} \\

$\rho_P$ & Planet density (cgs) &
$0.79^{+0.22}_{-0.20}$ &
$1.70^{+0.25}_{-0.22}$ &
{$0.70^{+0.12}_{-0.11}$} &
{$0.707^{+0.10}_{-0.090}$} &
{$1.36^{+0.19}_{-0.17}$} \\

$M_P\sin i$ & Minimum mass ($M_J$) &
$0.49\pm0.12$ &
$0.632\pm0.063$ &
{$0.626^{+0.078}_{-0.076}$} &
{$0.869^{+0.099}_{-0.095}$} &
{$1.012^{+0.073}_{-0.072}$} \\

$T_P$ & Time of Periastron (BJD$_{\rm TDB}$) &
$2460280.03873\pm0.00094$ &
$2459831.3^{+2.4}_{-1.9}$ &
$2460178.8^{+5.0}_{-3.1}$ &
$2460339.890^{+0.11}_{-0.096}$ &
$2459817.1723\pm0.0020$ \\
\hline
\end{tabular}%
}
\tablefoot{$^{1}$ We do not report error bars for the period of TOI-5217Ac, as its posterior distribution is essentially flat.}
\label{tab:toi_multi_systems}
\end{table*}

\subsection{Joint Modeling}
\label{join-model}
For the 14 systems, we performed a simultaneous fit of the planetary and stellar parameters using \texttt{EXOFASTv2} \citep{eastman2013,eastman2017,eastman2019}, combining all available photometric and spectroscopic datasets. The stellar modeling incorporated the SED and the MESA Isochrones and Stellar Tracks (MIST) models \citep{dotter2016mesa,choi2016mesa}. Gaussian priors were applied to $T_{\mathrm{eff}}$ and [Fe/H], centered on the values derived from our spectroscopic analysis (Sect.~\ref{stellar}). Their standard deviations were broadened relative to the quoted uncertainties to allow for additional model flexibility. A Gaussian prior was also applied to the \textit{Gaia}~DR3 parallax. To constrain extinction, $A_V$, we imposed an upper limit informed by the dust maps from \citet{Schlegel:1998}. Archival broadband photometry from multiple surveys was retrieved internally by \texttt{EXOFASTv2} for the SED modeling. Instrumental and stellar variability in the \textit{TESS} light curves were taken into account using a spline detrending method \citep{vanderburg2014technique}. The detrended light curves were visually inspected to ensure that the transit depths were not distorted by the detrending process. When the orbital eccentricity was not significantly constrained by the data, we additionally tested a circular-orbit model and adopted the preferred solution based on a comparison of the BIC.

For the four systems whose companions were found to be stellar
(TOI-3678AB, TOI-5621AB, TOI-5321AB, and TOI-5308AB), we estimated the
effective temperature of each secondary by interpolating its measured
mass in the main-sequence table of \citet{pecaut2013intrinsic}. The
resulting temperatures are reported in Table~\ref{tab:GP_params_complete2}.
Using these temperatures and the measured radii of both components, and
assuming blackbody emission, we calculated the secondary-to-primary flux
ratios in the TESS passband. The secondary flux contributions range from
$1.9\times10^{-4}$ to $3.5\times10^{-4}$, corresponding to relative
corrections of only $0.009$--$0.017\%$ to $R_B/R_*$. These corrections
are well below the uncertainties on the measured radius ratios;
therefore, we did not apply any correction for dilution by the secondary
flux. For TOI-5906AB, both stellar components were modeled
simultaneously, and the secondary flux contribution was therefore
included directly in the analysis.

The Markov Chain Monte Carlo (MCMC) fitting process using \texttt{EXOFASTv2} is considered converged when the Gelman–Rubin statistic ($\hat{R}$) for all parameters is less than 1.01 and the number of independent draws ($T_z$) exceeds 1000, following the convergence criteria outlined in \citet{eastman2019}. The resulting parameters are presented in Tables \ref{tab:GP_params_complete1}, \ref{tab:GP_params_complete2}, \ref{tab:toi5609} and Appendix \ref{tab:toi_bimodal}. The median model for the SED, RV, and photometric data of the TOI-5217A system is shown in Fig. \ref{image_5217}, while the corresponding plots for the remaining systems are presented in Appendix \ref{all_plot}. For some parameters, the posterior distributions deviate from a Gaussian shape; therefore, we present their full posterior distributions in Appendix \ref{all_plot}.

\subsection{Target specific notes}
\label{exofast_results}

\emph{TOI-5217A:} As discussed in Sect.~3, the \textit{WISE} photometry for this target is contaminated by nearby stars and was therefore excluded from the fit. Modeling the SED with a reduced number of photometric points while simultaneously fitting large additional free parameters prevented reliable convergence in a joint stellar and planetary fit. Consequently, we adopted Gaussian priors on the stellar mass and radius based on the stellar characterization described in Section~3.

\begin{table*}[h!]
\centering
\caption{Stellar parameters for the eclipsing binaries TOI-3678AB, TOI-5621AB, TOI-5321AB, TOI-5308AB.}
\resizebox{1.8\columnwidth}{!}{%
\begin{tabular}{l l l l l l}
\hline
Parameter & Definition & TOI-3678A & TOI-5621A & TOI-5321A & TOI-5308A$^{1}$ \\
\hline
$M_*$ & Mass ($M_\odot$) & $1.614^{+0.096}_{-0.095}$ & $1.082^{+0.084}_{-0.069}$ & $1.421^{+0.076}_{-0.075}$ & $1.226^{+0.041}_{-0.035}$ \\
$R_*$ & Radius ($R_\odot$) & $2.287\pm0.059$ & $1.429\pm0.044$ & $1.871^{+0.055}_{-0.056}$ & $1.590\pm0.038$ \\
$L_*$ & Luminosity ($L_\odot$) & $10.1^{+1.4}_{-1.2}$ & $2.264^{+0.091}_{-0.088}$ & $5.75^{+0.64}_{-0.58}$ & $2.27\pm0.11$ \\
$\rho_*$ & Density (cgs) & $0.190^{+0.012}_{-0.010}$ & $0.524^{+0.067}_{-0.058}$ & $0.305^{+0.027}_{-0.023}$ & $0.431^{+0.028}_{-0.026}$ \\
$\log g$ & Surface gravity (cgs) & $3.927^{+0.020}_{-0.019}$ & $4.163^{+0.042}_{-0.041}$ & $4.046^{+0.027}_{-0.025}$ & $4.125\pm0.018$ \\
$T_{\rm eff}$ & Effective temperature (K) & $6810^{+200}_{-190}$ & $5924^{+86}_{-85}$ & $6540\pm140$ & $5618\pm72$ \\
$[{\rm Fe/H}]$ & Metallicity (dex) & $-0.06\pm0.22$ & $0.043^{+0.070}_{-0.060}$ & $-0.22\pm0.13$ & $0.28^{+0.10}_{-0.11}$ \\
Age & Age (Gyr) & --- & $7.1^{+2.5}_{-2.4}$ & --- & $5.07^{+0.71}_{-0.66}$ \\
$EEP$ & Equal Evolutionary Phase$^{3}$ & --- & $434^{+11}_{-23}$ & --- & $418.6^{+7.0}_{-9.3}$ \\
$A_V$ & V-band extinction (mag) & --- & $0.064^{+0.025}_{-0.036}$ & --- & $0.085^{+0.022}_{-0.039}$ \\
$d$ & Distance (pc) & --- & $702.9^{+7.1}_{-7.0}$ &  ---  & $341.7\pm6.2$ \\
\hline
\\
&  & TOI-3678B & TOI-5621B & TOI-5321B & TOI-5308B \\
\\
$P$ & Period (days) & $4.8542845\pm0.0000024$ & $39.97545\pm0.00012$ & $10.1392043^{+0.0000078}_{-0.0000076}$ & $8.3403287^{+0.0000066}_{-0.0000067}$ \\
$R_B$ & Radius ($R_\odot$)&$0.1935^{+0.0055}_{-0.0057}$&$0.1378^{+0.0054}_{-0.0053}$&$0.1507^{+0.0053}_{-0.0054}$&$0.1470^{+0.0049}_{-0.0048}$\\

$M_B$ & Mass ($M_\odot$) & $0.1602\pm0.0062$ & $0.1071^{0.0053}_{0.0044}$& $0.1214\pm0.0042$ &$0.1088^{+0.0024}_{-0.0021}$\\
$T_C$ & Time of conjunction (BJD$_{\rm TDB}$) & $2459885.44761\pm0.00022$ & $2459594.4418\pm0.0014$ & $2459478.89509^{+0.00045}_{-0.00046}$ & $2459485.68489^{+0.00054}_{-0.00055}$ \\
$a$ & Semi-major axis (AU) & $0.0679\pm0.0013$ & $0.2424^{+0.0059}_{-0.0051}$ & $0.1059\pm0.0018$ & $0.08862^{+0.00094}_{-0.00082}$ \\
$i$ & Inclination (Degrees) & $85.49^{+0.37}_{-0.32}$ & $88.82^{+0.15}_{-0.14}$ & $84.54^{+0.20}_{-0.19}$ & $85.80\pm0.11$ \\
$e$ & Eccentricity & $0.0198^{+0.0021}_{-0.0020}$ & $0.47768\pm0.00083$ & $0.2535^{+0.0033}_{-0.0035}$ & 0 (fixed) \\
$\omega_*$ & Arg of periastron (Degrees) & $-5^{+15}_{-14}$ & $177.69^{+0.19}_{-0.18}$ & $105.92\pm0.44$ & 90 (fixed) \\

$T_{\rm eff,B}^{2}$ & Estimated secondary effective temperature (K) &$\sim3050$ & $\sim2840$ &
$\sim2920$ & $\sim2850$ \\

$K$ & RV semi-amplitude (m/s) & $13706^{+86}_{-87}$ & $6759^{+16}_{-15}$ & $9209\pm31$ & $9391^{+49}_{-41}$ \\
$R_B/R_*$ & Radius ratio &$0.08462\pm0.00046$&$0.0965\pm0.0014$&$0.08047^{+0.00091}_{-0.00090}$&$0.0924\pm0.0013$\\
$a/R_*$ & Semi-major axis in stellar radii & $6.38^{+0.13}_{-0.12}$ & $36.5^{+1.5}_{-1.4}$ & $12.16^{+0.34}_{-0.31}$ & $11.99^{+0.25}_{-0.24}$ \\
$\delta_{\rm TESS}$ & Transit depth in TESS (frac) & $0.007653\pm0.000081$ & $0.01015^{+0.00025}_{-0.00026}$ & $0.00604\pm0.00011$ & $0.00741^{+0.00016}_{-0.00015}$ \\
$\tau$ & In/egress transit duration (days) & $0.02407^{+0.00100}_{-0.0010}$ & $0.0354^{+0.0033}_{-0.0030}$ & $0.0358^{+0.0026}_{-0.0025}$ & $0.0466^{+0.0032}_{-0.0030}$ \\
$T_{14}$ & Total transit duration (days) & $0.2347^{+0.0010}_{-0.0011}$ & $0.2813\pm0.0037$ & $0.1325\pm0.0018$ & $0.1444^{+0.0021}_{-0.0020}$ \\
$b$ & Transit impact parameter & $0.502^{+0.026}_{-0.032}$ & $0.571^{+0.043}_{-0.052}$ & $0.8713^{+0.0075}_{-0.0088}$ & $0.8778^{+0.0060}_{-0.0065}$ \\
$\rho_B$ & Density (cgs) & $31.1^{+2.4}_{-2.1}$ & $57.8^{+8.3}_{-7.2}$ & $50.0^{+5.5}_{-4.7}$ & $48.4^{+4.8}_{-4.3}$ \\
$T_P$ & Time of Periastron (\tjdtdb) & $2459884.19^{+0.20}_{-0.19}$ & $2459558.450^{+0.019}_{-0.018}$ & $2459469.0150^{+0.0077}_{-0.0075}$ & $2459485.68442^{+0.00054}_{-0.00055}$ \\
\hline
\end{tabular}%
}
\tablefoot{$^{1}$ For TOI-5308A, the more probable solution (85\%) is presented here, while the less probable solution (15\%) is provided in the Appendix \ref{bimodal}. $2$ The secondary effective temperatures were estimated by interpolating their measured masses in the main-sequence table of \citet{pecaut2013intrinsic}.}
\label{tab:GP_params_complete2}
\end{table*}


Allowing the eccentricity of the inner planet to vary freely does not result in a significant detection ($e = 0.015^{+0.018}_{-0.011}$). We therefore also tested a circular orbital solution which was favored, with a BIC difference of $\Delta \mathrm{BIC} = 17$. The derived planetary parameters are consistent between the two solutions within $1\sigma$. Our final adopted parameters for TOI-5217A\,b are an orbital period of $P = 4.3712411 \pm 0.0000082$~d, a radius of $R_{\rm p} = 1.385^{+0.056}_{-0.055}\,R_{\mathrm{J}}$, and a mass of $M_{\rm p} = 1.282\pm0.060\,M_{\mathrm{J}}$. We note that the uncertainty on the transit mid-time ($T_{\rm c} = 2460315.3526 \pm 0.0010$) is larger than for the other systems in this work with multiple observed transits, except for the grazing system TOI-6649\,b (below). We do not investigate the possibility of transit timing variations (TTVs) in this work, leaving such an analysis to future studies.

In addition, our results show the long-period signal has a period of approximately $P \sim 1761$~d. The posterior distribution for this signal remains largely unconstrained, indicating that the true orbital period may differ. This limitation primarily arises from the relatively short RV baseline of about 1300~d. The corresponding RV semi-amplitude is $K = 195.6\pm3.7~\mathrm{m\,s^{-1}}$, implying a minimum mass of about $\sim 13\,M_{\mathrm{J}}$ for an eccentric orbit with $e = 0.529\pm0.017$. Because the orbital period is not well determined, the inferred minimum mass should be regarded as provisional. This companion therefore lies in the regime of either a massive planet or a brown dwarf. Additional RV observations are required to better constrain its orbital parameters. We note that the poorly constrained orbit of this outer companion does not significantly affect the derived parameters of the inner planet.

\emph{TOI-6649:} TOI-6649\,b exhibits a highly grazing transit geometry, leading to a strong degeneracy between the planet-to-star radius ratio and the impact parameter. To aid convergence, we imposed a Gaussian prior on $R_{\rm p}/R_\star$ based on the SPOC transit depth, adopting $R_{\rm p}/R_\star = 0.034$ with a conservative standard deviation of 0.1. Additionally, we consider a upper bound limit to roughly three times Jupiter's radius ($R_p / R_\star \approx 0.22$ for $R_\star = 1.374\,R_\odot$). This wide prior allows for a broad range of physically plausible radii while incorporating external information. Additional Gaussian priors were applied to $T_c$ and the orbital period, centered on the \textit{TESS} values with a standard deviation of 0.5~d—several orders of magnitude larger than the uncertainties reported by the SPOC pipeline. Finally, we tested both circular and eccentric orbital configurations. The eccentric model is strongly favored, with $\Delta\mathrm{BIC} = 77$. We therefore adopt the eccentric solution, while noting that the eccentricity remains consistent with zero within uncertainties ($0.039^{+0.039}_{-0.026}$).

The planet has an orbital period of $P = 5.9877995^{+0.000010}_{-0.0000098}$~d, a radius of $R_{\rm p} = 1.25^{+0.74}_{-0.45}\,R_{\mathrm{J}}$, and a mass of $M_{\rm p} = 0.645^{+0.032}_{-0.033}\,M_{\mathrm{J}}$. The large uncertainty on the radius is primarily driven by the grazing transit geometry. The inferred radius is nevertheless consistent with the empirical mass--radius relation of \citet{chen2016probabilistic}.

\emph{TOI-6883A:} TOI-6883A\,b was initially identified as a single-transit candidate by \citet{conzo2024toi} and later characterized by \citet{Sgro2024}. Here we use the \textit{TESS} and APF data from \citet{Sgro2024} together with our new SOPHIE and TRES data, as well as additional \textit{TESS} and ground-based photometry. One additional transit observed on 2024 August~19 was excluded from the analysis due to its poor data quality. Our refined joint fit yields $P = 16.2499265\pm0.0000077$~d, $R_{\rm p} = 1.073^{+0.021}_{-0.020}\,R_{\mathrm{J}}$, and $M_{\rm p} = 4.26 \pm 0.14\,M_{\mathrm{J}}$. These values are consistent with the radius and mass reported by \citet{Sgro2024}, who found $R_P = 1.087^{+0.023}_{-0.021}\,R_{\mathrm{J}}$ and $M_P = 4.34 \pm 0.15\,M_{\mathrm{J}}$. However, our derived orbital period differs by about 7$\sigma$ from that reported by \citet{Sgro2024} ($P = 16.24921^{+0.00010}_{-0.00011}$~d). This discrepancy is likely due to the improvement of our analysis that includes four complete transits, whereas \citet{Sgro2024} analyzed only a single full transit.



\emph{TOI-5893:} As mentioned earlier, only a single transit of this planet was detected in the \textit{TESS} data from Sector~55. We subsequently detected a second transit from the ground. We then modeled the system jointly using the ground-based transit, the SOPHIE RV measurements, and the \textit{TESS} Sector~55 photometry.

The eccentricity is not significantly constrained by the data ($e = 0.036^{+0.042}_{-0.025}$), and a circular solution is preferred based on the BIC ($\Delta\mathrm{BIC} > 17$). We therefore adopt the circular model as our final solution. TOI-5893\,b is a warm Jupiter with an orbital period of $40.369376 \pm 0.000100$~d, a mass of $M_{\rm p} = 1.012^{+0.073}_{-0.072}\,M_{\mathrm{J}}$, and a radius of $R_{\rm p} = 0.973^{+0.036}_{-0.035}\,R_{\mathrm{J}}$. These parameters are fully consistent with those derived from the eccentric model.

\emph{TOI-7025:} For this system, the posterior distributions of the stellar mass, equivalent evolutionary phase (EEP), and age exhibited bimodality, making convergence particularly challenging. Therefore, rather than fitting the stellar parameters simultaneously with the planetary parameters, we adopted Gaussian priors on $R_\star$, $M_\star$, [Fe/H], and $T_{\mathrm{eff}}$, based on the independent stellar analysis presented in Sect.~\ref{stellar}. To further improve convergence, we imposed a Gaussian prior on the time of conjunction, $T_{\rm c}$, centred on the value reported by SPOC, with a width of 0.01~d, approximately ten times larger than the SPOC uncertainty. We also imposed bounds on the eccentricity--argument-of-periastron parametrisation, adopting $-0.25 < \sqrt{e}\cos\omega_* < 0.50$ and $0.78 < \sqrt{e}\sin\omega_* < 0.98$. These bounds were guided by preliminary unconstrained runs and were chosen to encompass the high-probability region around the expected solution. The posterior distribution of the argument of periastron remains non-Gaussian (see Fig.~\ref{image_7025}), indicating that additional observations are required to better constrain the orbital solution. Finally, uniform priors were adopted for the SOPHIE and TRES RV jitter terms, with upper limits approximately five times larger than the mean RV uncertainties of the corresponding data sets.

TOI-7025\,b has a long orbital period of 
$P = 66.995436 \pm 0.000034$~d, 
a planetary radius of 
$R_{\rm p} = 1.152 \pm 0.029\,R_{\rm J}$, 
and a mass of 
$M_{\rm p} = 0.873^{+0.099}_{-0.095}\,M_{\rm J}$. 
The system exhibits a notably high eccentricity of 
$e = 0.815^{+0.022}_{-0.021}$, 
placing TOI-7025\,b among the most eccentric long-period giant planets characterized to date (see Sect.~\ref{discu} for further discussion).

\emph{TOI-7176:} The resulting best-fit parameters for TOI-7176\,b indicate that this is a warm Jupiter with orbital period of $P =47.880^{+0.059}_{-0.065}\,d$, a planetary radius of $R_{\rm p} = 0.772\pm0.026\,R_{\rm J}$, and a mass of $M_{\rm p} = 0.632\pm0.063\,M_{\rm J}$.

\emph{TOI-6457:} The circular orbit is strongly preferred ($\Delta$BIC = 17) over an eccentric orbit model. We therefore adopted a circular orbit. TOI-6457\,b has $P = 41.41833 \pm 0.00010$~d, $R_{\rm p} = 0.915^{+0.034}_{-0.029}\,R_{\mathrm{J}}$, and $M_{\rm p} = 0.49 \pm 0.12\,M_{\mathrm{J}}$.

\emph{TOI-6893A:} As discussed in Sect.~\ref{detection}, TOI-6893A\,b—similar to TOI-7176\,b—was identified from a single-transit event. In addition to the RV measurements and the single transit observed in Sector~70, we incorporated two additional available \textit{TESS} light curves to further constrain the orbital period through the non-detection of additional transits. 

We also identify bimodality in the posterior distributions of the stellar mass, age, and EEP, which prevents the fit from converging. Therefore, rather than modelling the star and its companion simultaneously, we applied Gaussian priors on $R_\star$, $M_\star$, [Fe/H], and $T_{\mathrm{eff}}$, adopting the values obtained from the independent stellar analysis presented in Sect.~\ref{stellar}.


TOI-6893A\,b is a warm Jupiter with an orbital period of $56.222^{+0.087}_{-0.10}$~d, a radius of $1.036^{+0.038}_{-0.035}\,R_{\rm J}$, and a mass of $0.627^{+0.078}_{-0.076}\,M_{\rm J}$. The posterior distribution of the orbital period is notably non-Gaussian, as part of the parameter space is excluded by the non-detection of additional transit events in the \textit{TESS} data. This indicates that further transit observations will be required to better constrain the orbital period.

\emph{TOI-6842A:} Allowing the eccentricity to vary yields a value consistent with zero ($e = 0.040^{+0.045}_{-0.028}$). A circular model is strongly favored ($\Delta\mathrm{BIC} \approx 16$), and we adopt it as the preferred solution. TOI-6842\,b has $P = 5.3917436 \pm 0.0000025$~d, $R_{\rm p} = 1.647 \pm 0.039\,R_{\mathrm{J}}$, and $M_{\rm p} = 0.302 \pm 0.035\,M_{\mathrm{J}}$. As discussed in Sect.~\ref{detection}, TOI-6842A is one of the most active stars in our sample, and its estimated stellar rotation period is close to the planetary orbital period, raising the possibility of activity-induced contamination of the RV signal. Therefore, the derived planetary parameters should be interpreted as the best estimates permitted by the current RV dataset, while acknowledging that residual stellar activity may still introduce a modest systematic uncertainty. Additionally, the relatively old isochronal age of TOI-6842A ($9.6\pm2.1$~Gyr) appears to be in tension with its comparatively high chromospheric activity level, as indicated by
$\log(R'_{\rm HK})=-4.5\pm0.1$. The age estimate should therefore be
interpreted with caution.

\emph{TOI-5621AB:} The companion has $P = 39.97547 \pm 0.00012$~d, $R = 1.341^{+0.053}_{-0.052}\,R_{\mathrm{J}}$ ($0.1378^{+0.0054}_{-0.0053}\,R_\odot$), $M = 112.4^{+5.5}_{-4.6}\,M_{\mathrm{J}}$ ($0.1071^{0.0053}_{0.0044}\,M_\odot$), and a high eccentricity of $e = 0.47768\pm 0.00083$, consistent with a brown dwarf or low-mass stellar companion.

\emph{TOI-5906AB:} As discussed in Sect.~\ref{detection}, the event detected in the \textit{TESS} data corresponds to an eclipse of a secondary star rather than a planetary transit. Consequently, we modified the default configuration of \texttt{EXOFASTv2} to jointly model the RVs and the eclipse, together with the SED and the MIST stellar evolution tracks for both stellar components, similar to the procedure outlined in \cite{2025arXiv251121817F}. Among the binary systems studied here, this is the only system for which we modeled both stars simultaneously, as the secondary is bright enough to produce a SB2, unlike the companions in the other systems. We imposed a uniform prior on the time of secondary eclipse, $\mathcal{U}(2459831.24, 2459831.29)$ BJD$_{\rm TDB}$. In addition, we applied a lower bound of $\cos i > 0.03$, corresponding to $i < 88.3^\circ$, consistent with the non-detection of the primary eclipse. Our analysis indicates that the binary has an orbital period of $30.116874\pm0.000022$~d. The primary star, TOI-5906A, has a mass of $0.968^{+0.058}_{-0.057}\,M_\odot$ and a radius of $1.057^{+0.040}_{-0.037}\,R_\odot$, while the secondary, TOI-5906B, has a mass of $0.647 \pm 0.021\,M_\odot$ and a radius of $0.625 \pm 0.024\,R_\odot$. TOI-5906B is also detected in the combined Hipparcos/\textit{Gaia} astrometric dataset (see Sect.~\ref{Astrometry}).

\begin{table}
\centering
\caption{Median values and 68\% confidence intervals for TOI-5906AB. The definitions of all parameters are provided in Tables~\ref{tab:GP_params_complete1} and \ref{tab:GP_params_complete2}. The only additional parameters introduced here are the eclipse-specific quantities: $b_{\mathrm{S}}$, the eclipse impact parameter; $\tau_{\mathrm{S}}$, the ingress/egress eclipse duration; and $T_{\mathrm{S},14}$, the total eclipse duration.}
\resizebox{0.4\textwidth}{!}{%
\begin{tabular}{lcc}
\hline
Parameters & \multicolumn{2}{c}{Values} \\
\hline
Stellar Parameters:&A&B \\
~~~~$M_*$$(M_\odot) $              & $0.967^{+0.058}_{-0.057}$&$0.647\pm0.021$  \\
~~~~$R_*$$(R_\odot) $               & $1.057^{+0.040}_{-0.038}$&$0.625\pm0.024$  \\
~~~~$L_*$ ($L_\odot$)             & $1.27^{+0.17}_{-0.10}$&$0.160^{+0.021}_{-0.018}$ \\
~~~~$\rho_*$ (cgs)           & $1.16^{+0.15}_{-0.14}$&$3.74^{+0.44}_{-0.38}$\\
~~~~$\log{g}$ (cgs)          & $4.377^{+0.038}_{-0.042}$&$4.657^{+0.032}_{-0.031}$\\
~~~~$T_{\rm eff}$(K)      & $5970^{+190}_{-160}$&$4618^{+100}_{-93}$\\
~~~~$[{\rm Fe/H}]$ (dex)    & $-0.202\pm0.047$&$-0.150^{+0.063}_{-0.064}$\\
~~~~$Age$ (Gyr)              & $6.0^{+3.3}_{-2.6}$&$6.0^{+3.3}_{-2.6}$ \\
~~~~$d$ (pc)               & $289.4\pm1.5$&$289.4\pm1.5$\\

\smallskip\\Orbital Parameters& \multicolumn{2}{c}{TOI-5906B}\\
~~~~$P$ (days)               & \multicolumn{2}{c}{$30.116874\pm0.000022$} \\
~~~~$T_C$ (BJD$_{TDB}$)             &  \multicolumn{2}{c}{$2459807.8329^{+0.0084}_{-0.0085}$}\\
~~~~$a$ (AU)               &  \multicolumn{2}{c}{$0.2222\pm0.0036$} \\
~~~~$i$ (Degrees)                &  \multicolumn{2}{c}{$87.46$} \\
~~~~$e$                & \multicolumn{2}{c}{$0.47138^{+0.00055}_{-0.00054}$} \\
~~~~$\omega$         & \multicolumn{2}{c}{$-18.16\pm0.14$}\\
~~~~$K$ (m/s)              &  \multicolumn{2}{c}{$36440^{+50}_{-49}$} \\


\smallskip\\\multicolumn{2}{l}{Secondary Eclipse Parameters:}\\
~~~~$b_S$              & \multicolumn{2}{c}{$2.296^{+0.085}_{-0.080}$}\\
~~~~$\tau_S$ (days)          &\multicolumn{2}{c}{$0.0710^{+0.0015}_{-0.0012}$}\\
~~~~$T_{S,14}$ (days)        & \multicolumn{2}{c}{$0.1419^{+0.0031}_{-0.0023}$} \\
~~~~$\delta$  & \multicolumn{2}{c}{$0.0124^{+0.0048}_{-0.0038}$} \\
\hline
\end{tabular}%
}
\label{tab:toi5609}
\end{table}

\begin{figure*}[h!]
    \centering
    \includegraphics[width=0.33\linewidth]{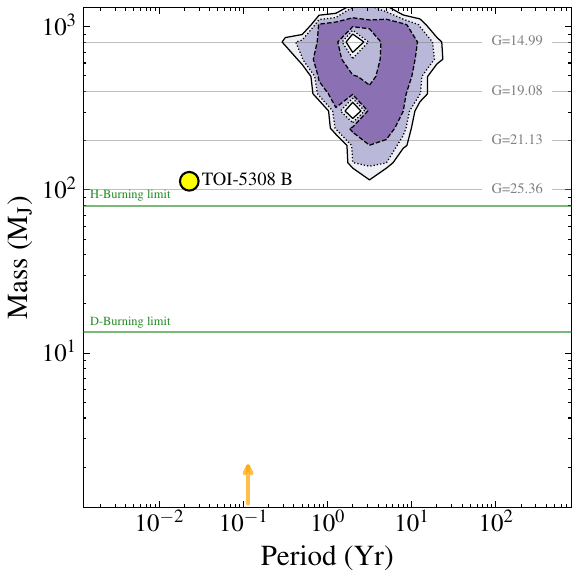}
    \includegraphics[width=0.33\linewidth]{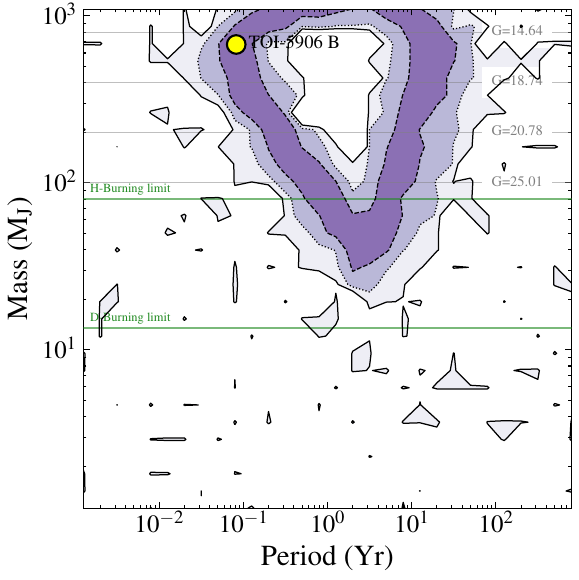}
    \includegraphics[width=0.33\linewidth]{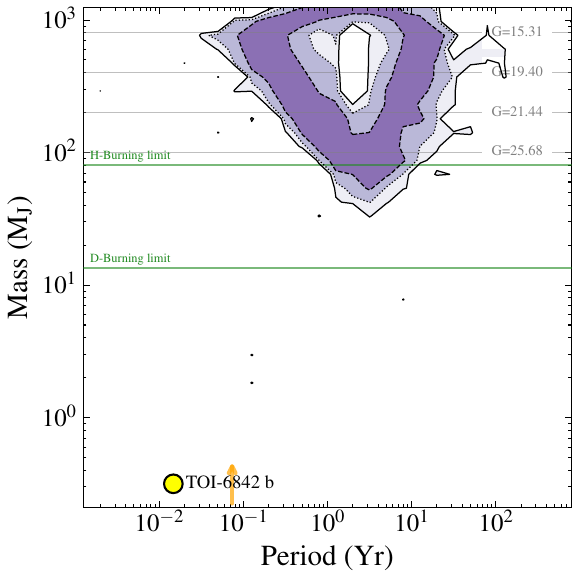}
   \caption{Astrometric detection maps of three stars with statistically significant astrometric signal ($\delta_{\texttt{ruwe}}\geq3\sigma$). The contours represent, from lightest to darkest shading, 68.3\%, 95.4\%, and 99.97\% confidence levels for the companion's mass and period. \textit{Left}\,: TOI-5308 ; \textit{center}\,: TOI-5906 ; \textit{right}\,: TOI-6842A. The yellow dots represent the solutions presented in Tables~\ref{tab:GP_params_complete1}, \ref{tab:GP_params_complete2}, and \ref{tab:toi5609}. The horizontal green lines are at 13.5\,M$_J$ and 80\,M$_J$, respectively the planetary and brown dwarfs upper limits. On the left and right plots, the orange arrow shows the minimum period for stability in the case of an S-type system ($P=5\times P_{in}$, where $P_{in}$ is the period of the companion detected in transit). The gray lines are isochrones of apparent \textit{Gaia} magnitudes computed using COND evolution models from \cite{baraffe03}.}
    \label{fig:pmex}
\end{figure*}

\emph{TOI-5308AB:} The orbital eccentricity is consistent with zero ($e = 0.0031^{+0.0041}_{-0.0022}$), and a circular model is strongly preferred ($\Delta\mathrm{BIC} \approx 14$). We also detect bimodality in the posterior distributions of several stellar and companion parameters. (Fig.~\ref{image_5308}). Following the approach applied to other systems with bimodal posteriors, we separate the two solutions and present the higher-probability solution (85\%) here, while the lower-probability solution (15\%) is provided in Appendix~\ref{bimodal}. Most planetary parameters are consistent within $1\sigma$, whereas moderate discrepancies ($\sim 2$--$3\sigma$) are observed in the stellar age, mass, EEP, planetary mass, and semi-major axis.  

The higher-probability solution indicates that the companion orbits TOI-5308A with a period of $P = 8.3403287^{+0.0000066}_{-0.0000067}$~d, a radius of $R = 1.405^{+0.046}_{-0.045}\,R_{\rm J}$ ($0.1470^{+0.0049}_{-0.0048}\,R_\odot$), and a mass of $M = 105.8^{+3.2}_{-3.6}\,M_{\rm J}$ ($0.1088^{+0.0024}_{-0.0021}\,M_\odot$).

\emph{TOI-5321AB:} Due to the rapid stellar rotation, spectroscopic measurements of $T_{\mathrm{eff}}$ and [Fe/H] could not be obtained. As a result, a simultaneous modeling of the star and its companion could not be reliably performed. We therefore applied Gaussian priors on $R_\star$, $M_\star$, [Fe/H], and $T_{\mathrm{eff}}$, adopting values from the independent stellar analysis presented in Sect.~\ref{stellar}.

In addition, an initial run without imposing any prior on the RV jitter term failed to converge, with the posterior distribution of the jitter exploring unrealistically large values. This behavior is primarily driven by the small number of RV measurements—only nine observations from SOPHIE and two from FEROS. We therefore adopted a uniform prior on the RV jitter for both RV datasets, between 0 and seven times the median RV uncertainty, which is conservatively large.

The companion has an orbital period of $P = 10.1392043^{+0.0000078}_{-0.0000076}$\,d, a radius of $R_{\mathrm{P}} = 1.466^{+0.052}_{-0.053}\,R_{\mathrm{J}}$ ($0.1507^{+0.0053}_{0.0054}\,R_\odot$), and a mass of $M_{\mathrm{P}} = 127.2\pm4.4\,M_{\mathrm{J}}$ ($0.1214\pm0.0042\,M_\odot$), consistent with a stellar object.

\emph{TOI-3678AB:} Similarly, TOI-3678A is a fast rotator, so we adopted the same stellar-modelling setup as for TOI-5321A. The best-fit parameters indicate a stellar companion with a $P =4.8542845\pm0.0000024$ d, $R = 1.884^{+0.053}_{-0.055}~ R_J$ ($0.1935^{+0.0055}_{-0.0057}\,R_\odot$), mass of $167.9^{+6.5}_{-6.6}~M_J$ ($0.1602\pm0.0062\,M_\odot$), and a low eccentricity of $e = 0.0198^{+0.0022}_{-0.0020}$.

\section{Combining RV and Hipparcos/\textit{Gaia} astrometry data}
\label{Astrometry}

\begin{figure*}
\centering
      \begin{tabular}{cc}
          \includegraphics[width=.45\textwidth]{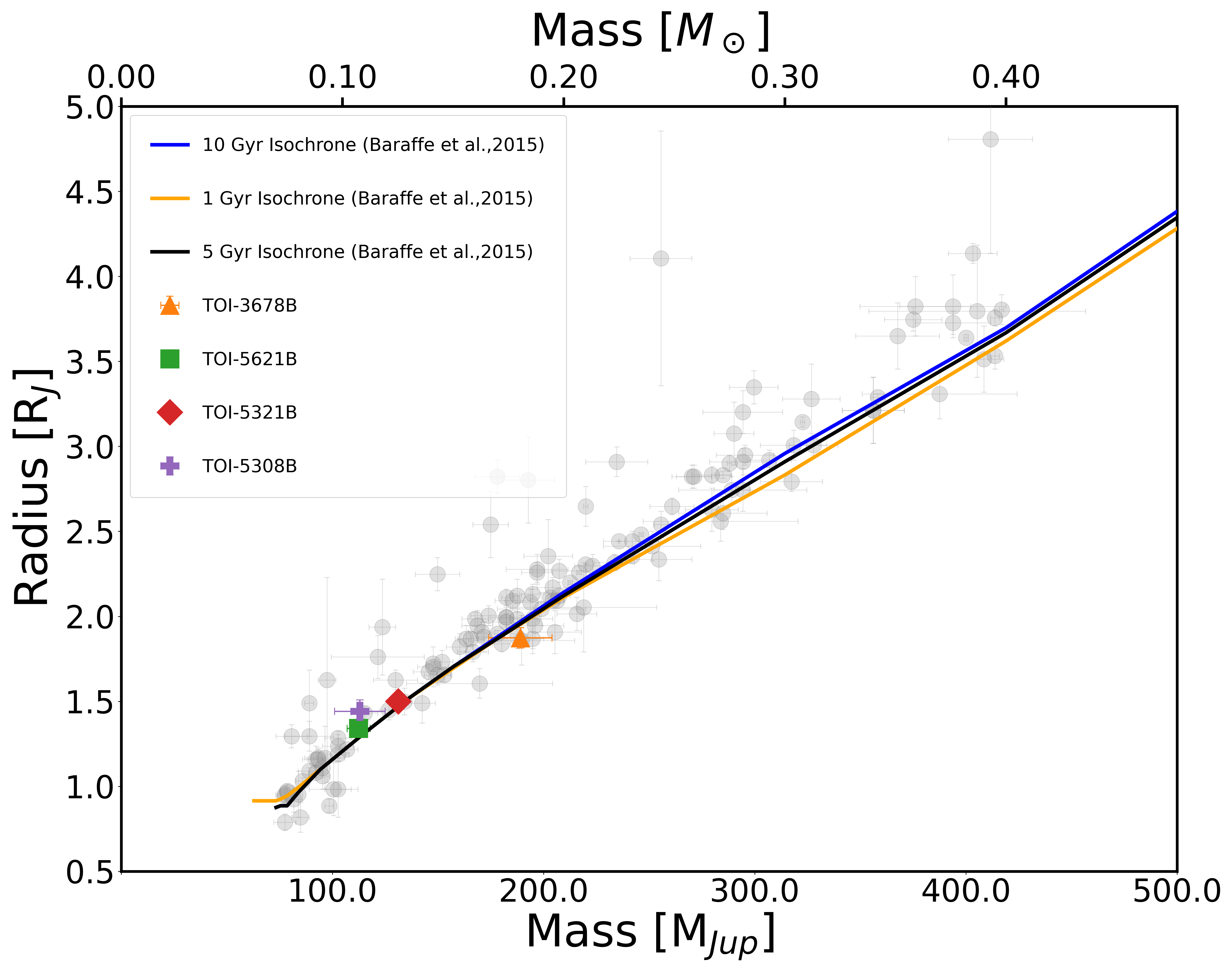} 
          \includegraphics[width=.45\textwidth]{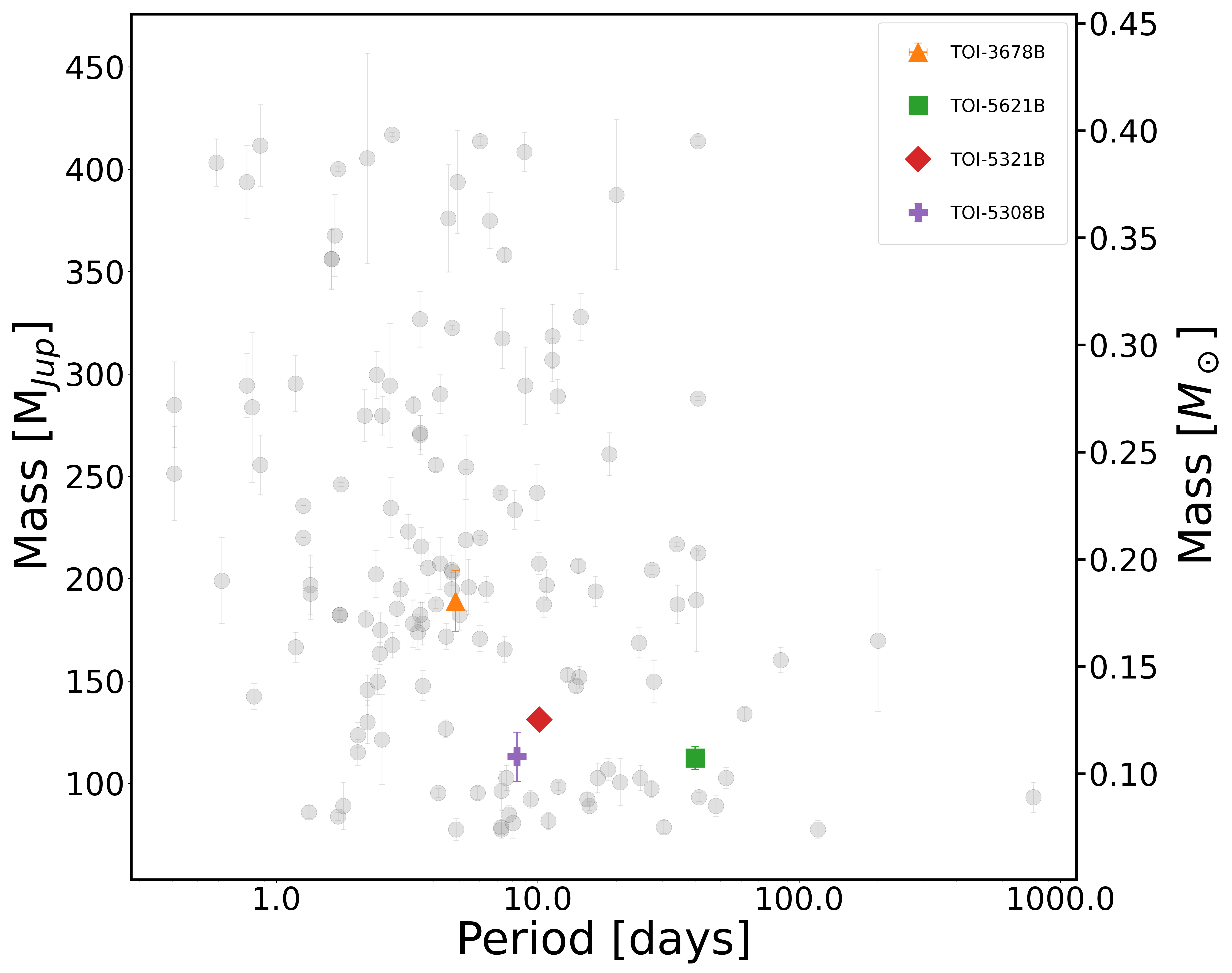}\\
          \includegraphics[width=.45\textwidth]{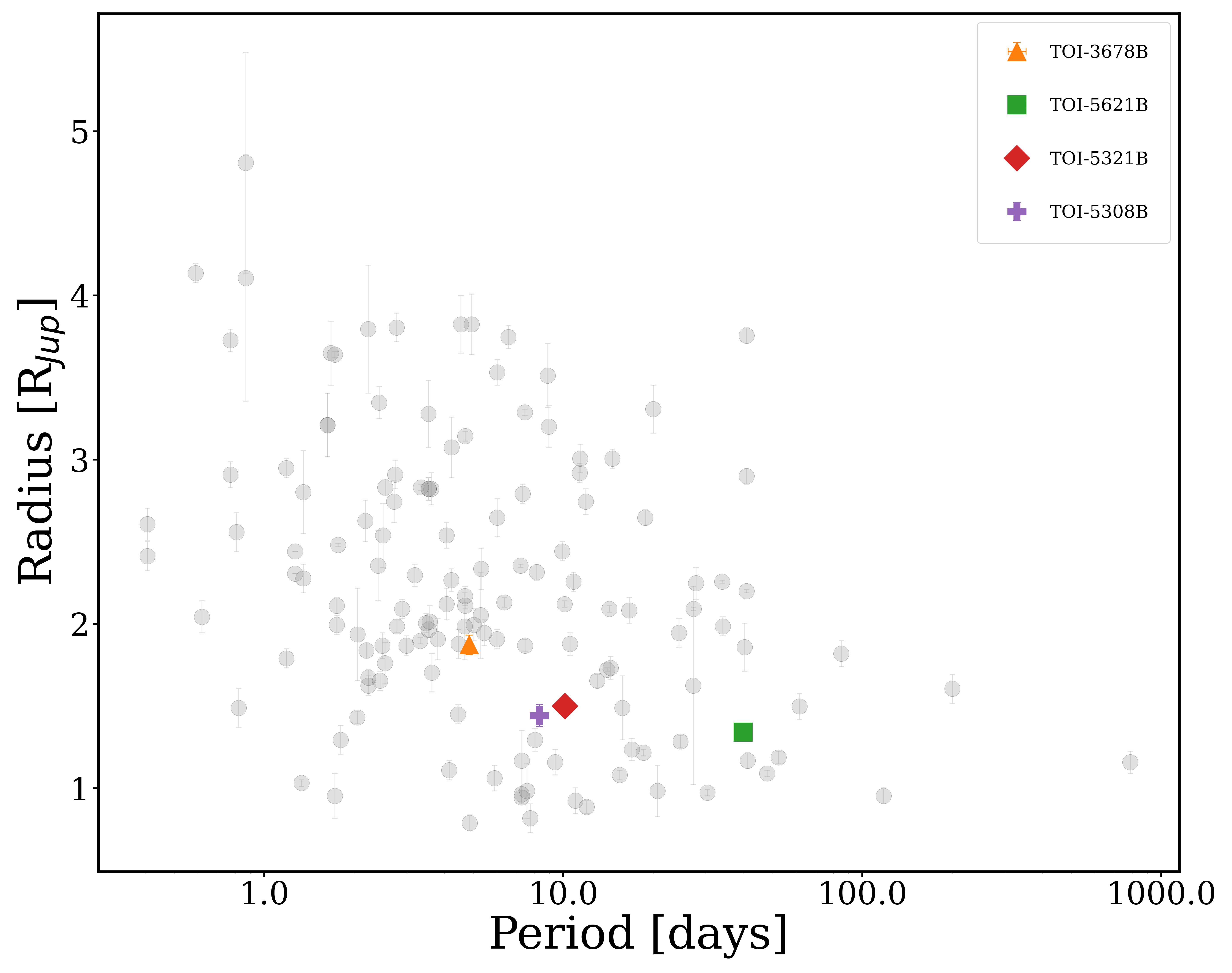}
          \includegraphics[width=.45\textwidth]{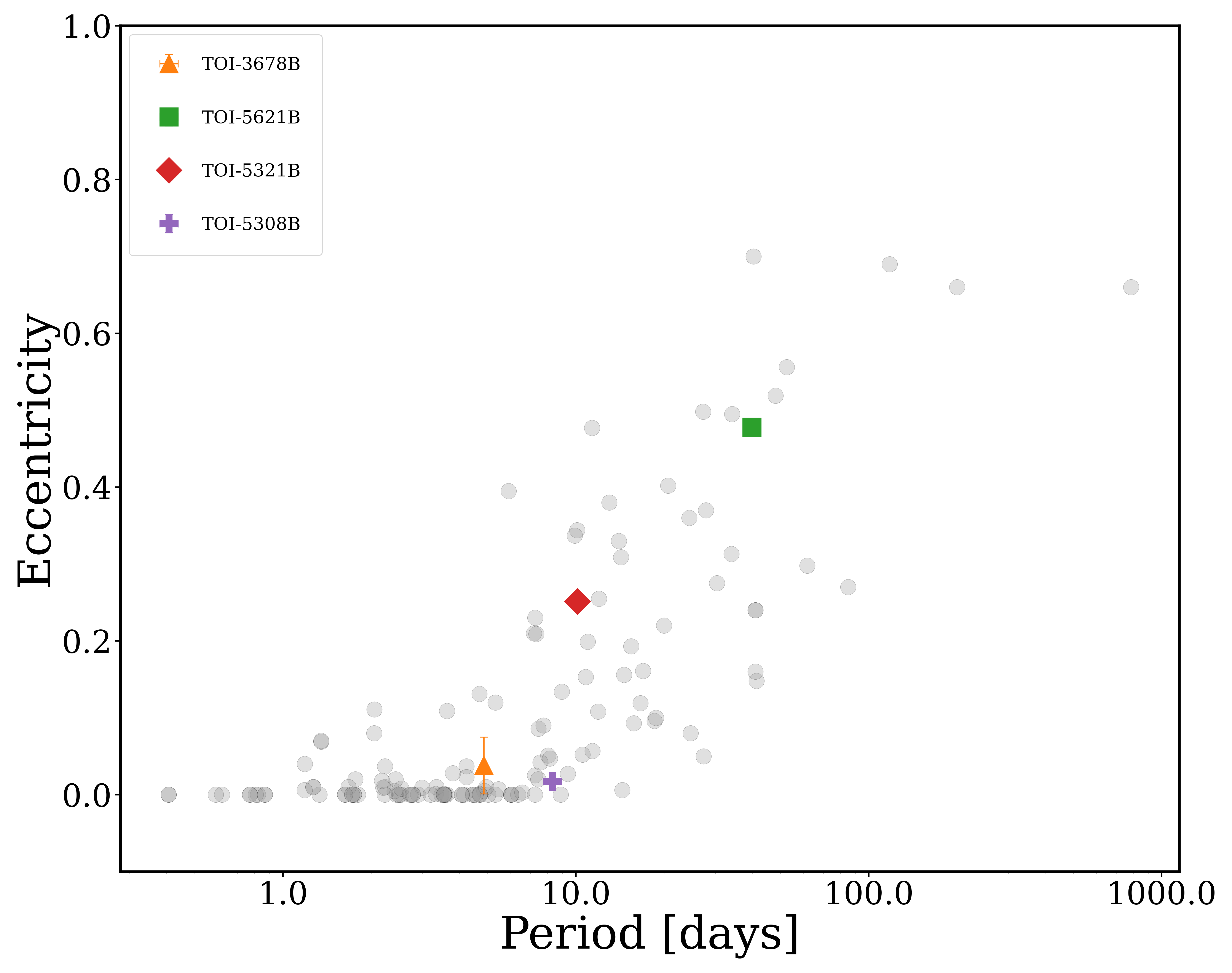}
    \end{tabular}
\caption{Properties of low-mass stars analyzed in this work, compared to known low-mass stellar companions from the EBLM catalog and other literature sources \citep{maxted2023eblm}. Top left: Mass–radius diagram showing the positions of TOI-5308~B, TOI-5621~B, TOI-3678~B, and TOI-5321~B relative to theoretical isochrones from \citet{baraffe2015new}. Top right: Orbital period versus mass. Bottom left: Radius–period diagram. Bottom right: Orbital eccentricity as a function of orbital period. The stars analyzed in this work are highlighted in color in all panels.}
\label{eblm}
\end{figure*}

In addition to the transit photometry and RV study, we used \textit{Gaia} astrometry to confirm the detections, or further constrain the system. To do so, we used the \texttt{GaiaPMEX} \citep{kiefer25a} tool, which uses the statistical significance of the re-calibrated renormalized unit weight error (\texttt{ruwe}) to determine the likelihood that a star has a bound companion. In short, as a first step, \texttt{GaiaPMEX} computes the significance of \textit{Gaia}'s astrometric excess noise. It then simulates multiple astrometric \textit{Gaia} orbits following the null hypothesis of the star being single, then performs a linear five-parameters fit (position, proper motion and parallax) and finally compares the distribution of the residuals, the \texttt{ruwe}, to the \textit{Gaia} DR3 values \citep{Gaia23}. The deviation between the simulated "single star" \texttt{ruwe} distribution and the \textit{Gaia} DR3 value is the astrometric significance $\delta_{\texttt{ruwe}}$. It represents in numbers of sigma, the probability that the \textit{Gaia} data does not follow the null hypothesis; i.e., the star is not single. Among our 15 stars, three show an astrometric significance above 3$\sigma$: TOI-5308A ($\delta_{\texttt{ruwe}} > 8\sigma$) and TOI-5906A ($\delta_{\texttt{ruwe}} = 3\sigma$), both hosting low-mass stellar companions (see Sect.~\ref{exofast_results}), and TOI-6842A ($\delta_{\texttt{ruwe}} = 4\sigma$), which hosts a planetary companion (see Sect.~\ref{exofast_results}) as well as a stellar companion (see Sect.~\ref{data_tess}).

The second step of the \texttt{GaiaPMEX} analysis is to characterize the companion detected in astrometry, by plotting the detection maps in the (period, mass) parameter space. These maps show the companion's orbital period and mass permitted by the astrometric data at the 1, 2, and 3$\sigma$ confidence levels (68.3\%, 95.4\%, and 99.7\% confidence, respectively). We show the \texttt{GaiaPMEX} maps in Fig.\,\ref{fig:pmex}, where we included the solution found in transit+RVs. The astrometric detection map for the companion of TOI-5906\,B is consistent with the transit+RVs solution, as seen on the center panel of Figure\,\ref{fig:pmex}. We thus consider that the astrometric signal is due to the transiting binary. 

For the two other cases, TOI-5308A and TOI-6842A, the solution found in transit+RVs is inconsistent with the solutions in astrometry, which suggests that the system is composed of a third object. TOI-5308\,C and TOI-6842\,C are likely to be stellar companions, with minimum masses of approximately 115\,M$_J$ and 32\,M$_J$\footnote{Although the minimum mass of TOI-6842\,C lies within the brown dwarf regime, most of the parameter space is above the H-burning limit at 80\,M$_J$.}, respectively. Both systems are in consequence S-type, i.e., an orbit at short period with an additional long period companion. We can use the criterion for stability on the period ratio to further characterize the orbit of the long-period astrometric companion. This ratio is $P_{out}/P_{in}\lesssim 5$ \citep{Holman99}, where $P_{out}$ is the period of the wide binary detected in astrometry and $P_{in}$ is the period of the inner companion detected in transit. However, all the possible periods for TOI-5308\,C and near all those for TOI-6842\,C are longer than five times the respective inner periods as seen on left and right panels of Fig.\,\ref{fig:pmex}, which means these two systems meet the stability criterion. No significant signatures of these companions are detected in our current RV datasets.

\section{Discussion}
\label{Discussion}

This section is divided into two parts based on the nature of the companion, either planetary or low-mass stellar. In each part, we place the systems in context and highlight their most interesting properties.

\subsection{Comparison to known very low mass stars}

Many studies have suggested that magnetic activity could be responsible for the so-called “radius inflation” observed in M dwarfs \citep[e.g.,][]{morales2022fundamental}. In particular, close binary systems are expected to show more pronounced inflation because tidal interactions with the primary star can increase the rotation rate of the secondary, enhancing its magnetic dynamo and potentially inflating its radius. Despite these theoretical expectations, the observational evidence remains mixed. Some studies \citep[e.g.,][]{von2019eblm} found no clear correlation between orbital period and radius inflation, while others \citep[e.g.,][]{swayne2024eblm} reported stronger inflation in short-period binaries. 

\begin{figure*}[h!]
\centering
      \begin{tabular}{cc}
          \includegraphics[width=.42\textwidth]{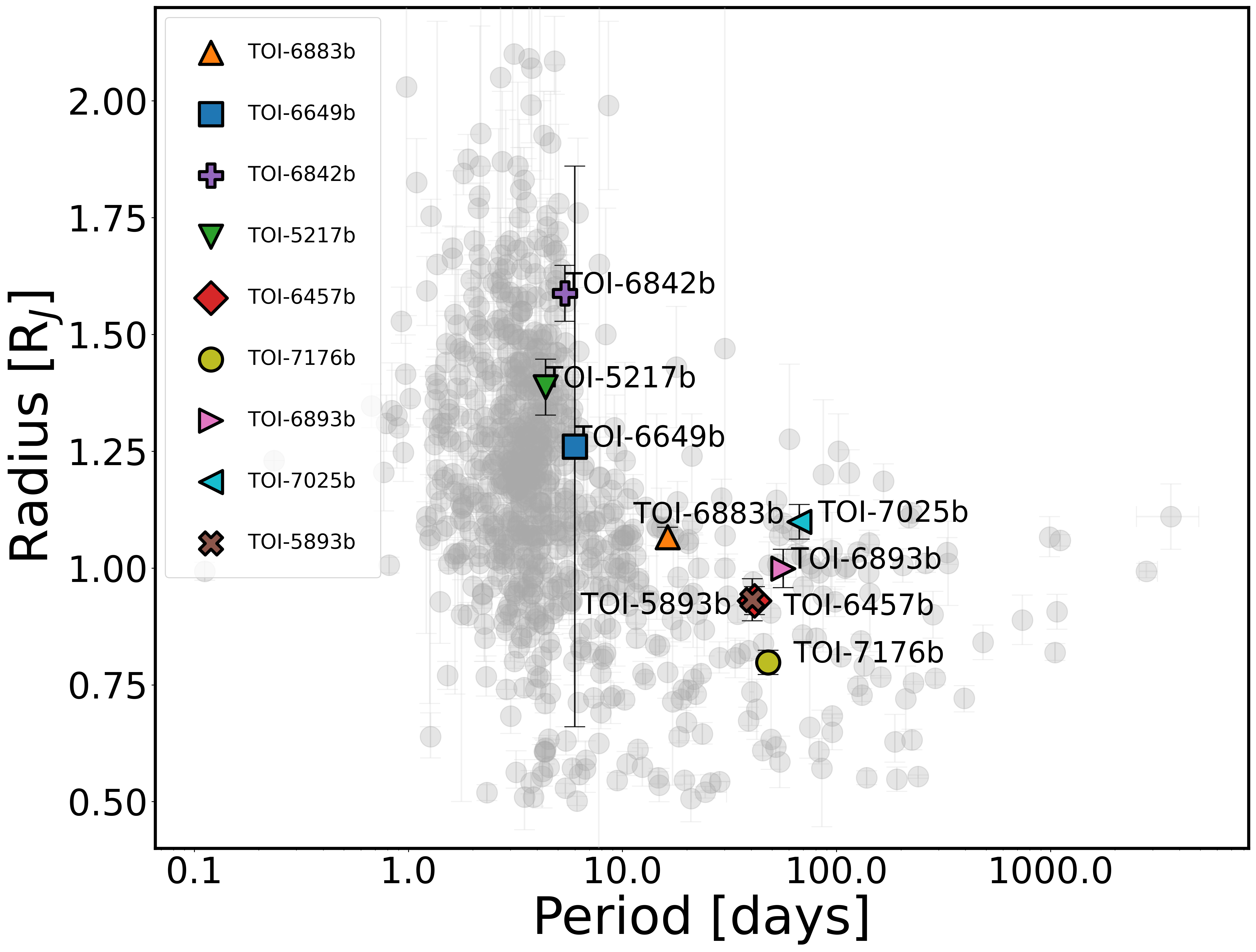} 
          \includegraphics[width=.42\textwidth]{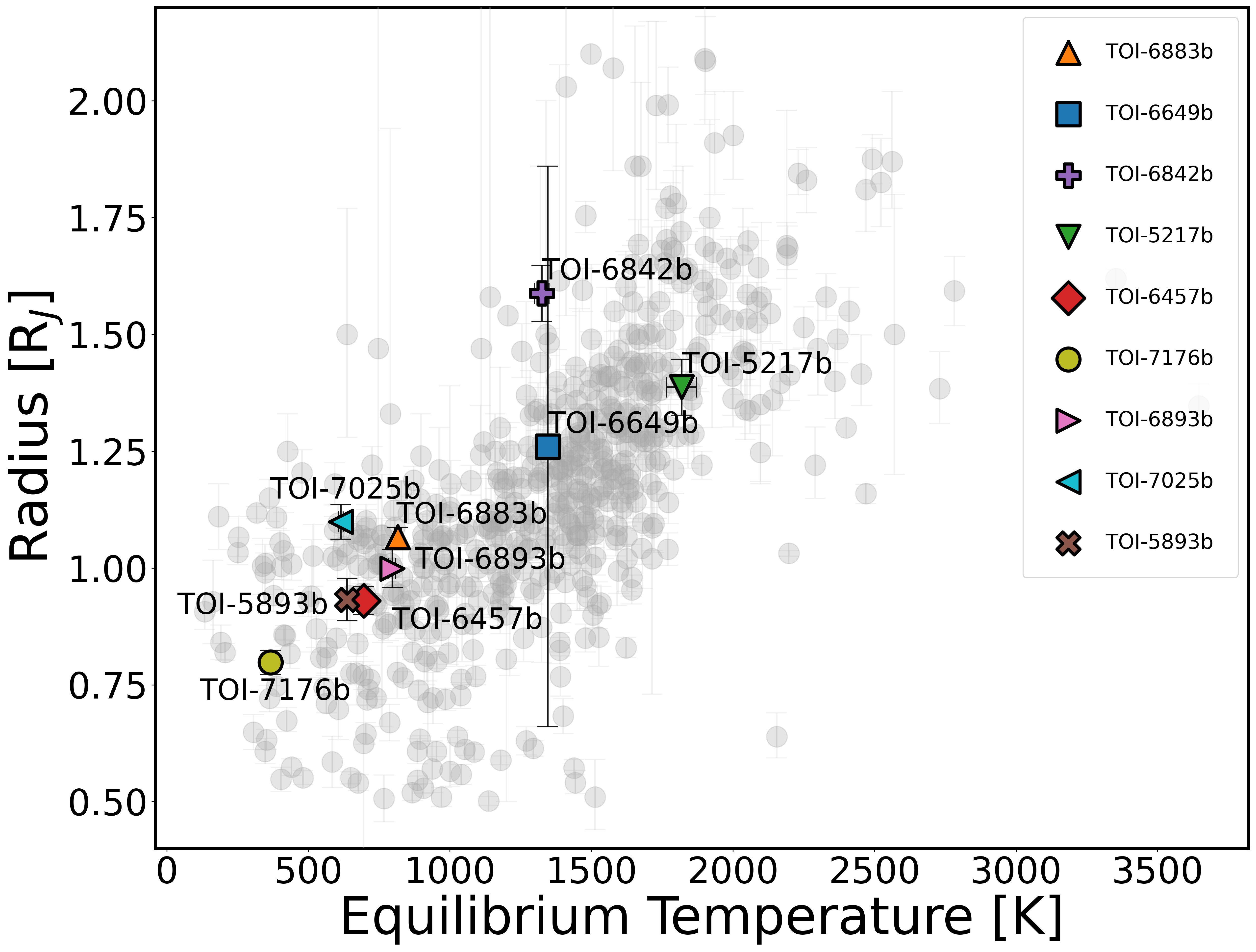}\\
          \includegraphics[width=.42\textwidth]{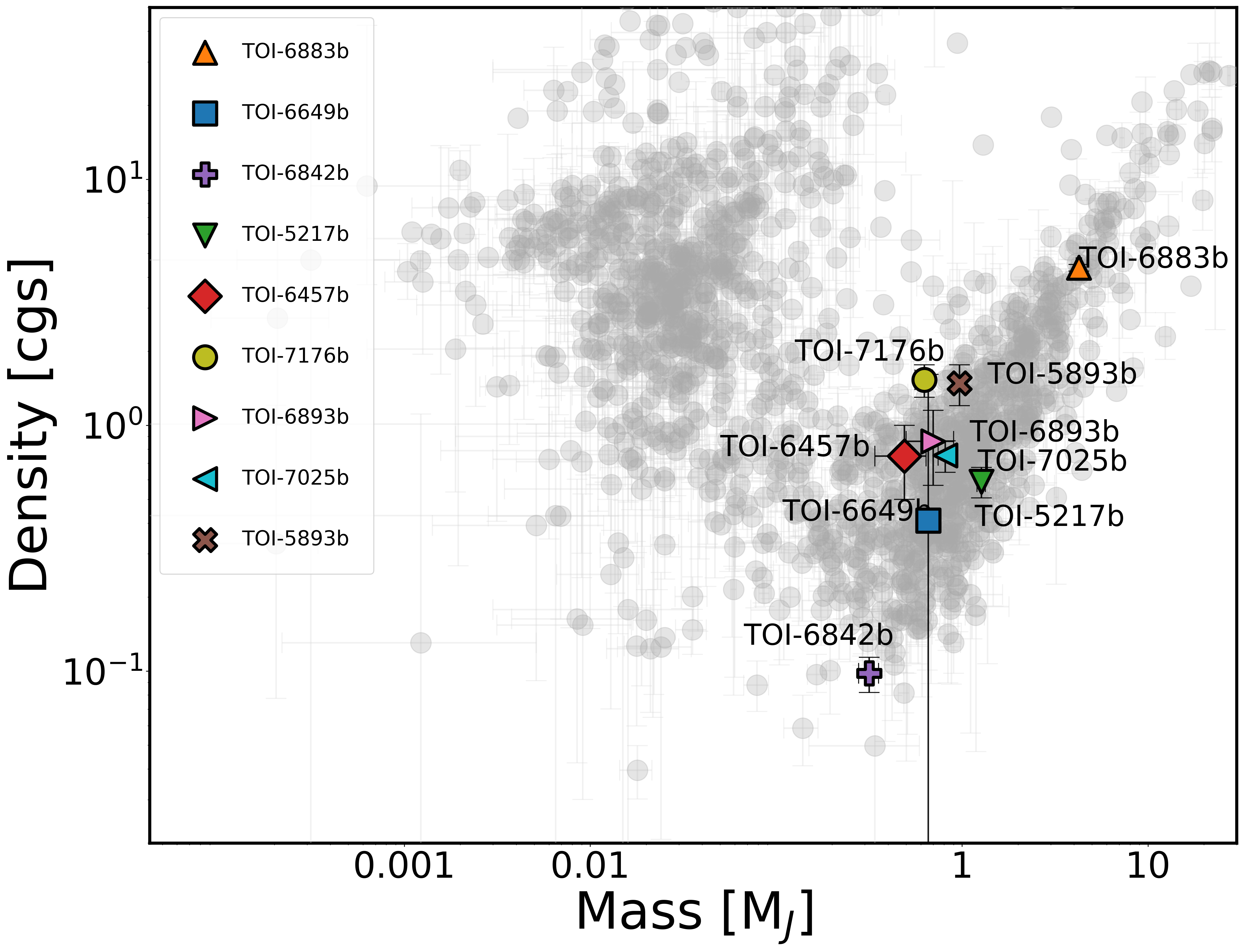}
          \includegraphics[width=.42\textwidth]{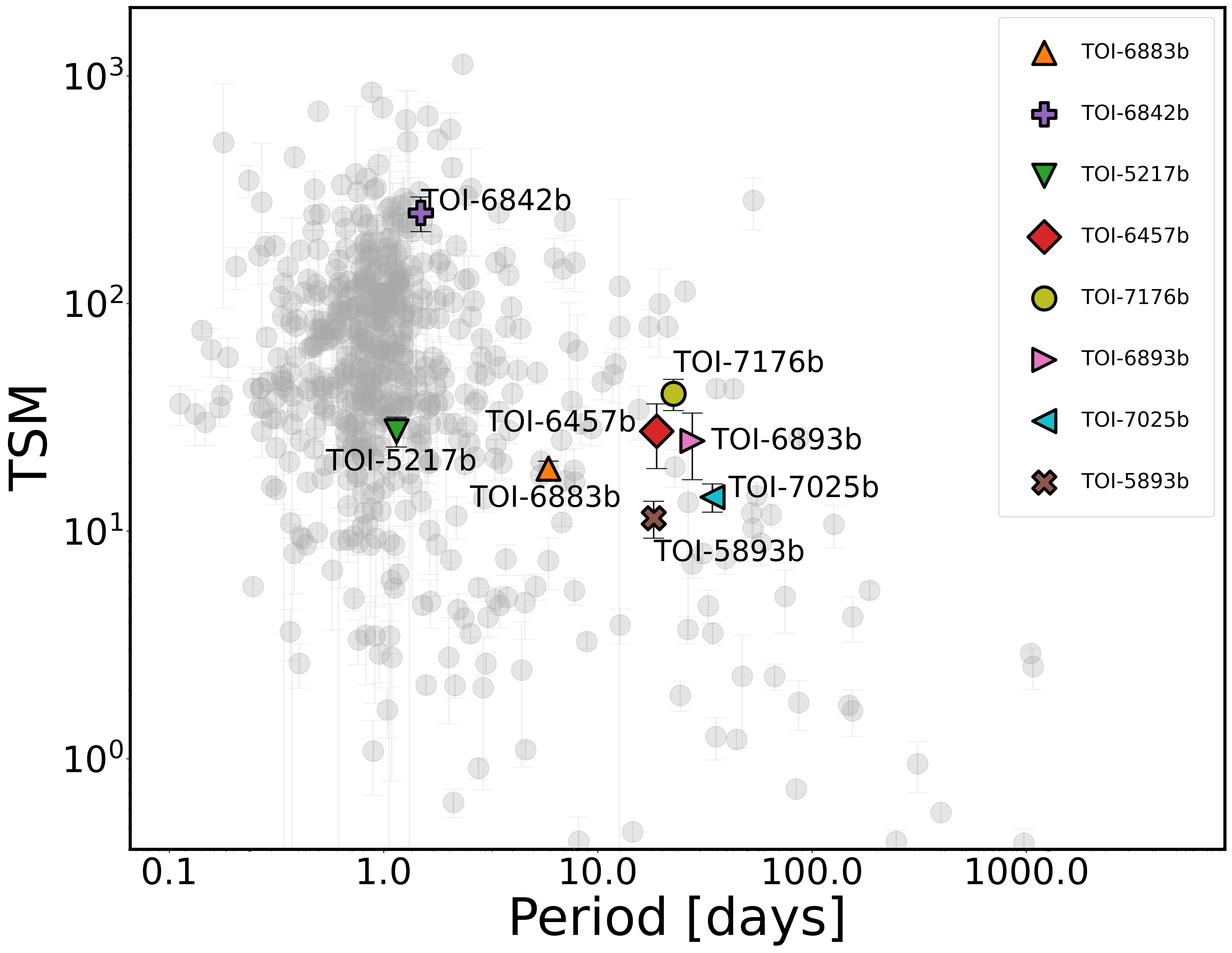}
    \end{tabular}
    \caption{Position of the planets analyzed in this work among the population of known transiting exoplanets with measured masses and radii, compiled from the NASA Exoplanet Archive (as of September~15,~2025), without applying any precision cut. Top left: radius–orbital period diagram for giant planets ($R_\mathrm{p} > 0.5\,R_\mathrm{J}$). Top right: planetary radius as a function of equilibrium temperature for giant planets, assuming zero Bond albedo and full heat redistribution. Bottom left: planetary density–mass diagram. Bottom right: TSM as a function of orbital period for giant exoplanets with known masses, radii, and equilibrium temperatures. Planets discussed in this study are highlighted in color.}
    \label{four_figures}
\end{figure*}

The top-left panel of Fig.~\ref{eblm} shows the mass–radius diagram for the low-mass stars detected in this work—TOI-5308~B, TOI-5621~B, TOI-3678~B, and TOI-5321~B—compared with other low-mass stars from the catalog of \citet{maxted2023eblm}, which compiles systems from the EBLM project and other binaries with radii measured to better than 5\% precision. Isochrone models from \citet{baraffe2015new} are overplotted. Within the uncertainties, we find no significant evidence for radius inflation. We note that TOI-5609~B was excluded from this comparison because its mass ($\sim$0.6~$M_\odot$) is substantially higher than those of the stars in the \citet{maxted2023eblm} sample, which are all below 0.5~$M_\odot$.

In the top-right and bottom-left panels of Fig.~\ref{eblm}, we compare the masses and radii of our low mass targets as a function of orbital period with the same comparison sample. As shown in these panels, TOI-5621~B, with an orbital period of 40~d, is among the longest-period low-mass binaries currently known. The detection and characterization of such long-period systems, which are potentially less affected by proximity effects from the host star, are particularly valuable for improving our understanding of the role of magnetic activity and other processes in shaping the observed radii of low-mass stars.

Tidal interactions are expected to circularize the orbits of close binary stars through angular momentum exchange and energy dissipation within the stellar interiors \citep{goupil2008tidal}. The canonical circularization period is typically found to lie between 7 and 9~d \citep{triaud2017eblm}, although this range depends on factors such as the stellar mass ratio and the age of the system. The bottom-right panel of Fig.~\ref{eblm} shows orbital eccentricity as a function of period for the same sample. Consistent with previous studies, TOI-3678~B, with its short period of $P = 4.8542845 \pm 0.0000024$~d, exhibits a nearly circular orbit. The eccentricity generally increases with orbital period: TOI-5321~B, with $P = 10.1392041^{+0.0000078}_{-0.0000076}$~d, and TOI-5621~B, with $P = 39.97547 \pm 0.00012$~d, both display eccentric orbits. We note that although the eccentricity of TOI-5308~B, which has a period of $P = 8.3403287^{+0.0000066}_{-0.0000067}$~d, is not tightly constrained by our data, it remains very small, with a nominal value of $e = 0.0031^{+0.0041}_{-0.0022}$, in agreement with the findings of \citet{triaud2017eblm}.

\subsection{Comparison to known planets}
\label{discu}

Figure~\ref{four_figures} (top panel, left) presents the radius–period diagram of transiting giant exoplanets (R~>~0.5~R$_\mathrm{J}$) with known masses and radii, taken from the NASA Exoplanet Archive as of September~15,~2025. In particular, TOI-6842Ab, TOI-5217Ab, and TOI-6649b are located in the hot-Jupiter region, while TOI-6883Ab, TOI-7176b, TOI-6457b, TOI-6893Ab, TOI-7025b, and TOI-5893b, with longer orbital periods, fall within the warm-Jupiter regime, a region that remains sparsely populated. Indeed, only about 5\% of transiting planets with measured masses are warm Jupiters with periods longer than 40 d, comparable to those of the warm Jupiters characterized in this work.

In the top-right panel of Fig.~\ref{four_figures}, we show the same sample of planets in the radius–equilibrium temperature plane, assuming zero albedo and uniform heat redistribution throughout the planetary photosphere. The figure clearly illustrates the phenomenon of hot-Jupiter radius inflation—the unexpectedly large radii of giant planets with equilibrium temperatures above roughly 1000~K. This phenomenon remains poorly understood \citep[e.g.,][]{miller2011heavy, demory2011lack,thorngren2024hot}. Several mechanisms have been proposed to explain this anomaly, typically invoking the deposition of additional energy into the planetary interior. Such processes heat the convective layers, slow the planet’s contraction, and lead to radii exceeding those predicted by standard evolutionary models \citep[e.g.,][]{batygin2011evolution,arras2010thermal,sarkis2021evidence}.

The planets TOI-6842Ab, TOI-5217Ab, and TOI-6649b are all highly irradiated ($T_{\rm eq} > 1000$~K) and follow the well-established trend of increasing radius inflation with higher equilibrium temperatures. Among them, TOI-6842Ab appears more inflated than most of the known giant-planet population. In contrast, the warm Jupiters TOI-6883Ab, TOI-7176b, TOI-6457b, TOI-6893Ab, TOI-7025b, and TOI-5893b, with equilibrium temperatures between 377 and 846~K, do not exhibit significant inflation. These planets therefore provide valuable control cases for testing and constraining models proposed to explain the hot-Jupiter radius anomaly.

In the bottom-left panel of Fig.~\ref{four_figures}, we present the density–mass diagram for all known transiting exoplanets with both measured masses and radii. Without applying any precision cut to the comparison sample, TOI-6842Ab lies among the nine lowest-density planets detected to date, with a bulk density of $\rho_\mathrm{p}= 0.084^{+0.011}_{-0.010}~\mathrm{g\,cm^{-3}}$, placing it within the so-called ``super-puff'' population (planets with $\rho_\mathrm{p} \leq 0.1~\mathrm{g\,cm^{-3}}$).

Within this regime, TOI-6842Ab ($R_\mathrm{p} = 1.647 \pm 0.039~R_\mathrm{J}$, $M_\mathrm{p} = 0.302 \pm 0.035~M_\mathrm{J}$, $T_\mathrm{eq} = 1327^{+19}_{-18}~\mathrm{K}$, age $= 9.6 \pm 2.1~\mathrm{Gyr}$, G-type host) is comparable to HAT-P-67b ($M_\mathrm{p} = 0.45 \pm 0.15~M_\mathrm{J}$, $R_\mathrm{p} = 2.140 \pm 0.025~R_\mathrm{J}$, $\rho_\mathrm{p} = 0.061^{+0.021}_{-0.020}~\mathrm{g\,cm^{-3}}$, $T_\mathrm{eq} = 1903\pm 25~\mathrm{K}$, stellar age $= 1.46^{+0.29}_{-0.26}~\mathrm{Gyr}$, evolved F-type host; \citealt{wang2025revised}), Kepler-12b ($M_\mathrm{p} = 0.431 \pm 0.041~M_\mathrm{J}$, $R_\mathrm{p} = 1.695 \pm 0.030~R_\mathrm{J}$, $\rho_\mathrm{p} = 0.111 \pm 0.010~\mathrm{g\,cm^{-3}}$, $T_\mathrm{eq} = 1480.0\pm 30~\mathrm{K}$, age $= 4.0^{+0.3}_{-0.4}~\mathrm{Gyr}$, evolved G0 host; \citealt{fortney2011discovery,esteves2015changing}), and WASP-17b ($M\mathrm{p} = 0.486 \pm 0.032~M_\mathrm{J}$, $R_\mathrm{p} = 1.991 \pm 0.081~R_\mathrm{J}$, $T_\mathrm{eq} = 2106\pm 41~\mathrm{K}$, age $= 3.0 \pm 2.6~\mathrm{Gyr}$, F4 host; \citealt{anderson2011thermal}).

These planets orbit stars younger than the host star of TOI-6842A\,b. Therefore, TOI-6842A\,b provides a valuable benchmark for investigating how host-star properties influence planetary inflation.


\begin{figure}[h!]
\includegraphics[width=0.9\columnwidth]{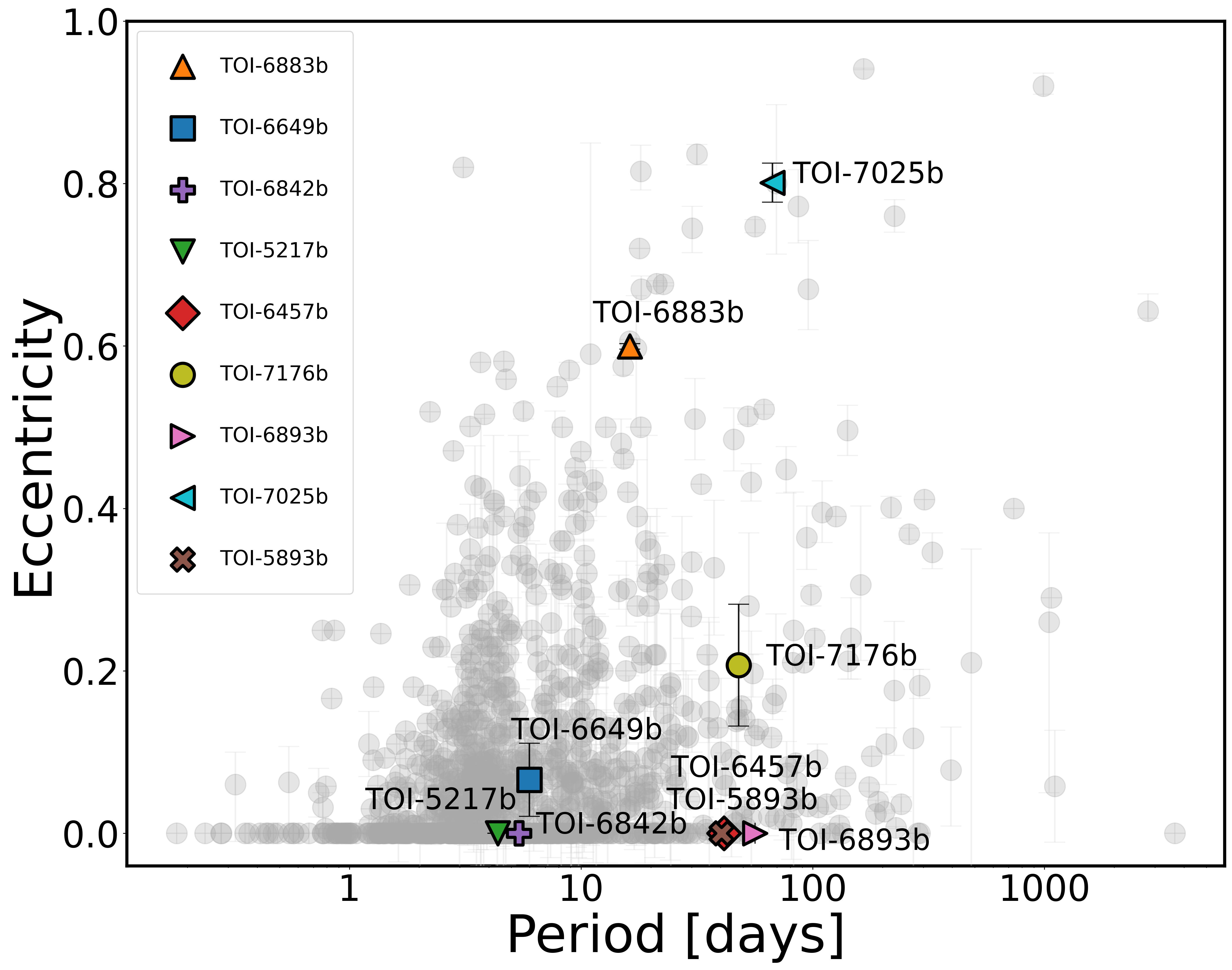}
\caption{Eccentricity-period diagrams of characterized transiting planet in this study together with other exoplanets with known mass and radius from the NASA Exoplanet Data Archive (as of September~15,~2025). The panel highlights the unique position of TOI-7025~b and TOI-6883A~b, as one of the most eccentric planets known to date.}
\label{period_eccentricity}
\end{figure}

Additionally, we investigated the prospects for atmospheric characterization of these planets using the Transmission Spectroscopy Metric (TSM), defined by Equation~5 in \citet{kempton2018framework}. This metric includes a scale factor that depends on the planetary radius: 1.28 for planets with radii between 2.75 and 4~R$_{\oplus}$, and 1.15 for those between 4 and 10~R$_{\oplus}$. For planets larger than 10~R$_{\oplus}$, the scale factor is not formally defined and we thus assumed to be 1. Using this framework, we computed the TSM values and report them in Table~\ref{tab:tsm}. The uncertainty for TOI-6649\,b is extremely large due to its grazing transit geometry.

\begin{table}
\centering
\caption{TSM for the planets in this work.}
\label{tab:tsm}
\begin{tabular}{l c}
\hline\hline
Planet & TSM \\
\hline
TOI-5893b & $11 \pm 2$ \\
TOI-7025b & $14 \pm 2$ \\
TOI-6893Ab & $25 \pm 8$ \\
TOI-7176b & $40\pm 6$ \\
TOI-6457b & $28\pm 9$ \\
TOI-6883Ab & $19 \pm 2$ \\
TOI-6649b & $136 \pm 192$ \\
TOI-6842Ab & $250 \pm 43$ \\
TOI-5217Ab & $28\pm 4$ \\
\hline
\end{tabular}
\end{table}

The bottom-right panel of the Fig. ~\ref{four_figures} displays the TSM as a function of orbital period for all giant exoplanets with measured masses and radii, and equilibrium temperatures. Among the planets studied here, TOI-6842Ab stands out as one of the most favorable targets for atmospheric characterization with both ground-based facilities and space missions such as \textit{JWST} \citep{sabelhaus2004overview} and \textit{Ariel} \citep{tinetti2018chemical}. This planet, along with TOI-5217Ab, is already listed as a candidate in the \textit{Ariel} mission sample \citep{edwards2019updated,zingales2018ariel}. This list is not final and is continuously being updated.

Figure \ref{period_eccentricity} shows the eccentricity-period diagrams. TOI-6883A~b and TOI-7025~b exhibits a high orbital eccentricity of $0.604 \pm 0.002$ and $0.82 \pm 0.02$, respectively, placing them among the most eccentric planets currently known, in particular transiting ones. Moderate eccentricities in giant planets are often explained by interactions with the protoplanetary disk \citep{Duffell2015} or by dynamical scattering with other planets after disk dispersal \citep{Rasio1996,Juric2008}, but these processes generally do not produce eccentricities as extreme as observed here \citep{Goldreich2004,Ida2013,Petrovich2014}. The large eccentricity of TOI-7025~b and TOI-6883A~b therefore points toward a different evolutionary history, such as formation at a larger orbital distance followed by inward migration via high eccentricity migration. Similar extreme eccentric warm Jupiters, including HD80606~b \citep{Naef2001,hebrard2010observation} and TOI-3362~b \citep{Dong2021,espinoza2023aligned}, are often described as “proto-hot Jupiters,” caught in the process of migrating toward close-in circular orbits \citep[e.g.,][]{2003ApJ...589..605W}. While the presence of additional planets in the TOI-7025 and TOI-6883A~b system cannot be ruled out, the current data reveal no evidence for them; more precise observations may be required to detect such companions. Additionally, we detect no significant long-term RV trend in either system.

\section{Summary}
\label{summary}

In this study, we present the analysis of fourteen \textit{TESS} Objects of Interest using a combination of space- and ground-based photometry, high-resolution spectroscopy, and high-spatial-resolution imaging. Our results can be grouped into three categories: hot Jupiters, warm Jupiters, and low-mass stellar companions.

\subsection*{Hot Jupiters:}

We identify three hot Jupiters, TOI-6649\,b, TOI-5217A\,b, and TOI-6842A\,b. These planets have orbital periods ranging from 4.4 to 6.0 d, radii of 1.3–1.6~$R_{\rm J}$, masses of 0.3–1.3~$M_{\rm J}$, and inferred bulk densities of 0.1–0.6~g~cm$^{-3}$. Their equilibrium temperatures range from 1327 to 1817~K, and all three planets follow the radius inflation trend observed for highly irradiated giant planets.

The TOI-5217A system also hosts an outer companion, TOI-5217A\,c, with an orbital period of approximately 1761 d, and a significantly eccentric orbit ($e = 0.53 \pm 0.02$). However, the orbital period remains poorly constrained by the current RV baseline, and its posterior distribution is flat across the explored parameter space. Consequently, the inferred minimum mass ($\sim$ 13~$M_{\rm J}$) is likewise not constrained by the current data. Additional RV observations are required to refine both the orbital period and the companion mass. TOI-6842A\,b stands out as an extremely low-density giant planet, placing it in the super-puff regime. We note, however, that the derived parameters of TOI-6842A\,b may be modestly affected by stellar activity. With a transmission spectroscopy metric of 250, it is an excellent target for atmospheric characterization.

\subsection*{Warm Jupiters:}

We characterize six planets—TOI-5893\,b, TOI-6893A\,b, TOI-6457\,b, TOI-7176\,b, TOI-6883A\,b, and TOI-7025\,b—as warm Jupiters. All of these planets are transiting. The transiting warm Jupiters have orbital periods ranging from 16.2 to 67.0~d, radii of 0.8–1.2~$R_{\rm J}$, and masses of 0.5–4.3~$M_{\rm J}$. Their equilibrium temperatures span 377-843~K, making them valuable low-irradiation benchmarks for testing models of hot-Jupiter inflation.

TOI-7025\,b and TOI-6883A\,b are particularly notable for their high orbital eccentricities of $0.82 \pm 0.02$ and $0.604 \pm 0.002$, respectively, placing them among the most eccentric giant planets known to date.

\subsection*{Low-mass stellar companions:}

Finally, five systems—TOI-5621\,B, TOI-5321\,B, TOI-3678\,B, TOI-5906\,B, and TOI-5308\,B—are identified as eclipsing binaries. These companion have masses in the range 0.1--0.6~$M_\odot$ and radii of 0.1–0.6~$R_\odot$. Using \textit{Gaia} astrometric excess noise, we detect TOI-5906\,B, marking the first discovery of an object identified using all three techniques: transit, RV, and astrometry.

\bibliography{aanda}

@ARTICLE{ps1,
       author = {{Zak}, J. and {Kabath}, P. and {Boffin}, H.~M.~J. and {Liptak}, J. and {Skarka}, M. and {Brahm}, R. and {Gajdo{\v{s}}}, P. and {Bocchieri}, A. and {Itrich}, D. and {Vanzi}, L. and {Pintr}, P. and {Janik}, J. and {Hatzes}, A.},
        title = "{PLATOSpec's first results: Planets WASP-35b and TOI-622b are on aligned orbits, and K2-237b is on a polar orbit}",
      journal = {\aap},
         year = 2025,
        month = oct,
       volume = {702},
          eid = {A266},
        pages = {A266},
          doi = {10.1051/0004-6361/202556115},
archivePrefix = {arXiv},
       eprint = {2508.10145},
 primaryClass = {astro-ph.EP},
       adsurl = {https://ui.adsabs.harvard.edu/abs/2025A\&A...702A.266Z}
}

@ARTICLE{ps2,
       author = {{Rojas}, Felipe I. and {Brahm}, Rafael and {Jones}, Mat{\'\i}as I. and {Catelan}, M{\'a}rcio and {Liptak}, Jozef and {Acu{\~n}a}, Lorena and {Eberhardt}, Jan and {Espinoza}, N{\'e}stor and {Henning}, Thomas and {Jord{\'a}n}, Andr{\'e}s and {Reinarz}, Yared and {Tala Pinto}, Marcelo and {Trifonov}, Trifon and {V{\'\i}tkov{\'a}}, Michaela and {Antonucci}, Luca and {Bakos}, Gaspar and {B{\'o}di}, Attila and {Boyle}, Gavin and {Csubry}, Zolt{\'a}n and {Hartman}, Joel and {Jan{\'\i}k}, Jan and {Kab{\'a}th}, Petr and {Keyes}, Anthony and {Roth}, Markus and {{\v{S}}koda}, Petr and {Spencer}, Alton and {Suc}, Vincent and {Talens}, Geert Jan and {Vaclavik}, Jan and {Vanzi}, Leonardo},
        title = "{TIC65910228b: A single-transit discovery of a massive long-period warm Jupiter with TESS}",
      journal = {arXiv e-prints},
         year = 2026,
        month = feb,
          eid = {arXiv:2602.13125},
        pages = {arXiv:2602.13125},
          doi = {10.48550/arXiv.2602.13125},
archivePrefix = {arXiv},
       eprint = {2602.13125},
 primaryClass = {astro-ph.EP},
       adsurl = {https://ui.adsabs.harvard.edu/abs/2026arXiv260213125R}
}

@article{pecaut2013intrinsic,
  title={Intrinsic colors, temperatures, and bolometric corrections of pre-main-sequence stars},
  author={Pecaut, Mark J and Mamajek, Eric E},
  journal={The Astrophysical Journal Supplement Series},
  volume={208},
  number={1},
  pages={9},
  year={2013},
  publisher={The American Astronomical Society}
}

@ARTICLE{ps3,
       author = {{Gajdo{\v{s}}}, Pavol and {Brahm}, Rafael and {Acu{\~n}a-Aguirre}, Lorena and {Jones}, Mat{\'\i}as I. and {Salinas}, Helem and {Lipt{\'a}k}, Jozef and {Jord{\'a}n}, Andr{\'e}s and {Henning}, Thomas and {Srba}, Ji{\v{r}}{\'\i} and {{\v{Z}}d'{\'a}rsk{\'a}}, Eva and {Balk{\'o}ov{\'a}}, Zuzana and {V{\'\i}tkov{\'a}}, Michaela and {Jan{\'\i}k}, Jan and {{\v{S}}koda}, Petr and {{\v{Z}}{\'a}k}, Ji{\v{r}}{\'\i} and {Mekarnia}, Djamel and {Suarez}, Olga and {Abe}, Lyu and {Beltrame}, Matteo and {Triaud}, Amaury H.~M.~J. and {Guillot}, Tristan and {Collins}, Karen A. and {Barkaoui}, Khalid and {Boyle}, Gavin and {Suc}, Vincent and {Antonucci}, Luca and {Pinto}, Marcelo Tala and {Vostretcova}, Elizaveta and {Eberhardt}, Jan and {Espinoza}, N{\'e}stor and {Mireles}, Ismael and {Pintr}, Pavel and {Rojas}, Felipe I. and {Schaffenroth}, Veronika and {Vanzi}, Leo and {Kab{\'a}th}, Petr},
        title = "{PLATOSpec first results: Three new transiting warm Jupiters from the WINE survey, TIC 147027702, TIC 245076932, and TIC 87422071}",
      journal = {\aap},
         year = 2026,
        month = apr,
       volume = {708},
          eid = {A189},
        pages = {A189},
          doi = {10.1051/0004-6361/202558698},
archivePrefix = {arXiv},
       eprint = {2602.20654},
 primaryClass = {astro-ph.EP},
       adsurl = {https://ui.adsabs.harvard.edu/abs/2026A\&A...708A.189G}
}

@ARTICLE{platospec,
       author = {{Kab{\'a}th}, P. and {Skarka}, M. and {Hatzes}, A. and {Guenther}, E. and {Vanzi}, L. and {Brahm}, R. and {Jan{\'\i}k}, J. and {Pintr}, P. and {Gajdo{\v{s}}}, P. and {Lipt{\'a}k}, J. and {{\v{Z}}{\'a}k}, J. and {Boffin}, H.~M.~J. and {Antonucci}, L. and {Avila}, G. and {Balk{\'o}ov{\'a}}, Z. and {Ball}, M.~E. and {Flores}, M. and {Fuentes}, A. and {Fuchs}, J. and {Greimel}, R. and {Gajardo}, A. and {Ivanov}, V.~D. and {K{\"o}hler}, J. and {Leitzinger}, M. and {Ludwig}, M.~R. and {Morav{\v{c}}{\'\i}k}, T. and {Ne{\v{c}}{\'a}sek}, J. and {Neubert}, R.~J.~U. and {Odert}, P. and {Olgu{\'\i}n}, G. and {Pinto}, M. Tala and {Roth}, M. and {{\v{R}}ezba}, L. and {Schaffenroth}, V. and {Sigwarth}, M. and {Srba}, J. and {Su{\'a}rez}, A. and {{\v{S}}koda}, P. and {{\v{S}}ubjak}, J. and {V{\'a}clav{\'\i}k}, J. and {Vesel{\'y}}, M. and {Vesel{\'y}}, R. and {V{\'\i}tkov{\'a}}, M. and {Winkler}, J.~U. and {Zummer}, M. and {{\v{Z}}{\v{d}}{\'a}rsk{\'a}}, E.},
        title = "{PLATOSpec: a precise spectrograph in support of space missions}",
      journal = {\mnras},
         year = 2026,
        month = jan,
       volume = {545},
       number = {1},
          eid = {staf1972},
        pages = {staf1972},
          doi = {10.1093/mnras/staf1972},
archivePrefix = {arXiv},
       eprint = {2510.11961},
 primaryClass = {astro-ph.IM},
       adsurl = {https://ui.adsabs.harvard.edu/abs/2026MNRAS.545f1972K}
}

@article{sousa2011spectroscopic,
  title={Spectroscopic characterization of a sample of metal-poor solar-type stars from the HARPS planet search program-Precise spectroscopic parameters and mass estimation},
  author={Sousa, S{\'e}rgio G and Santos, Nuno C and Israelian, Garik and Lovis, C and Mayor, Michel and Silva, Pedro B and Udry, Stephane},
  journal={Astronomy \& Astrophysics},
  volume={526},
  pages={A99},
  year={2011},
  publisher={EDP Sciences}
}

@article{santos2013sweet,
  title={SWEET-Cat: A catalogue of parameters for Stars With ExoplanETs-I. New atmospheric parameters and masses for 48 stars with planets},
  author={Santos, NC and Sousa, SG and Mortier, A and Neves, V and Adibekyan, V and Tsantaki, M and Mena, E Delgado and Bonfils, X and Israelian, G and Mayor, M and others},
  journal={A \& A},
  volume={556},
  pages={A150},
  year={2013},
  publisher={EDP Sciences}
}

@article{sousa2018sweet,
  title={SWEET-Cat updated-New homogenous spectroscopic parameters},
  author={Sousa, SG and Adibekyan, V and Delgado-Mena, E and Santos, NC and Andreasen, DT and Ferreira, ACS and Tsantaki, M and Barros, SCC and Demangeon, O and Israelian, G and others},
  journal={A \& A},
  volume={620},
  pages={A58},
  year={2018},
  publisher={EDP Sciences}
}

@ARTICLE{SWEET-Cat,
       author = {{Sousa}, S.~G. and {Adibekyan}, V. and {Delgado-Mena}, E. and {Santos}, N.~C. and {Rojas-Ayala}, B. and {Soares}, B.~M.~T.~B. and {Legoinha}, H. and {Ulmer-Moll}, S. and {Camacho}, J.~D. and {Barros}, S.~C.~C. and {Demangeon}, O.~D.~S. and {Hoyer}, S. and {Israelian}, G. and {Mortier}, A. and {Tsantaki}, M. and {Monteiro}, M.~A.},
        title = "{SWEET-Cat 2.0: The Cat just got SWEETer. Higher quality spectra and precise parallaxes from Gaia eDR3}",
      journal = {\aap},
         year = 2021,
        month = dec,
       volume = {656},
          eid = {A53},
        pages = {A53},
          doi = {10.1051/0004-6361/202141584},
archivePrefix = {arXiv},
       eprint = {2109.04781},
 primaryClass = {astro-ph.EP},
       adsurl = {https://ui.adsabs.harvard.edu/abs/2021A\&A...656A..53S}
}

@article{eastman2019,
  title={EXOFASTv2: A public, generalized, publication-quality exoplanet modeling code},
  author={Eastman, Jason D and Rodriguez, Joseph E and Agol, Eric and Stassun, Keivan G and Beatty, Thomas G and Vanderburg, Andrew and Gaudi, B Scott and Collins, Karen A and Luger, Rodrigo},
  journal={arXiv preprint arXiv:1907.09480},
  year={2019}
}

@article{dotter2016mesa,
  title={MESA Isochrones and Stellar Tracks (MIST) 0: Methods for the construction of stellar isochrones},
  author={Dotter, Aaron},
  journal={The Astrophysical Journal Supplement Series},
  volume={222},
  number={1},
  pages={8},
  year={2016},
  publisher={IOP Publishing}
}

@article{choi2016mesa,
  title={Mesa isochrones and stellar tracks (MIST). I. Solar-scaled models},
  author={Choi, Jieun and Dotter, Aaron and Conroy, Charlie and Cantiello, Matteo and Paxton, Bill and Johnson, Benjamin D},
  journal={The Astrophysical Journal},
  volume={823},
  number={2},
  pages={102},
  year={2016},
  publisher={IOP Publishing}
}

@ARTICLE{buchhave2010,
       author = {{Buchhave}, L.~A. and {Bakos}, G. {\'A}. and {Hartman},
J.~D. and {Torres}, G. and {Kov{\'a}cs}, G. and {Latham}, D.~W. and
{Noyes}, R.~W. and {Esquerdo}, G.~A. and {Everett}, M. and {Howard}, A.~W.
and {Marcy}, G.~W. and {Fischer}, D.~A. and {Johnson}, J.~A. and
{Andersen}, J. and {F{\H{u}}r{\'e}sz}, G. and {Perumpilly}, G. and
{Sasselov}, D.~D. and {Stefanik}, R.~P. and {B{\'e}ky}, B. and
{L{\'a}z{\'a}r}, J. and {Papp}, I. and {S{\'a}ri}, P.},
        title = "{HAT-P-16b: A 4 M $_{J}$ Planet Transiting a Bright Star
on an Eccentric Orbit}",
      journal = {\apj},
         year = 2010,
        month = sep,
       volume = {720},
       number = {2},
        pages = {1118-1125},
          doi = {10.1088/0004-637X/720/2/1118},
archivePrefix = {arXiv},
       eprint = {1005.2009},
 primaryClass = {astro-ph.EP},
       adsurl = {https://ui.adsabs.harvard.edu/abs/2010ApJ...720.1118B}
}

@article{henry1996survey,
  title={A survey of Ca II H and K chromospheric emission in southern solar-type stars},
  author={Henry, Todd J and Soderblom, David R and Donahue, Robert A and Baliunas, Sallie L},
  journal={Astronomical Journal, 111, 439},
  volume={111},
  year={1996}
}

@article{hojjatpanah2020correlation,
  title={The correlation between photometric variability and radial velocity jitter: Based on TESS and HARPS observations},
  author={Hojjatpanah, S and Oshagh, M and Figueira, P and Santos, NC and Amazo-G{\'o}mez, EM and Sousa, SG and Adibekyan, V and Akinsanmi, B and Demangeon, O and Faria, J and others},
  journal={Astronomy \& Astrophysics},
  volume={639},
  pages={A35},
  year={2020},
  publisher={EDP Sciences}
}

@article{boisse2011disentangling,
  title={Disentangling between stellar activity and planetary signals},
  author={Boisse, Isabelle and Bouchy, Fran{\c{c}}ois and H{\'e}brard, Guillaume and Bonfils, Xavier and Santos, Nuno and Vauclair, Sylvie},
  journal={Astronomy \& Astrophysics},
  volume={528},
  pages={A4},
  year={2011},
  publisher={EDP Sciences}
}

@article{mamajek2008improved,
  title={Improved age estimation for solar-type dwarfs using activity-rotation diagnostics},
  author={Mamajek, Eric E and Hillenbrand, Lynne A},
  journal={The Astrophysical Journal},
  volume={687},
  number={2},
  pages={1264--1293},
  year={2008}
}

@article{buchhave2012,
  title={An abundance of small exoplanets around stars with a wide range of
metallicities},
  author={Lars A. Buchhave and David Latham and Anders Johansen and Martin
Bizzarro and Guillermo Torres and Jason F. Rowe and Natalie M. Batalha and
William J. Borucki and Erik J. Brugamyer and Caroline Caldwell and Stephen
T. Bryson and David R. Ciardi and William D. Cochran and Michael Endl and
Gilbert A. Esquerdo and Eric B. Ford and John C. Geary and Ronald L.
Gilliland and Terese Holst Hansen and Howard Isaacson and John Laird and
Philip W. Lucas and Geoffrey W. Marcy and Jon A. Morse and Paul Robertson
and Avi Shporer and Robert P. Stefanik and Martin D. Still and Samuel N.
Quinn},
  journal={Nature},
  year={2012},
  volume={486},
  pages={375-377}
}

@article{noyes1984rotation,
  title={Rotation, convection, and magnetic activity in lower main-sequence stars},
  author={Noyes, RW and Hartmann, LW and Baliunas, StLt and Duncan, DK and Vaughan, AH},
  journal={The Astrophysical Journal},
  volume={279},
  pages={763--777},
  year={1984}
}

@article{Pecaut2013,
  author  = {Pecaut, M. J. and Mamajek, E. E.},
  title   = {Intrinsic Colors, Temperatures, and Bolometric Corrections
             of Pre-Main-Sequence Stars},
  journal = {The Astrophysical Journal Supplement Series},
  volume  = {208},
  pages   = {9},
  year    = {2013},
  doi     = {10.1088/0067-0049/208/1/9}
}

@ARTICLE{Espinoza2019,
  author = {{Espinoza}, N. and others},
  title = {juliet: a versatile modelling tool for transiting exoplanets and radial-velocity observations},
  journal = {MNRAS},
  volume = {490},
  pages = {2262--2283},
  year = {2019}
}

@article{foreman2017fast,
  title={Fast and scalable Gaussian process modeling with applications to astronomical time series},
  author={Foreman-Mackey, Daniel and Agol, Eric and Ambikasaran, Sivaram and Angus, Ruth},
  journal={The Astronomical Journal},
  volume={154},
  number={6},
  pages={220},
  year={2017},
  publisher={IOP Publishing}
}

@article{ambikasaran2015fast,
  title={Fast direct methods for Gaussian processes},
  author={Ambikasaran, Sivaram and Foreman-Mackey, Daniel and Greengard, Leslie and Hogg, David W and O’Neil, Michael},
  journal={IEEE transactions on pattern analysis and machine intelligence},
  volume={38},
  number={2},
  pages={252--265},
  year={2015},
  publisher={IEEE}
}

@phdthesis{gaborthesis,
Author = {G\'abor F\H{u}r\'esz},
School = {University of Szeged, Hungary},
Year = {2008},
Title = {Design and Application of High Resolution and Multiobject
Spectrographs: Dynamical Studies of Open Clusters}}

@ARTICLE{2020RNAAS...4..201C,
       author = {{Caldwell}, Douglas A. and {Tenenbaum}, Peter and {Twicken}, Joseph D. and {Jenkins}, Jon M. and {Ting}, Eric and {Smith}, Jeffrey C. and {Hedges}, Christina and {Fausnaugh}, Michael M. and {Rose}, Mark and {Burke}, Christopher},
        title = "{TESS Science Processing Operations Center FFI Target List Products}",
      journal = {Research Notes of the American Astronomical Society},
         year = 2020,
        month = nov,
       volume = {4},
       number = {11},
          eid = {201},
        pages = {201},
          doi = {10.3847/2515-5172/abc9b3},
archivePrefix = {arXiv},
       eprint = {2011.05495},
 primaryClass = {astro-ph.EP},
       adsurl = {https://ui.adsabs.harvard.edu/abs/2020RNAAS...4..201C}
}

@ARTICLE{2020RNAAS...4..206H,
       author = {{Huang}, Chelsea X. and {Vanderburg}, Andrew and {P{\'a}l}, Andras and {Sha}, Lizhou and {Yu}, Liang and {Fong}, Willie and {Fausnaugh}, Michael and {Shporer}, Avi and {Guerrero}, Natalia and {Vanderspek}, Roland and {Ricker}, George},
        title = "{Photometry of 10 Million Stars from the First Two Years of TESS Full Frame Images: Part II}",
      journal = {Research Notes of the American Astronomical Society},
         year = 2020,
        month = nov,
       volume = {4},
       number = {11},
          eid = {206},
        pages = {206},
          doi = {10.3847/2515-5172/abca2d},
       adsurl = {https://ui.adsabs.harvard.edu/abs/2020RNAAS...4..206H}
}

@article{miller2011heavy,
  title={The heavy-element masses of extrasolar giant planets, revealed},
  author={Miller, Neil and Fortney, Jonathan J},
  journal={The Astrophysical Journal Letters},
  volume={736},
  number={2},
  pages={L29},
  year={2011},
  publisher={IOP Publishing}
}

@article{demory2011lack,
  title={Lack of inflated radii for Kepler giant planet candidates receiving modest stellar irradiation},
  author={Demory, Brice-Olivier and Seager, Sara},
  journal={The Astrophysical Journal Supplement Series},
  volume={197},
  number={1},
  pages={12},
  year={2011},
  publisher={IOP Publishing}
}

@article{thorngren2024hot,
  title={The Hot Jupiter Radius Anomaly and Stellar Connections},
  author={Thorngren, Daniel P},
  journal={arXiv preprint arXiv:2405.05307},
  year={2024}
}

@article{batygin2011evolution,
  title={Evolution of Ohmically heated hot Jupiters},
  author={Batygin, Konstantin and Stevenson, David J and Bodenheimer, Peter H},
  journal={The Astrophysical Journal},
  volume={738},
  number={1},
  pages={1},
  year={2011},
  publisher={IOP Publishing}
}

@article{arras2010thermal,
  title={Thermal tides in fluid extrasolar planets},
  author={Arras, Phil and Socrates, Aristotle},
  journal={The Astrophysical Journal},
  volume={714},
  number={1},
  pages={1},
  year={2010},
  publisher={IOP Publishing}
}

@article{sarkis2021evidence,
  title={Evidence of three mechanisms explaining the radius anomaly of hot Jupiters},
  author={Sarkis, Paula and Mordasini, Christoph and Henning, Th and Marleau, Gabriel D and Molli{\`e}re, Paul},
  journal={A \& A},
  volume={645},
  pages={A79},
  year={2021},
  publisher={EDP Sciences}
}

@article{kempton2018framework,
  title={A framework for prioritizing the TESS planetary candidates most amenable to atmospheric characterization},
  author={Kempton, Eliza M-R and Bean, Jacob L and Louie, Dana R and Deming, Drake and Koll, Daniel DB and Mansfield, Megan and Christiansen, Jessie L and L{\'o}pez-Morales, Mercedes and Swain, Mark R and Zellem, Robert T and others},
  journal={Publications of the Astronomical Society of the Pacific},
  volume={130},
  number={993},
  pages={114401},
  year={2018},
  publisher={IOP Publishing}
}

@ARTICLE{Duffell2015,
       author = {{Duffell}, Paul C. and {Chiang}, Eugene},
        title = "{Eccentric Jupiters via Disk-Planet Interactions}",
      journal = {\apj},
         year = 2015,
        month = oct,
       volume = {812},
       number = {2},
          eid = {94},
        pages = {94},
          doi = {10.1088/0004-637X/812/2/94},
archivePrefix = {arXiv},
       eprint = {1507.08667},
 primaryClass = {astro-ph.EP},
       adsurl = {https://ui.adsabs.harvard.edu/abs/2015ApJ...812...94D}
}

@ARTICLE{Rasio1996,
       author = {{Rasio}, Frederic A. and {Ford}, Eric B.},
        title = "{Dynamical instabilities and the formation of extrasolar planetary systems}",
      journal = {Science},
         year = 1996,
        month = nov,
       volume = {274},
        pages = {954-956},
          doi = {10.1126/science.274.5289.954},
       adsurl = {https://ui.adsabs.harvard.edu/abs/1996Sci...274..954R}
}

@ARTICLE{Juric2008,
       author = {{Juri{\'c}}, Mario and {Tremaine}, Scott},
        title = "{Dynamical Origin of Extrasolar Planet Eccentricity Distribution}",
      journal = {\apj},
         year = 2008,
        month = oct,
       volume = {686},
       number = {1},
        pages = {603-620},
          doi = {10.1086/590047},
archivePrefix = {arXiv},
       eprint = {astro-ph/0703160},
 primaryClass = {astro-ph},
       adsurl = {https://ui.adsabs.harvard.edu/abs/2008ApJ...686..603J}
}

@ARTICLE{Goldreich2004,
       author = {{Goldreich}, Peter and {Lithwick}, Yoram and {Sari}, Re'em},
        title = "{Final Stages of Planet Formation}",
      journal = {\apj},
         year = 2004,
        month = oct,
       volume = {614},
       number = {1},
        pages = {497-507},
          doi = {10.1086/423612},
archivePrefix = {arXiv},
       eprint = {astro-ph/0404240},
 primaryClass = {astro-ph},
       adsurl = {https://ui.adsabs.harvard.edu/abs/2004ApJ...614..497G}
}

@ARTICLE{Ida2013,
       author = {{Ida}, S. and {Lin}, D.~N.~C. and {Nagasawa}, M.},
        title = "{Toward a Deterministic Model of Planetary Formation. VII. Eccentricity Distribution of Gas Giants}",
      journal = {\apj},
         year = 2013,
        month = sep,
       volume = {775},
       number = {1},
          eid = {42},
        pages = {42},
          doi = {10.1088/0004-637X/775/1/42},
archivePrefix = {arXiv},
       eprint = {1307.6450},
 primaryClass = {astro-ph.EP},
       adsurl = {https://ui.adsabs.harvard.edu/abs/2013ApJ...775...42I}
}

@ARTICLE{Petrovich2014,
       author = {{Petrovich}, Cristobal and {Tremaine}, Scott and {Rafikov}, Roman},
        title = "{Scattering Outcomes of Close-in Planets: Constraints on Planet Migration}",
      journal = {\apj},
         year = 2014,
        month = may,
       volume = {786},
       number = {2},
          eid = {101},
        pages = {101},
          doi = {10.1088/0004-637X/786/2/101},
archivePrefix = {arXiv},
       eprint = {1401.4457},
 primaryClass = {astro-ph.EP},
       adsurl = {https://ui.adsabs.harvard.edu/abs/2014ApJ...786..101P}
}

@ARTICLE{Naef2001,
       author = {{Naef}, D. and {Latham}, D.~W. and {Mayor}, M. and {Mazeh}, T. and {Beuzit}, J.~L. and {Drukier}, G.~A. and {Perrier-Bellet}, C. and {Queloz}, D. and {Sivan}, J.~P. and {Torres}, G. and {Udry}, S. and {Zucker}, S.},
        title = "{HD 80606 b, a planet on an extremely elongated orbit}",
      journal = {A \& A},
         year = 2001,
        month = aug,
       volume = {375},
        pages = {L27-L30},
          doi = {10.1051/0004-6361:20010853},
archivePrefix = {arXiv},
       eprint = {astro-ph/0106256},
 primaryClass = {astro-ph}
}

@article{hebrard2010observation,
  title={Observation of the full 12-hour-long transit of the exoplanet HD 80606b-Warm-Spitzer photometry and SOPHIE spectroscopy},
  author={H{\'e}brard, G and D{\'e}sert, J-M and D{\'\i}az, RF and Boisse, I and Bouchy, F and Des Etangs, A Lecavelier and Moutou, C and Ehrenreich, D and Arnold, L and Bonfils, X and others},
  journal={A \& A},
  volume={516},
  pages={A95},
  year={2010},
  publisher={EDP Sciences}
}

@ARTICLE{Dong2021,
       author = {{Dong}, Jiayin and {Huang}, Chelsea X. and {Zhou}, George and {Dawson}, Rebekah I. and {Rodriguez}, Joseph E. and {Eastman}, Jason D. and {Collins}, Karen A. and {Quinn}, Samuel N. and {Shporer}, Avi and {Triaud}, Amaury H.~M.~J. and {Wang}, Songhu and {Beatty}, Thomas and {Jackson}, Jonathon M. and {Collins}, Kevin I. and {Abe}, Lyu and {Suarez}, Olga and {Crouzet}, Nicolas and {M{\'e}karnia}, Djamel and {Dransfield}, Georgina and {Jensen}, Eric L.~N. and {Stockdale}, Chris and {Barkaoui}, Khalid and {Heitzmann}, Alexis and {Wright}, Duncan J. and {Addison}, Brett C. and {Wittenmyer}, Robert A. and {Okumura}, Jack and {Bowler}, Brendan P. and {Horner}, Jonathan and {Kane}, Stephen R. and {Kielkopf}, John and {Liu}, Huigen and {Plavchan}, Peter and {Mengel}, Matthew W. and {Ricker}, George R. and {Vanderspek}, Roland and {Latham}, David W. and {Seager}, S. and {Winn}, Joshua N. and {Jenkins}, Jon M. and {Christiansen}, Jessie L. and {Paegert}, Martin},
        title = "{TOI-3362b: A Proto Hot Jupiter Undergoing High-eccentricity Tidal Migration}",
      journal = {\apjl},
         year = 2021,
        month = oct,
       volume = {920},
       number = {1},
          eid = {L16},
        pages = {L16},
          doi = {10.3847/2041-8213/ac2600},
archivePrefix = {arXiv},
       eprint = {2109.03771},
 primaryClass = {astro-ph.EP},
       adsurl = {https://ui.adsabs.harvard.edu/abs/2021ApJ...920L..16D}
}

@article{espinoza2023aligned,
  title={The Aligned Orbit of the Eccentric Proto Hot Jupiter TOI-3362b},
  author={Espinoza-Retamal, Juan I and Brahm, Rafael and Petrovich, Cristobal and Jord{\'a}n, Andr{\'e}s and Stef{\'a}nsson, Gu{\dj}mundur and Sedaghati, Elyar and Hobson, Melissa J and Mu{\~n}oz, Diego J and Boyle, Gavin and Leiva, Rodrigo and others},
  journal={The Astrophysical Journal Letters},
  volume={958},
  number={2},
  pages={L20},
  year={2023},
  publisher={IOP Publishing}
}

@ARTICLE{2003ApJ...589..605W,
       author = {{Wu}, Y. and {Murray}, N.},
        title = "{Planet Migration and Binary Companions: The Case of HD 80606b}",
      journal = {\apj},
         year = 2003,
        month = may,
       volume = {589},
       number = {1},
        pages = {605-614},
          doi = {10.1086/374598},
archivePrefix = {arXiv},
       eprint = {astro-ph/0303010},
 primaryClass = {astro-ph},
       adsurl = {https://ui.adsabs.harvard.edu/abs/2003ApJ...589..605W}
}

@article{zingales2018ariel,
  title={The ARIEL mission reference sample},
  author={Zingales, Tiziano and Tinetti, Giovanna and Pillitteri, Ignazio and Leconte, J{\'e}r{\'e}my and Micela, Giuseppina and Sarkar, Subhajit},
  journal={Experimental Astronomy},
  volume={46},
  number={1},
  pages={67--100},
  year={2018},
  publisher={Springer}
}

@article{sabelhaus2004overview,
  title={An overview of the James Webb space telescope (JWST) project},
  author={Sabelhaus, Phillip A and Decker, John E},
  journal={Optical, Infrared, and Millimeter Space Telescopes},
  volume={5487},
  pages={550--563},
  year={2004},
  publisher={SPIE}
}

@article{tinetti2018chemical,
  title={A chemical survey of exoplanets with ARIEL},
  author={Tinetti, Giovanna and Drossart, Pierre and Eccleston, Paul and Hartogh, Paul and Heske, Astrid and Leconte, J{\'e}r{\'e}my and Micela, Giusi and Ollivier, Marc and Pilbratt, G{\"o}ran and Puig, Ludovic and others},
  journal={Experimental astronomy},
  volume={46},
  number={1},
  pages={135--209},
  year={2018},
  publisher={Springer}
}

@article{edwards2019updated,
  title={An updated study of potential targets for Ariel},
  author={Edwards, Billy and Mugnai, Lorenzo and Tinetti, Giovanna and Pascale, Enzo and Sarkar, Subhajit},
  journal={The Astronomical Journal},
  volume={157},
  number={6},
  pages={242},
  year={2019},
  publisher={American Astronomical Society}
}

@article{esteves2015changing,
  title={Changing phases of alien worlds: Probing atmospheres of Kepler planets with high-precision photometry},
  author={Esteves, Lisa J and De Mooij, Ernst JW and Jayawardhana, Ray},
  journal={The Astrophysical Journal},
  volume={804},
  number={2},
  pages={150},
  year={2015},
  publisher={IOP Publishing}
}

@article{eeles2025high,
  title={A high stellar multiplicity rate amongst TESS planet candidates in the Neptunian desert using Gaia DR3 astrometry},
  author={Eeles-Nolle, Fintan and Armstrong, David J},
  journal={Monthly Notices of the Royal Astronomical Society},
  volume={541},
  number={2},
  pages={1419--1433},
  year={2025},
  publisher={Oxford University Press}
}

@article{ricker2015transiting,
  title={Transiting exoplanet survey satellite},
  author={Ricker, George R and Winn, Joshua N and Vanderspek, Roland and Latham, David W and Bakos, G{\'a}sp{\'a}r {\'A} and Bean, Jacob L and Berta-Thompson, Zachory K and Brown, Timothy M and Buchhave, Lars and Butler, Nathaniel R and others},
  journal={Journal of Astronomical Telescopes, Instruments, and Systems},
  volume={1},
  number={1},
  pages={014003--014003},
  year={2015},
  publisher={Society of Photo-Optical Instrumentation Engineers}
}

@article{fortney2021hot,
  title={Hot Jupiters: origins, structure, atmospheres},
  author={Fortney, Jonathan J and Dawson, Rebekah I and Komacek, Thaddeus D},
  journal={Journal of Geophysical Research: Planets},
  volume={126},
  number={3},
  pages={e2020JE006629},
  year={2021},
  publisher={Wiley Online Library}
}

@article{dawson2018origins,
  title={Origins of hot Jupiters},
  author={Dawson, Rebekah I and Johnson, John Asher},
  journal={Annual Review of Astronomy and Astrophysics},
  volume={56},
  number={1},
  pages={175--221},
  year={2018},
  publisher={Annual Reviews}
}

@article{eastman2013,
  title={EXOFAST: A fast exoplanetary fitting suite in IDL},
  author={Eastman, Jason and Gaudi, B Scott and Agol, Eric},
  journal={Publications of the Astronomical Society of the Pacific},
  volume={125},
  number={923},
  pages={83},
  year={2013},
  publisher={IOP Publishing}
}

@article{vanderburg2014technique,
  title={A technique for extracting highly precise photometry for the two-wheeled Kepler mission},
  author={Vanderburg, Andrew and Johnson, John Asher},
  journal={Publications of the Astronomical Society of the Pacific},
  volume={126},
  number={944},
  pages={948--958},
  year={2014},
  publisher={The Astronomical Society of the Pacific}
}

@article{collaboration2018lightkurve,
  title={Lightkurve: Kepler and TESS time series analysis in Python},
  author={Collaboration, Lightkurve and Cardoso, JVdM and Hedges, C and Gully-Santiago, M and Saunders, N and Cody, AM and Barclay, T and Hall, O and Sagear, S and Turtelboom, E and others},
  journal={Astrophysics Source Code Library (ascl: 1812.013)},
  year={2018}
}

@ARTICLE{Stumpe2012,
       author = {{Stumpe}, Martin C. and {Smith}, Jeffrey C. and {Van Cleve}, Jeffrey E. and
         {Twicken}, Joseph D. and {Barclay}, Thomas S. and
         {Fanelli}, Michael N. and {Girouard}, Forrest R. and {Jenkins}, Jon M. and
         {Kolodziejczak}, Jeffery J. and {McCauliff}, Sean D. and
         {Morris}, Robert L.},
        title = "{Kepler Presearch Data Conditioning I{\textemdash}Architecture and Algorithms for Error Correction in Kepler Light Curves}",
      journal = {\pasp},
         year = "2012",
        month = "Sep",
       volume = {124},
       number = {919},
        pages = {985},
          doi = {10.1086/667698},
archivePrefix = {arXiv},
       eprint = {1203.1382},
primaryClass = {astro-ph.IM},
       adsurl = {https://ui.adsabs.harvard.edu/abs/2012PASP..124..985S}
}

@article{birkby2012discovery,
  title={Discovery and characterization of detached M dwarf eclipsing binaries in the WFCAM Transit Survey},
  author={Birkby, Jayne and Nefs, Bas and Hodgkin, Simon and Kov{\'a}cs, G{\'a}bor and Sip{\H{o}}cz, Brigitta and Pinfield, David and Snellen, Ignas and Mislis, Dimitris and Murgas, Felipe and Lodieu, Nicolas and others},
  journal={Monthly Notices of the Royal Astronomical Society},
  volume={426},
  number={2},
  pages={1507--1532},
  year={2012},
  publisher={Blackwell Science Ltd Oxford, UK}
}

@article{chabrier2007heat,
  title={Heat transport in giant (exo) planets: a new perspective},
  author={Chabrier, Gilles and Baraffe, Isabelle},
  journal={The Astrophysical Journal},
  volume={661},
  number={1},
  pages={L81},
  year={2007},
  publisher={IOP Publishing}
}

@article{burrows2007possible,
  title={Possible solutions to the radius anomalies of transiting giant planets},
  author={Burrows, Adam and Hubeny, I and Budaj, J and Hubbard, WB},
  journal={The Astrophysical Journal},
  volume={661},
  number={1},
  pages={502},
  year={2007},
  publisher={IOP Publishing}
}

@article{dittmann2017discovery,
  title={Discovery and Precise Characterization by the MEarth Project of LP 661-13, an Eclipsing Binary Consisting of Two Fully Convective Low-mass Stars},
  author={Dittmann, Jason A and Irwin, Jonathan M and Charbonneau, David and Berta-Thompson, Zachory K and Newton, Elisabeth R and Latham, David W and Latham, Christian A and Esquerdo, Gilbert and Berlind, Perry and Calkins, Michael L},
  journal={The Astrophysical Journal},
  volume={836},
  number={1},
  pages={124},
  year={2017},
  publisher={IOP Publishing}
}

@article{kraus2011mass,
  title={The mass--radius (--rotation?) relation for low-mass stars},
  author={Kraus, Adam L and Tucker, Roy A and Thompson, Michael I and Craine, Eric R and Hillenbrand, Lynne A},
  journal={The Astrophysical Journal},
  volume={728},
  number={1},
  pages={48},
  year={2011},
  publisher={IOP Publishing}
}

@article{leconte2010tidal,
  title={Is tidal heating sufficient to explain bloated exoplanets? Consistent calculations accounting for finite initial eccentricity},
  author={Leconte, J{\'e}r{\'e}my and Chabrier, Gilles and Baraffe, Isabelle and Levrard, Benjamin},
  journal={Astronomy \& Astrophysics},
  volume={516},
  pages={A64},
  year={2010},
  publisher={EDP Sciences}
}

@article{boisse2010sophie,
  title={The SOPHIE search for northern extrasolar planets-III. A Jupiter-mass companion around HD 109246},
  author={Boisse, I and Eggenberger, A and Santos, NC and Lovis, C and Bouchy, F and H{\'e}brard, G and Arnold, L and Bonfils, X and Delfosse, X and Desort, M and others},
  journal={A \& A},
  volume={523},
  pages={A88},
  year={2010},
  publisher={EDP Sciences}
}

@ARTICLE{StassunTorres:2021,
       author = {{Stassun}, Keivan G. and {Torres}, Guillermo},
        title = "{Parallax Systematics and Photocenter Motions of Benchmark Eclipsing Binaries in Gaia EDR3}",
      journal = {\apjl},
         year = 2021,
        month = feb,
       volume = {907},
       number = {2},
          eid = {L33},
        pages = {L33},
          doi = {10.3847/2041-8213/abdaad},
archivePrefix = {arXiv},
       eprint = {2101.03425},
 primaryClass = {astro-ph.SR},
       adsurl = {https://ui.adsabs.harvard.edu/abs/2021ApJ...907L..33S}
}

@ARTICLE{Stassun:2016,
       author = {{Stassun}, Keivan G. and {Torres}, Guillermo},
        title = "{Eclipsing Binaries as Benchmarks for Trigonometric Parallaxes in the Gaia Era}",
      journal = {\aj},
         year = 2016,
        month = dec,
       volume = {152},
       number = {6},
          eid = {180},
        pages = {180},
          doi = {10.3847/0004-6256/152/6/180},
archivePrefix = {arXiv},
       eprint = {1609.02579},
 primaryClass = {astro-ph.SR},
       adsurl = {https://ui.adsabs.harvard.edu/abs/2016AJ....152..180S}
}

@ARTICLE{Stassun:2017,
       author = {{Stassun}, Keivan G. and {Collins}, Karen A. and {Gaudi}, B. Scott},
        title = "{Accurate Empirical Radii and Masses of Planets and Their Host Stars with Gaia Parallaxes}",
      journal = {\aj},
         year = 2017,
        month = mar,
       volume = {153},
       number = {3},
          eid = {136},
        pages = {136},
          doi = {10.3847/1538-3881/aa5df3},
archivePrefix = {arXiv},
       eprint = {1609.04389},
 primaryClass = {astro-ph.EP},
       adsurl = {https://ui.adsabs.harvard.edu/abs/2017AJ....153..136S}
}

@ARTICLE{Stassun:2018,
       author = {{Stassun}, Keivan G. and {Corsaro}, Enrico and {Pepper}, Joshua A. and
         {Gaudi}, B. Scott},
        title = "{Empirical Accurate Masses and Radii of Single Stars with TESS and Gaia}",
      journal = {\aj},
         year = 2018,
        month = jan,
       volume = {155},
       number = {1},
          eid = {22},
        pages = {22},
          doi = {10.3847/1538-3881/aa998a},
archivePrefix = {arXiv},
       eprint = {1710.01460},
 primaryClass = {astro-ph.SR},
       adsurl = {https://ui.adsabs.harvard.edu/abs/2018AJ....155...22S}
}

@ARTICLE{Husser:2013,
       author = {{Husser}, T. -O. and {Wende-von Berg}, S. and {Dreizler}, S. and {Homeier}, D. and {Reiners}, A. and {Barman}, T. and {Hauschildt}, P.~H.},
        title = "{A new extensive library of PHOENIX stellar atmospheres and synthetic spectra}",
      journal = {A \& A},
         year = 2013,
        month = may,
       volume = {553},
          eid = {A6},
        pages = {A6},
          doi = {10.1051/0004-6361/201219058},
archivePrefix = {arXiv},
       eprint = {1303.5632},
 primaryClass = {astro-ph.SR}
}

@ARTICLE{Schlegel:1998,
       author = {{Schlegel}, David J. and {Finkbeiner}, Douglas P. and {Davis}, Marc},
        title = "{Maps of Dust Infrared Emission for Use in Estimation of Reddening and Cosmic Microwave Background Radiation Foregrounds}",
      journal = {\apj},
         year = 1998,
        month = jun,
       volume = {500},
       number = {2},
        pages = {525-553},
          doi = {10.1086/305772},
archivePrefix = {arXiv},
       eprint = {astro-ph/9710327},
 primaryClass = {astro-ph},
       adsurl = {https://ui.adsabs.harvard.edu/abs/1998ApJ...500..525S}
}

@article{baraffe2015new,
  title={New evolutionary models for pre-main sequence and main sequence low-mass stars down to the hydrogen-burning limit},
  author={Baraffe, Isabelle and Homeier, Derek and Allard, France and Chabrier, Gilles},
  journal={Astronomy \& Astrophysics},
  volume={577},
  pages={A42},
  year={2015},
  publisher={EDP Sciences}
}

@ARTICLE{feros,
       author = {{Kaufer}, A. and {Stahl}, O. and {Tubbesing}, S. and
{N{\o}rregaard}, P. and {Avila}, G. and {Francois}, P. and {Pasquini}, L.
and {Pizzella}, A.},
        title = "{Commissioning FEROS, the new high-resolution spectrograph
at La-Silla.}",
      journal = {The Messenger},
         year = 1999,
        month = mar,
       volume = {95},
        pages = {8-12},
       adsurl = {https://ui.adsabs.harvard.edu/abs/1999Msngr..95....8K}
}

@ARTICLE{wine1,
       author = {{Brahm}, Rafael and {Espinoza}, N{\'e}stor and
{Jord{\'a}n}, Andr{\'e}s and {Henning}, Thomas and {Sarkis}, Paula and
{Jones}, Mat{\'\i}as I. and {D{\'\i}az}, Mat{\'\i}as R. and {Jenkins},
James S. and {Vanzi}, Leonardo and {Zapata}, Abner and {Petrovich},
Cristobal and {Kossakowski}, Diana and {Rabus}, Markus and {Rojas}, Felipe
and {Torres}, Pascal},
        title = "{HD 1397b: A Transiting Warm Giant Planet Orbiting A V =
7.8 mag Subgiant Star Discovered by TESS}",
      journal = {\aj},
         year = 2019,
        month = jul,
       volume = {158},
       number = {1},
          eid = {45},
        pages = {45},
          doi = {10.3847/1538-3881/ab279a},
archivePrefix = {arXiv},
       eprint = {1811.02156},
 primaryClass = {astro-ph.EP},
       adsurl = {https://ui.adsabs.harvard.edu/abs/2019AJ....158...45B}
}

@ARTICLE{wine2,
       author = {{Trifonov}, Trifon and {Brahm}, Rafael and {Jord{\'a}n},
Andr{\'e}s and {Hartogh}, Christian and {Henning}, Thomas and {Hobson},
Melissa J. and {Schlecker}, Martin and {Howard}, Saburo and {Reichardt},
Finja and {Espinoza}, Nestor and {Lee}, Man Hoi and {Nesvorny}, David and
{Rojas}, Felipe I. and {Barkaoui}, Khalid and {Kossakowski}, Diana and
{Boyle}, Gavin and {Dreizler}, Stefan and {K{\"u}rster}, Martin and
{Heller}, Ren{\'e} and {Guillot}, Tristan and {Triaud}, Amaury H.~M.~J. and
{Abe}, Lyu and {Agabi}, Abdelkrim and {Bendjoya}, Philippe and {Crouzet},
Nicolas and {Dransfield}, Georgina and {Gasparetto}, Thomas and
{G{\"u}nther}, Maximilian N. and {Marie-Sainte}, Wenceslas and
{M{\'e}karnia}, Djamel and {Suarez}, Olga and {Teske}, Johanna and
{Butler}, R. Paul and {Crane}, Jeffrey D. and {Shectman}, Stephen and
{Ricker}, George R. and {Shporer}, Avi and {Vanderspek}, Roland and
{Jenkins}, Jon M. and {Wohler}, Bill and {Collins}, Karen A. and {Collins},
Kevin I. and {Ciardi}, David R. and {Barclay}, Thomas and {Mireles}, Ismael
and {Seager}, Sara and {Winn}, Joshua N.},
        title = "{TOI-2525 b and c: A Pair of Massive Warm Giant Planets
with Strong Transit Timing Variations Revealed by TESS}",
      journal = {\aj},
         year = 2023,
        month = apr,
       volume = {165},
       number = {4},
          eid = {179},
        pages = {179},
          doi = {10.3847/1538-3881/acba9b},
archivePrefix = {arXiv},
       eprint = {2302.05694},
 primaryClass = {astro-ph.EP},
       adsurl = {https://ui.adsabs.harvard.edu/abs/2023AJ....165..179T}
}

@ARTICLE{wine3,
       author = {{Brahm}, Rafael and {Trifonov}, Trifon and {Jord{\'a}n},
Andr{\'e}s and {Henning}, Thomas and {Espinoza}, N{\'e}stor and {Rojas},
Felipe I. and {Tala Pinto}, Marcelo and {Jones}, Mat{\'\i}as I. and
{Thorngren}, Daniel and {Acu{\~n}a}, Lorena and {Eberhardt}, Jan and
{Reinarz}, Yared and {Salinas}, Helem and {V{\'\i}tkov{\'a}}, Michaela and
{Espinoza-Retamal}, Juan I. and {Bakos}, Gaspar and {B{\'o}di}, Attila and
{Boyle}, Gavin and {Csubry}, Zoltan and {Hartman}, Joel and {Keyes},
Anthony and {Suc}, Vincent and {Talens}, Geert Jan},
        title = "{A Compact Multi-Planet System of Three Transiting Giant
Planets Around TIC118798035}",
      journal = {arXiv e-prints},
         year = 2025,
        month = oct,
          eid = {arXiv:2510.19162},
        pages = {arXiv:2510.19162},
          doi = {10.48550/arXiv.2510.19162},
archivePrefix = {arXiv},
       eprint = {2510.19162},
 primaryClass = {astro-ph.EP},
       adsurl = {https://ui.adsabs.harvard.edu/abs/2025arXiv251019162B}
}

@phdthesis{heidari2022overcoming,
  title={Overcoming the stellar activity impact on exoplanet detection and characterization: application to the follow-up of TESS candidates with the SOPHIE spectrograph},
  author={Heidari, Neda},
  year={2022},
  school={Universit{\'e} C{\^o}te d'Azur; Shahid Beheshti University (Tehran)}
}

@article{heidari2024sophie,
  title={The SOPHIE search for northern extrasolar planets-XIX. A system including a cold sub-Neptune potentially transiting a V= 6.5 star HD 88986},
  author={Heidari, N and Boisse, I and Hara, NC and Wilson, TG and Kiefer, F and H{\'e}brard, G and Philipot, F and Hoyer, S and Stassun, KG and Henry, GW and others},
  journal={Astronomy \& Astrophysics},
  volume={681},
  pages={A55},
  year={2024},
  publisher={EDP Sciences}
}

@inproceedings{perruchot2008sophie,
  title={The SOPHIE spectrograph: design and technical key-points for high throughput and high stability},
  author={Perruchot, Sandrine and Kohler, D and Bouchy, F and Richaud, Y and Richaud, P e and Moreaux, G and Merzougui, M and Sottile, R and Hill, L and Knispel, G and others},
  booktitle={Ground-based and Airborne Instrumentation for Astronomy II},
  volume={7014},
  pages={235--246},
  year={2008},
  organization={SPIE}
}

@article{skrutskie2006two,
  title={The two micron all sky survey (2MASS)},
  author={Skrutskie, Michael F and Cutri, RM and Stiening, R and Weinberg, MD and Schneider, S and Carpenter, JM and Beichman, Capps and Capps, R and Chester, T and Elias, J and others},
  journal={The Astronomical Journal},
  volume={131},
  number={2},
  pages={1163},
  year={2006},
  publisher={IOP Publishing}
}

@article{gaia2023gaia,
  title={Gaia Data Release 3: Summary of the content and survey properties},
  author={Gaia Collaboration and others},
  journal={Astronomy \& Astrophysics},
  volume={674},
  pages={A1},
  year={2023},
  publisher={EDP SCIENCES SA}
}

@article{wright2010wide,
  title={The Wide-field Infrared Survey Explorer (WISE): mission description and initial on-orbit performance},
  author={Wright, Edward L and Eisenhardt, Peter RM and Mainzer, Amy K and Ressler, Michael E and Cutri, Roc M and Jarrett, Thomas and Kirkpatrick, J Davy and Padgett, Deborah and McMillan, Robert S and Skrutskie, Michael and others},
  journal={The Astronomical Journal},
  volume={140},
  number={6},
  pages={1868},
  year={2010},
  publisher={IOP Publishing}
}

@article{heidari2025characterization,
  title={Characterization of seven transiting systems, including four warm Jupiters from SOPHIE and TESS},
  author={Heidari, N and H{\'e}brard, G and Martioli, E and Eastman, JD and Jackson, JM and Delfosse, X and Jord{\'a}n, A and Correia, ACM and Sousa, S and Dragomir, D and others},
  journal={Astronomy \& Astrophysics},
  volume={694},
  pages={A36},
  year={2025},
  publisher={EDP Sciences}
}

@ARTICLE{Brown:2013,
        author = {{Brown}, T.~M. and {Baliber}, N. and {Bianco}, F.~B. and
 {Bowman}, M. and {Burleson}, B. and {Conway}, P. and {Crellin}, M. and
 {Depagne}, {\'E}. and {De Vera}, J. and {Dilday}, B. and {Dragomir}, D. and
 {Dubberley}, M. and {Eastman}, J.~D. and {Elphick}, M. and {Falarski}, M.
 and {Foale}, S. and {Ford}, M. and {Fulton}, B.~J. and {Garza}, J. and
 {Gomez}, E.~L. and {Graham}, M. and {Greene}, R. and {Haldeman}, B. and
 {Hawkins}, E. and {Haworth}, B. and {Haynes}, R. and {Hidas}, M. and
 {Hjelstrom}, A.~E. and {Howell}, D.~A. and {Hygelund}, J. and {Lister},
 T.~A. and {Lobdill}, R. and {Martinez}, J. and {Mullins}, D.~S. and
 {Norbury}, M. and {Parrent}, J. and {Paulson}, R. and {Petry}, D.~L. and
 {Pickles}, A. and {Posner}, V. and {Rosing}, W.~E. and {Ross}, R. and
 {Sand}, D.~J. and {Saunders}, E.~S. and {Shobbrook}, J. and {Shporer}, A.
 and {Street}, R.~A. and {Thomas}, D. and {Tsapras}, Y. and {Tufts}, J.~R.
 and {Valenti}, S. and {Vander Horst}, K. and {Walker}, Z. and {White}, G.
 and {Willis}, M.},
         title = "{Las Cumbres Observatory Global Telescope Network}",
       journal = {\pasp},
          year = 2013,
         month = sep,
        volume = {125},
        number = {931},
         pages = {1031},
          doi = {10.1086/673168},
archivePrefix = {arXiv},
       eprint = {1305.2437},
 primaryClass = {astro-ph.IM},
       adsurl = {https://ui.adsabs.harvard.edu/abs/2013PASP..125.1031B}
 }

@ARTICLE{Collins:2017,
    author = {{Collins}, K.~A. and {Kielkopf}, J.~F. and {Stassun}, K.~G.
 and
    {Hessman}, F.~V.},
     title = "{AstroImageJ: Image Processing and Photometric Extraction for
 Ultra-precise Astronomical Light Curves}",
   journal = {\aj},
 archivePrefix = "arXiv",
    eprint = {1601.02622},
  primaryClass = "astro-ph.IM",
      year = 2017,
     month = feb,
    volume = 153,
       eid = {77},
     pages = {77},
       doi = {10.3847/1538-3881/153/2/77},
    adsurl = {http://adsabs.harvard.edu/abs/2017AJ....153...77C}
 }

@INPROCEEDINGS{McCully:2018,
        author = {{McCully}, Curtis and {Volgenau}, Nikolaus H. and
          {Harbeck}, Daniel-Rolf and {Lister}, Tim A. and {Saunders}, Eric
 S. and
        {Turner}, Monica L. and {Siiverd}, Robert J. and {Bowman}, Mark},
        title = "{Real-time processing of the imaging data from the
network of Las Cumbres Observatory Telescopes using BANZAI}",
    booktitle = {\procspie},
         year = "2018",
       series = {Society of Photo-Optical Instrumentation Engineers (SPIE)
Conference Series},
       volume = {10707},
        month = "Jul",
          eid = {107070K},
        pages = {107070K},
          doi = {10.1117/12.2314340},
archivePrefix = {arXiv},
       eprint = {1811.04163},
 primaryClass = {astro-ph.IM},
       adsurl = {https://ui.adsabs.harvard.edu/abs/2018SPIE10707E..0KM}
}

@INPROCEEDINGS{collins:2019,
       author = {{Collins}, Karen},
        title = "{TESS Follow-up Observing Program Working Group (TFOP WG)
Sub Group 1 (SG1): Ground-based Time-series Photometry}",
    booktitle = {American Astronomical Society Meeting Abstracts \#233},
         year = 2019,
       series = {American Astronomical Society Meeting Abstracts},
       volume = {233},
        month = jan,
          eid = {140.05},
        pages = {140.05},
       adsurl = {https://ui.adsabs.harvard.edu/abs/2019AAS...23314005C}
}

@article{santerne2013contribution,
  title={The contribution of secondary eclipses as astrophysical false positives to exoplanet transit surveys},
  author={Santerne, A and Fressin, F and D{\'\i}az, RF and Figueira, P and Almenara, J-M and Santos, NC},
  journal={Astronomy \& Astrophysics},
  volume={557},
  pages={A139},
  year={2013},
  publisher={EDP Sciences}
}

@article{santerne2012sophie,
  title={SOPHIE velocimetry of Kepler transit candidates-VII. A false-positive rate of 35\% for Kepler close-in giant candidates},
  author={Santerne, A and D{\'\i}az, RF and Moutou, C and Bouchy, F and H{\'e}brard, G and Almenara, J-M and Bonomo, AS and Deleuil, M and Santos, NC},
  journal={Astronomy \& Astrophysics},
  volume={545},
  pages={A76},
  year={2012},
  publisher={EDP Sciences}
}

@article{bouchy2009sophie,
  title={The SOPHIE search for northern extrasolar planets-I. A companion around HD 16760 with mass close to the planet/brown-dwarf transition},
  author={Bouchy, F and H{\'e}brard, G and Udry, S and Delfosse, X and Boisse, I and Desort, M and Bonfils, X and Eggenberger, A and Ehrenreich, D and Forveille, T and others},
  journal={Astronomy \& Astrophysics},
  volume={505},
  number={2},
  pages={853--858},
  year={2009},
  publisher={EDP Sciences}
}

@article{bouchy2013sophie+,
  title={SOPHIE+: First results of an octagonal-section fiber for high-precision radial velocity measurements},
  author={Bouchy, F and D{\'\i}az, RF and H{\'e}brard, G and Arnold, L and Boisse, I and Delfosse, X and Perruchot, S and Santerne, A},
  journal={Astronomy \& Astrophysics},
  volume={549},
  pages={A49},
  year={2013},
  publisher={EDP Sciences}
}

@article{pepe2002coralie,
  title={The coralie survey for southern extra-solar planets vii-two short-period saturnian companions to hd 108147 and hd 168746},
  author={Pepe, F and Mayor, M and Galland, F and Naef, D and Queloz, D and Santos, NC and Udry, S and Burnet, M},
  journal={Astronomy \& Astrophysics},
  volume={388},
  number={2},
  pages={632--638},
  year={2002},
  publisher={EDP Sciences}
}

@article{baranne1996elodie,
  title={ELODIE: A spectrograph for accurate radial velocity measurements},
  author={Baranne, Andr{\'e} and Queloz, D and Mayor, Michel and Adrianzyk, G and Knispel, G and Kohler, D and Lacroix, D and Meunier, J-P and Rimbaud, G and Vin, A},
  journal={Astronomy and Astrophysics Supplement Series},
  volume={119},
  number={2},
  pages={373--390},
  year={1996},
  publisher={EDP Sciences}
}

@ARTICLE{ceres,
       author = {{Brahm}, Rafael and {Jord{\'a}n}, Andr{\'e}s and
{Espinoza}, N{\'e}stor},
        title = "{CERES: A Set of Automated Routines for Echelle Spectra}",
      journal = {\pasp},
         year = 2017,
        month = mar,
       volume = {129},
       number = {973},
        pages = {034002},
          doi = {10.1088/1538-3873/aa5455},
archivePrefix = {arXiv},
       eprint = {1609.02279},
 primaryClass = {astro-ph.IM},
       adsurl = {https://ui.adsabs.harvard.edu/abs/2017PASP..129c4002B}
}

@article{conzo2024toi,
  title={TOI-6883.01: A Single-transit Planet Candidate Detected from TESS},
  author={Conzo, G and Moriconi, M},
  journal={Research Notes of the AAS},
  volume={8},
  number={2},
  pages={53},
  year={2024},
  publisher={The American Astronomical Society}
}

@article{Sgro2024,
  title={Confirmation and characterization of the eccentric, warm Jupiter TIC 393818343 b with a network of citizen scientists},
  author={Sgro, Lauren A and Dalba, Paul A and Esposito, Thomas M and Marchis, Franck and Dragomir, Diana and Villanueva, Steven and Fulton, Benjamin and Billiani, Mario and Loose, Margaret and Meneghelli, Nicola and others},
  journal={The Astronomical Journal},
  volume={168},
  number={1},
  pages={26},
  year={2024},
  publisher={IOP Publishing}
}

@article{swayne2024eblm,
  title={The EBLM Project--XI. Mass, radius, and effective temperature measurements for 23 M-dwarf companions to solar-type stars observed with CHEOPS},
  author={Swayne, Matthew I and Maxted, Pierre FL and Triaud, Amaury HMJ and Sousa, S{\'e}rgio G and Deline, A and Ehrenreich, David and Hoyer, Sergio and Olofsson, G{\"o}ran and Boisse, I and Duck, Alison and others},
  journal={Monthly Notices of the Royal Astronomical Society},
  volume={528},
  number={4},
  pages={5703--5722},
  year={2024},
  publisher={Oxford University Press}
}

@article{von2019eblm,
  title={The EBLM Project-V. Physical properties of ten fully convective, very-low-mass stars},
  author={von Boetticher, Alexander and Triaud, Amaury HMJ and Queloz, Didier and Gill, Sam and Maxted, Pierre FL and Almleaky, Yaseen and Anderson, David R and Bouchy, Fran{\c{c}}ois and Burdanov, Artem and Cameron, Andrew Collier and others},
  journal={Astronomy \& Astrophysics},
  volume={625},
  pages={A150},
  year={2019},
  publisher={EDP Sciences}
}

@article{maxted2023eblm,
  title={The EBLM Project—From False Positives to Benchmark Stars and Circumbinary Exoplanets},
  author={Maxted, Pierre FL and Triaud, Amaury HMJ and Martin, David V},
  journal={Universe},
  volume={9},
  number={12},
  pages={498},
  year={2023},
  publisher={MDPI}
}

@article{triaud2017eblm,
  title={The EBLM Project-IV. Spectroscopic orbits of over 100 eclipsing M dwarfs masquerading as transiting hot Jupiters},
  author={Triaud, Amaury HMJ and Martin, David V and S{\'e}gransan, Damien and Smalley, Barry and Maxted, Pierre FL and Anderson, David R and Bouchy, Fran{\c{c}}ois and Cameron, Andrew Collier and Faedi, Francesca and Chew, Yilen G{\'o}mez Maqueo and others},
  journal={Astronomy \& Astrophysics},
  volume={608},
  pages={A129},
  year={2017},
  publisher={EDP Sciences}
}

@article{goupil2008tidal,
  title={Tidal dissipation in binary systems},
  author={Goupil, M-J and Zahn, J-P},
  journal={European Astronomical Society Publications Series},
  volume={29},
  pages={67--90},
  year={2008},
  publisher={EDP Sciences}
}

@ARTICLE{Torres:2010,
       author = {{Torres}, G. and {Andersen}, J. and {Gim{\'e}nez}, A.},
        title = "{Accurate masses and radii of normal stars: modern results and applications}",
      journal = {\aapr},
         year = 2010,
        month = feb,
       volume = {18},
       number = {1-2},
        pages = {67-126},
          doi = {10.1007/s00159-009-0025-1},
archivePrefix = {arXiv},
       eprint = {0908.2624},
 primaryClass = {astro-ph.SR}
}

@article{kroupa2001variation,
  title={On the variation of the initial mass function},
  author={Kroupa, Pavel},
  journal={Monthly Notices of the Royal Astronomical Society},
  volume={322},
  number={2},
  pages={231--246},
  year={2001},
  publisher={Blackwell Science Ltd Oxford, UK}
}

@article{chabrier2007evolution,
  title={Evolution of low-mass star and brown dwarf eclipsing binaries},
  author={Chabrier, Gilles and Gallardo, Jos{\'e} and Baraffe, Isabelle},
  journal={Astronomy \& Astrophysics},
  volume={472},
  number={2},
  pages={L17--L20},
  year={2007},
  publisher={EDP Sciences}
}

@article{lopez2007correlation,
  title={On the correlation between the magnetic activity levels, metallicities, and radii of low-mass stars},
  author={L{\'o}pez-Morales, Mercedes},
  journal={The Astrophysical Journal},
  volume={660},
  number={1},
  pages={732},
  year={2007},
  publisher={IOP Publishing}
}

@article{berger2006first,
  title={First results from the CHARA Array. IV. The interferometric radii of low-mass stars},
  author={Berger, David H and Gies, Douglas R and McAlister, Harold A and ten Brummelaar, TA and Henry, TJ and Sturmann, Judit and Sturmann, Laszlo and Turner, Nils H and Ridgway, ST and Aufdenberg, JP and others},
  journal={The Astrophysical Journal},
  volume={644},
  number={1},
  pages={475},
  year={2006},
  publisher={IOP Publishing}
}

@article{anderson2011thermal,
  title={Thermal emission at 4.5 and 8 $\mu$m of WASP-17b, an extremely large planet in a slightly eccentric orbit},
  author={Anderson, DR and Smith, AMS and Lanotte, AA and Barman, TS and Cameron, A Collier and Campo, CJ and Gillon, Micha{\"e}l and Harrington, Joseph and Hellier, C and Maxted, PFL and others},
  journal={Monthly Notices of the Royal Astronomical Society},
  volume={416},
  number={3},
  pages={2108--2122},
  year={2011},
  publisher={Blackwell Publishing Ltd Oxford, UK}
}

@article{fortney2011discovery,
  title={Discovery and atmospheric characterization of giant planet Kepler-12b: An inflated radius outlier},
  author={Fortney, Jonathan J and Demory, Brice-Olivier and D{\'e}sert, Jean-Michel and Rowe, Jason and Marcy, Geoffrey W and Isaacson, Howard and Buchhave, Lars A and Ciardi, David and Gautier, Thomas N and Batalha, Natalie M and others},
  journal={The Astrophysical Journal Supplement Series},
  volume={197},
  number={1},
  pages={9},
  year={2011},
  publisher={IOP Publishing}
}

@article{wang2025revised,
  title={A Revised Density Estimate for the Largest Known Exoplanet, HAT-P-67 b},
  author={Wang, Gavin and Balmer, William O and Pueyo, Laurent and Thorngren, Daniel and Schmidt, Stephen P and Wang, Le-Chris and Schlaufman, Kevin C and Stef{\'a}nsson, Gu{\dh}mundur and Rustamkulov, Zafar and Sing, David K},
  journal={The Astronomical Journal},
  volume={169},
  number={6},
  pages={336},
  year={2025},
  publisher={IOP Publishing}
}

@inproceedings{jenkins2016tess,
  title={The TESS science processing operations center},
  author={Jenkins, Jon M and Twicken, Joseph D and McCauliff, Sean and Campbell, Jennifer and Sanderfer, Dwight and Lung, David and Mansouri-Samani, Masoud and Girouard, Forrest and Tenenbaum, Peter and Klaus, Todd and others},
  booktitle={Software and Cyberinfrastructure for Astronomy IV},
  volume={9913},
  pages={99133E},
  year={2016},
  organization={International Society for Optics and Photonics}
}

@article{stumpe2014multiscale,
  title={Multiscale systematic error correction via wavelet-based bandsplitting in Kepler data},
  author={Stumpe, Martin C and Smith, Jeffrey C and Catanzarite, Joseph H and Van Cleve, Jeffrey E and Jenkins, Jon M and Twicken, Joseph D and Girouard, Forrest R},
  journal={Publications of the Astronomical Society of the Pacific},
  volume={126},
  number={935},
  pages={100},
  year={2014},
  publisher={IOP Publishing}
}

@article{chen2016probabilistic,
  title={Probabilistic forecasting of the masses and radii of other worlds},
  author={Chen, Jingjing and Kipping, David},
  journal={The Astrophysical Journal},
  volume={834},
  number={1},
  pages={17},
  year={2016},
  publisher={IOP Publishing}
}

@article{eastman2017,
  title={EXOFASTv2: Generalized publication-quality exoplanet modeling code},
  author={Eastman, Jason},
  journal={Astrophysics Source Code Library},
  pages={ascl--1710},
  year={2017}
}

@ARTICLE{Kiefer25a,
       author = {{Kiefer}, F. and {Lagrange}, A.-M. and {Rubini}, P. and {Philipot}, F.},
        title = "{Searching for substellar companion candidates with Gaia: I. Introducing the GaiaPMEX tool}",
      journal = {\aap},
         year = 2025,
        month = oct,
       volume = {702},
          eid = {A76},
        pages = {A76},
          doi = {10.1051/0004-6361/202449335},
archivePrefix = {arXiv},
       eprint = {2409.16992},
 primaryClass = {astro-ph.EP},
       adsurl = {https://ui.adsabs.harvard.edu/abs/2025A\&A...702A..76K}
}

@ARTICLE{Holman99,
       author = {{Holman}, Matthew J. and {Wiegert}, Paul A.},
        title = "{Long-Term Stability of Planets in Binary Systems}",
      journal = {\aj},
         year = 1999,
        month = jan,
       volume = {117},
       number = {1},
        pages = {621-628},
          doi = {10.1086/300695},
archivePrefix = {arXiv},
       eprint = {astro-ph/9809315},
 primaryClass = {astro-ph},
       adsurl = {https://ui.adsabs.harvard.edu/abs/1999AJ....117..621H}
}

@ARTICLE{Gaia23,
	author = {{Gaia Collaboration} and {Vallenari, A.} and {Brown, A. G. A.} and {Prusti, T.} and {de Bruijne, J. H. J.} and {Arenou, F.} and {Babusiaux, C.} and {Biermann, M.} and {Creevey, O. L.} and {Ducourant, C.} and {Evans, D. W.} and {Eyer, L.} and {Guerra, R.} and {Hutton, A.} and {Jordi, C.} and {Klioner, S. A.} and {Lammers, U. L.} and {Lindegren, L.} and {Luri, X.} and {Mignard, F.} and {Panem, C.} and {Pourbaix, D.} and {Randich, S.} and {Sartoretti, P.} and {Soubiran, C.} and {Tanga, P.} and {Walton, N. A.} and {Bailer-Jones, C. A. L.} and {Bastian, U.} and {Drimmel, R.} and {Jansen, F.} and {Katz, D.} and {Lattanzi, M. G.} and {van Leeuwen, F.} and {Bakker, J.} and {Cacciari, C.} and {Castañeda, J.} and {De Angeli, F.} and {Fabricius, C.} and {Fouesneau, M.} and {Frémat, Y.} and {Galluccio, L.} and {Guerrier, A.} and {Heiter, U.} and {Masana, E.} and {Messineo, R.} and {Mowlavi, N.} and {Nicolas, C.} and {Nienartowicz, K.} and {Pailler, F.} and {Panuzzo, P.} and {Riclet, F.} and {Roux, W.} and {Seabroke, G. M.} and {Sordo, R.} and {Thévenin, F.} and {Gracia-Abril, G.} and {Portell, J.} and {Teyssier, D.} and {Altmann, M.} and {Andrae, R.} and {Audard, M.} and {Bellas-Velidis, I.} and {Benson, K.} and {Berthier, J.} and {Blomme, R.} and {Burgess, P. W.} and {Busonero, D.} and {Busso, G.} and {Cánovas, H.} and {Carry, B.} and {Cellino, A.} and {Cheek, N.} and {Clementini, G.} and {Damerdji, Y.} and {Davidson, M.} and {de Teodoro, P.} and {Nuñez Campos, M.} and {Delchambre, L.} and {Dell’Oro, A.} and {Esquej, P.} and {Fernández-Hernández, J.} and {Fraile, E.} and {Garabato, D.} and {García-Lario, P.} and {Gosset, E.} and {Haigron, R.} and {Halbwachs, J.-L.} and {Hambly, N. C.} and {Harrison, D. L.} and {Hernández, J.} and {Hestroffer, D.} and {Hodgkin, S. T.} and {Holl, B.} and {Janßen, K.} and {Jevardat de Fombelle, G.} and {Jordan, S.} and {Krone-Martins, A.} and {Lanzafame, A. C.} and {Löffler, W.} and {Marchal, O.} and {Marrese, P. M.} and {Moitinho, A.} and {Muinonen, K.} and {Osborne, P.} and {Pancino, E.} and {Pauwels, T.} and {Recio-Blanco, A.} and {Reylé, C.} and {Riello, M.} and {Rimoldini, L.} and {Roegiers, T.} and {Rybizki, J.} and {Sarro, L. M.} and {Siopis, C.} and {Smith, M.} and {Sozzetti, A.} and {Utrilla, E.} and {van Leeuwen, M.} and {Abbas, U.} and {Ábrahám, P.} and {Abreu Aramburu, A.} and {Aerts, C.} and {Aguado, J. J.} and {Ajaj, M.} and {Aldea-Montero, F.} and {Altavilla, G.} and {Álvarez, M. A.} and {Alves, J.} and {Anders, F.} and {Anderson, R. I.} and {Anglada Varela, E.} and {Antoja, T.} and {Baines, D.} and {Baker, S. G.} and {Balaguer-Núñez, L.} and {Balbinot, E.} and {Balog, Z.} and {Barache, C.} and {Barbato, D.} and {Barros, M.} and {Barstow, M. A.} and {Bartolomé, S.} and {Bassilana, J.-L.} and {Bauchet, N.} and {Becciani, U.} and {Bellazzini, M.} and {Berihuete, A.} and {Bernet, M.} and {Bertone, S.} and {Bianchi, L.} and {Binnenfeld, A.} and {Blanco-Cuaresma, S.} and {Blazere, A.} and {Boch, T.} and {Bombrun, A.} and {Bossini, D.} and {Bouquillon, S.} and {Bragaglia, A.} and {Bramante, L.} and {Breedt, E.} and {Bressan, A.} and {Brouillet, N.} and {Brugaletta, E.} and {Bucciarelli, B.} and {Burlacu, A.} and {Butkevich, A. G.} and {Buzzi, R.} and {Caffau, E.} and {Cancelliere, R.} and {Cantat-Gaudin, T.} and {Carballo, R.} and {Carlucci, T.} and {Carnerero, M. I.} and {Carrasco, J. M.} and {Casamiquela, L.} and {Castellani, M.} and {Castro-Ginard, A.} and {Chaoul, L.} and {Charlot, P.} and {Chemin, L.} and {Chiaramida, V.} and {Chiavassa, A.} and {Chornay, N.} and {Comoretto, G.} and {Contursi, G.} and {Cooper, W. J.} and {Cornez, T.} and {Cowell, S.} and {Crifo, F.} and {Cropper, M.} and {Crosta, M.} and {Crowley, C.} and {Dafonte, C.} and {Dapergolas, A.} and {David, M.} and {David, P.} and {de Laverny, P.} and {De Luise, F.} and {De March, R.} and {De Ridder, J.} and {de Souza, R.} and {de Torres, A.} and {del Peloso, E. F.} and {del Pozo, E.} and {Delbo, M.} and {Delgado, A.} and {Delisle, J.-B.} and {Demouchy, C.} and {Dharmawardena, T. E.} and {Di Matteo, P.} and {Diakite, S.} and {Diener, C.} and {Distefano, E.} and {Dolding, C.} and {Edvardsson, B.} and {Enke, H.} and {Fabre, C.} and {Fabrizio, M.} and {Faigler, S.} and {Fedorets, G.} and {Fernique, P.} and {Fienga, A.} and {Figueras, F.} and {Fournier, Y.} and {Fouron, C.} and {Fragkoudi, F.} and {Gai, M.} and {Garcia-Gutierrez, A.} and {Garcia-Reinaldos, M.} and {García-Torres, M.} and {Garofalo, A.} and {Gavel, A.} and {Gavras, P.} and {Gerlach, E.} and {Geyer, R.} and {Giacobbe, P.} and {Gilmore, G.} and {Girona, S.} and {Giuffrida, G.} and {Gomel, R.} and {Gomez, A.} and {González-Núñez, J.} and {González-Santamaría, I.} and {González-Vidal, J. J.} and {Granvik, M.} and {Guillout, P.} and {Guiraud, J.} and {Gutiérrez-Sánchez, R.} and {Guy, L. P.} and {Hatzidimitriou, D.} and {Hauser, M.} and {Haywood, M.} and {Helmer, A.} and {Helmi, A.} and {Sarmiento, M. H.} and {Hidalgo, S. L.} and {Hilger, T.} and {Hładczuk, N.} and {Hobbs, D.} and {Holland, G.} and {Huckle, H. E.} and {Jardine, K.} and {Jasniewicz, G.} and {Jean-Antoine Piccolo, A.} and {Jiménez-Arranz, Ó.} and {Jorissen, A.} and {Juaristi Campillo, J.} and {Julbe, F.} and {Karbevska, L.} and {Kervella, P.} and {Khanna, S.} and {Kontizas, M.} and {Kordopatis, G.} and {Korn, A. J.} and {Kóspál, Á} and {Kostrzewa-Rutkowska, Z.} and {Kruszyńska, K.} and {Kun, M.} and {Laizeau, P.} and {Lambert, S.} and {Lanza, A. F.} and {Lasne, Y.} and {Le Campion, J.-F.} and {Lebreton, Y.} and {Lebzelter, T.} and {Leccia, S.} and {Leclerc, N.} and {Lecoeur-Taibi, I.} and {Liao, S.} and {Licata, E. L.} and {Lindstrøm, H. E. P.} and {Lister, T. A.} and {Livanou, E.} and {Lobel, A.} and {Lorca, A.} and {Loup, C.} and {Madrero Pardo, P.} and {Magdaleno Romeo, A.} and {Managau, S.} and {Mann, R. G.} and {Manteiga, M.} and {Marchant, J. M.} and {Marconi, M.} and {Marcos, J.} and {Marcos Santos, M. M. S.} and {Marín Pina, D.} and {Marinoni, S.} and {Marocco, F.} and {Marshall, D. J.} and {Martin Polo, L.} and {Martín-Fleitas, J. M.} and {Marton, G.} and {Mary, N.} and {Masip, A.} and {Massari, D.} and {Mastrobuono-Battisti, A.} and {Mazeh, T.} and {McMillan, P. J.} and {Messina, S.} and {Michalik, D.} and {Millar, N. R.} and {Mints, A.} and {Molina, D.} and {Molinaro, R.} and {Molnár, L.} and {Monari, G.} and {Monguió, M.} and {Montegriffo, P.} and {Montero, A.} and {Mor, R.} and {Mora, A.} and {Morbidelli, R.} and {Morel, T.} and {Morris, D.} and {Muraveva, T.} and {Murphy, C. P.} and {Musella, I.} and {Nagy, Z.} and {Noval, L.} and {Ocaña, F.} and {Ogden, A.} and {Ordenovic, C.} and {Osinde, J. O.} and {Pagani, C.} and {Pagano, I.} and {Palaversa, L.} and {Palicio, P. A.} and {Pallas-Quintela, L.} and {Panahi, A.} and {Payne-Wardenaar, S.} and {Peñalosa Esteller, X.} and {Penttilä, A.} and {Pichon, B.} and {Piersimoni, A. M.} and {Pineau, F.-X.} and {Plachy, E.} and {Plum, G.} and {Poggio, E.} and {Prša, A.} and {Pulone, L.} and {Racero, E.} and {Ragaini, S.} and {Rainer, M.} and {Raiteri, C. M.} and {Rambaux, N.} and {Ramos, P.} and {Ramos-Lerate, M.} and {Re Fiorentin, P.} and {Regibo, S.} and {Richards, P. J.} and {Rios Diaz, C.} and {Ripepi, V.} and {Riva, A.} and {Rix, H.-W.} and {Rixon, G.} and {Robichon, N.} and {Robin, A. C.} and {Robin, C.} and {Roelens, M.} and {Rogues, H. R. O.} and {Rohrbasser, L.} and {Romero-Gómez, M.} and {Rowell, N.} and {Royer, F.} and {Ruz Mieres, D.} and {Rybicki, K. A.} and {Sadowski, G.} and {Sáez Núñez, A.} and {Sagristà Sellés, A.} and {Sahlmann, J.} and {Salguero, E.} and {Samaras, N.} and {Sanchez Gimenez, V.} and {Sanna, N.} and {Santoveña, R.} and {Sarasso, M.} and {Schultheis, M.} and {Sciacca, E.} and {Segol, M.} and {Segovia, J. C.} and {Ségransan, D.} and {Semeux, D.} and {Shahaf, S.} and {Siddiqui, H. I.} and {Siebert, A.} and {Siltala, L.} and {Silvelo, A.} and {Slezak, E.} and {Slezak, I.} and {Smart, R. L.} and {Snaith, O. N.} and {Solano, E.} and {Solitro, F.} and {Souami, D.} and {Souchay, J.} and {Spagna, A.} and {Spina, L.} and {Spoto, F.} and {Steele, I. A.} and {Steidelmüller, H.} and {Stephenson, C. A.} and {Süveges, M.} and {Surdej, J.} and {Szabados, L.} and {Szegedi-Elek, E.} and {Taris, F.} and {Taylor, M. B.} and {Teixeira, R.} and {Tolomei, L.} and {Tonello, N.} and {Torra, F.} and {Torra, J.} and {Torralba Elipe, G.} and {Trabucchi, M.} and {Tsounis, A. T.} and {Turon, C.} and {Ulla, A.} and {Unger, N.} and {Vaillant, M. V.} and {van Dillen, E.} and {van Reeven, W.} and {Vanel, O.} and {Vecchiato, A.} and {Viala, Y.} and {Vicente, D.} and {Voutsinas, S.} and {Weiler, M.} and {Wevers, T.} and {Wyrzykowski, Ł.} and {Yoldas, A.} and {Yvard, P.} and {Zhao, H.} and {Zorec, J.} and {Zucker, S.} and {Zwitter, T.}},
	title = {Gaia Data Release 3 - Summary of the content and survey properties},
	journal={Astronomy \& Astrophysics},
	year = 2023,
	volume = 674,
	pages = "A1",
}

@ARTICLE{Baraffe03,
       author = {{Baraffe}, I. and {Chabrier}, G. and {Barman}, T.~S. and {Allard}, F. and {Hauschildt}, P.~H.},
        title = "{Evolutionary models for cool brown dwarfs and extrasolar giant planets. The case of HD 209458}",
      journal = {\aap},
         year = 2003,
        month = may,
       volume = {402},
        pages = {701-712},
          doi = {10.1051/0004-6361:20030252},
archivePrefix = {arXiv},
       eprint = {astro-ph/0302293},
 primaryClass = {astro-ph},
       adsurl = {https://ui.adsabs.harvard.edu/abs/2003A\&A...402..701B}
}

@ARTICLE{2019ApJ...881L..19V,
       author = {{Vanderburg}, Andrew and {Huang}, Chelsea X. and {Rodriguez}, Joseph E. and {Becker}, Juliette C. and {Ricker}, George R. and {Vanderspek}, Roland K. and {Latham}, David W. and {Seager}, Sara and {Winn}, Joshua N. and {Jenkins}, Jon M. and {Addison}, Brett and {Bieryla}, Allyson and {Brice{\~n}o}, Cesar and {Bowler}, Brendan P. and {Brown}, Timothy M. and {Burke}, Christopher J. and {Burt}, Jennifer A. and {Caldwell}, Douglas A. and {Clark}, Jake T. and {Crossfield}, Ian and {Dittmann}, Jason A. and {Dynes}, Scott and {Fulton}, Benjamin J. and {Guerrero}, Natalia and {Harbeck}, Daniel and {Horner}, Jonathan and {Kane}, Stephen R. and {Kielkopf}, John and {Kraus}, Adam L. and {Kreidberg}, Laura and {Law}, Nicolas and {Mann}, Andrew W. and {Mengel}, Matthew W. and {Morton}, Timothy D. and {Okumura}, Jack and {Pearce}, Logan A. and {Plavchan}, Peter and {Quinn}, Samuel N. and {Rabus}, Markus and {Rose}, Mark E. and {Rowden}, Pam and {Shporer}, Avi and {Siverd}, Robert J. and {Smith}, Jeffrey C. and {Stassun}, Keivan and {Tinney}, C.~G. and {Wittenmyer}, Rob and {Wright}, Duncan J. and {Zhang}, Hui and {Zhou}, George and {Ziegler}, Carl A.},
        title = "{TESS Spots a Compact System of Super-Earths around the Naked-eye Star HR 858}",
      journal = {\apjl},
         year = 2019,
        month = aug,
       volume = {881},
       number = {1},
          eid = {L19},
        pages = {L19},
          doi = {10.3847/2041-8213/ab322d},
archivePrefix = {arXiv},
       eprint = {1905.05193},
 primaryClass = {astro-ph.EP},
       adsurl = {https://ui.adsabs.harvard.edu/abs/2019ApJ...881L..19V}
}

@ARTICLE{Thorngren_mass_metallicity_giants_2016,
       author = {{Thorngren}, Daniel P. and {Fortney}, Jonathan J. and {Murray-Clay}, Ruth A. and {Lopez}, Eric D.},
        title = "{The Mass-Metallicity Relation for Giant Planets}",
      journal = {\apj},
         year = 2016,
        month = nov,
       volume = {831},
       number = {1},
          eid = {64},
        pages = {64},
          doi = {10.3847/0004-637X/831/1/64},
archivePrefix = {arXiv},
       eprint = {1511.07854},
 primaryClass = {astro-ph.EP},
       adsurl = {https://ui.adsabs.harvard.edu/abs/2016ApJ...831...64T}
}

@ARTICLE{Johnson_Giant_Occurrence_Metallicity_2010,
       author = {{Johnson}, John Asher and {Aller}, Kimberly M. and {Howard}, Andrew W. and {Crepp}, Justin R.},
        title = "{Giant Planet Occurrence in the Stellar Mass-Metallicity Plane}",
      journal = {\pasp},
         year = 2010,
        month = aug,
       volume = {122},
       number = {894},
        pages = {905},
          doi = {10.1086/655775},
archivePrefix = {arXiv},
       eprint = {1005.3084},
 primaryClass = {astro-ph.EP},
       adsurl = {https://ui.adsabs.harvard.edu/abs/2010PASP..122..905J}
}

@ARTICLE{Johnson_Soderblom_galactic_space_velocity_1987,
author = {{Johnson}, Dean R.~H. and {Soderblom}, David R.}, title = "{Calculating
Galactic Space Velocities and Their Uncertainties, with an Application to the Ursa
Major Group}", journal = {\aj}, year = 1987, month = apr, volume = {93}, pages =
{864}, doi = {10.1086/114370}, adsurl =
{https://ui.adsabs.harvard.edu/abs/1987AJ.....93..864J} }

@ARTICLE{Bensby_lsr_2003, author = {{Bensby}, T. and {Feltzing}, S. and
{Lundstr{\"o}m}, I.}, title = "{Elemental abundance trends in the Galactic thin and
thick disks as traced by nearby F and G dwarf stars}", journal = {\aap}, year = 2003, month = nov,
volume = {410}, pages = {527-551}, doi = {10.1051/0004-6361:20031213}, adsurl =
{https://ui.adsabs.harvard.edu/abs/2003A\&A...410..527B} }

@article{morales2022fundamental,
  title={Fundamental properties of late-type stars in eclipsing binaries},
  author={Morales, Juan Carlos and Ribas, Ignasi and Gim{\'e}nez, {\'A}lvaro and Baroch, David},
  journal={Galaxies},
  volume={10},
  number={5},
  pages={98},
  year={2022},
  publisher={MDPI}
}

@ARTICLE{Gaia_DR3_Release_Summary_2023, author = {{Gaia Collaboration} and
{Vallenari}, A. and {Brown}, A.~G.~A. and {Prusti}, T. and {de Bruijne}, J.~H.~J.
and {Arenou}, F. and {Babusiaux}, C. and {Biermann}, M. and {Creevey}, O.~L. and
{Ducourant}, C. and {Evans}, D.~W. and {Eyer}, L. and {Guerra}, R. and {Hutton}, A.
and {Jordi}, C. and {Klioner}, S.~A. and {Lammers}, U.~L. and {Lindegren}, L. and
{Luri}, X. and {Mignard}, F. and {Panem}, C. and {Pourbaix}, D. and {Randich}, S.
and {Sartoretti}, P. and {Soubiran}, C. and {Tanga}, P. and {Walton}, N.~A. and
{Bailer-Jones}, C.~A.~L. and {Bastian}, U. and {Drimmel}, R. and {Jansen}, F. and
{Katz}, D. and {Lattanzi}, M.~G. and {van Leeuwen}, F. and {Bakker}, J. and
{Cacciari}, C. and {Casta{\~n}eda}, J. and {De Angeli}, F. and {Fabricius}, C. and
{Fouesneau}, M. and {Fr{\'e}mat}, Y. and {Galluccio}, L. and {Guerrier}, A. and
{Heiter}, U. and {Masana}, E. and {Messineo}, R. and {Mowlavi}, N. and {Nicolas}, C.
and {Nienartowicz}, K. and {Pailler}, F. and {Panuzzo}, P. and {Riclet}, F. and
{Roux}, W. and {Seabroke}, G.~M. and {Sordo}, R. and {Th{\'e}venin}, F. and
{Gracia-Abril}, G. and {Portell}, J. and {Teyssier}, D. and {Altmann}, M. and
{Andrae}, R. and {Audard}, M. and {Bellas-Velidis}, I. and {Benson}, K. and
{Berthier}, J. and {Blomme}, R. and {Burgess}, P.~W. and {Busonero}, D. and {Busso},
G. and {C{\'a}novas}, H. and {Carry}, B. and {Cellino}, A. and {Cheek}, N. and
{Clementini}, G. and {Damerdji}, Y. and {Davidson}, M. and {de Teodoro}, P. and
{Nu{\~n}ez Campos}, M. and {Delchambre}, L. and {Dell'Oro}, A. and {Esquej}, P. and
{Fern{\'a}ndez-Hern{\'a}ndez}, J. and {Fraile}, E. and {Garabato}, D. and
{Garc{\'\i}a-Lario}, P. and {Gosset}, E. and {Haigron}, R. and {Halbwachs}, J. -L.
and {Hambly}, N.~C. and {Harrison}, D.~L. and {Hern{\'a}ndez}, J. and {Hestroffer},
D. and {Hodgkin}, S.~T. and {Holl}, B. and {Jan{\ss}en}, K. and {Jevardat de
Fombelle}, G. and {Jordan}, S. and {Krone-Martins}, A. and {Lanzafame}, A.~C. and
{L{\"o}ffler}, W. and {Marchal}, O. and {Marrese}, P.~M. and {Moitinho}, A. and
{Muinonen}, K. and {Osborne}, P. and {Pancino}, E. and {Pauwels}, T. and
{Recio-Blanco}, A. and {Reyl{\'e}}, C. and {Riello}, M. and {Rimoldini}, L. and
{Roegiers}, T. and {Rybizki}, J. and {Sarro}, L.~M. and {Siopis}, C. and {Smith}, M.
and {Sozzetti}, A. and {Utrilla}, E. and {van Leeuwen}, M. and {Abbas}, U. and
{{\'A}brah{\'a}m}, P. and {Abreu Aramburu}, A. and {Aerts}, C. and {Aguado}, J.~J.
and {Ajaj}, M. and {Aldea-Montero}, F. and {Altavilla}, G. and {{\'A}lvarez}, M.~A.
and {Alves}, J. and {Anders}, F. and {Anderson}, R.~I. and {Anglada Varela}, E. and
{Antoja}, T. and {Baines}, D. and {Baker}, S.~G. and {Balaguer-N{\'u}{\~n}ez}, L.
and {Balbinot}, E. and {Balog}, Z. and {Barache}, C. and {Barbato}, D. and {Barros},
M. and {Barstow}, M.~A. and {Bartolom{\'e}}, S. and {Bassilana}, J. -L. and
{Bauchet}, N. and {Becciani}, U. and {Bellazzini}, M. and {Berihuete}, A. and
{Bernet}, M. and {Bertone}, S. and {Bianchi}, L. and {Binnenfeld}, A. and
{Blanco-Cuaresma}, S. and {Blazere}, A. and {Boch}, T. and {Bombrun}, A. and
{Bossini}, D. and {Bouquillon}, S. and {Bragaglia}, A. and {Bramante}, L. and
{Breedt}, E. and {Bressan}, A. and {Brouillet}, N. and {Brugaletta}, E. and
{Bucciarelli}, B. and {Burlacu}, A. and {Butkevich}, A.~G. and {Buzzi}, R. and
{Caffau}, E. and {Cancelliere}, R. and {Cantat-Gaudin}, T. and {Carballo}, R. and
{Carlucci}, T. and {Carnerero}, M.~I. and {Carrasco}, J.~M. and {Casamiquela}, L.
and {Castellani}, M. and {Castro-Ginard}, A. and {Chaoul}, L. and {Charlot}, P. and
{Chemin}, L. and {Chiaramida}, V. and {Chiavassa}, A. and {Chornay}, N. and
{Comoretto}, G. and {Contursi}, G. and {Cooper}, W.~J. and {Cornez}, T. and
{Cowell}, S. and {Crifo}, F. and {Cropper}, M. and {Crosta}, M. and {Crowley}, C.
and {Dafonte}, C. and {Dapergolas}, A. and {David}, M. and {David}, P. and {de
Laverny}, P. and {De Luise}, F. and {De March}, R.}, title = "{Gaia Data Release 3.
Summary of the content and survey properties}", journal = {\aap}, year =
2023, month = jun, volume = {674}, eid = {A1}, pages = {A1}, doi =
{10.1051/0004-6361/202243940}, archivePrefix = {arXiv}, eprint = {2208.00211},
primaryClass = {astro-ph.GA}, adsurl =
{https://ui.adsabs.harvard.edu/abs/2023A\&A...674A...1G} }

@ARTICLE{Almeida_Fernandes_Stellar_Ages_LSR_2018, author = {{Almeida-Fernandes}, F.
and {Rocha-Pinto}, H.~J.}, title = "{A method to estimate stellar ages from
kinematical data}", journal = {\mnras}, year = 2018, month = may,
volume = {476}, number = {1}, pages = {184-197}, doi = {10.1093/mnras/sty119},
archivePrefix = {arXiv}, eprint = {1801.04046}, primaryClass = {astro-ph.SR}, adsurl
= {https://ui.adsabs.harvard.edu/abs/2018MNRAS.476..184A} }

@ARTICLE{Reddy_Thick_Disk_2006, author = {{Reddy}, Bacham E. and {Lambert}, David L.
and {Allende Prieto}, Carlos}, title = "{Elemental abundance survey of the Galactic
thick disc}", journal = {\mnras}, year = 2006, month = apr,
volume = {367}, number = {4}, pages = {1329-1366}, doi =
{10.1111/j.1365-2966.2006.10148.x}, archivePrefix = {arXiv}, eprint =
{astro-ph/0512505}, primaryClass = {astro-ph}, adsurl =
{https://ui.adsabs.harvard.edu/abs/2006MNRAS.367.1329R} }

@ARTICLE{Chen_PAST_I_2021, author = {{Chen}, Di-Chang and {Xie}, Ji-Wei and {Zhou},
Ji-Lin and {Dong}, Subo and {Liu}, Chao and {Wang}, Hai-Feng and {Xiang}, Mao-Sheng
and {Huang}, Yang and {Luo}, Ali and {Zheng}, Zheng}, title = "{Planets Across Space
and Time (PAST). I. Characterizing the Memberships of Galactic Components and
Stellar Ages: Revisiting the Kinematic Methods and Applying to Planet Host Stars}",
journal = {\apj}, year = 2021, month = mar, volume = {909}, number =
{2}, eid = {115}, pages = {115}, doi = {10.3847/1538-4357/abd5be}, archivePrefix =
{arXiv}, eprint = {2102.09424}, primaryClass = {astro-ph.EP}, adsurl =
{https://ui.adsabs.harvard.edu/abs/2021ApJ...909..115C} }

@INPROCEEDINGS{Wilson2003,

       author = {{Wilson}, John C. and {Eikenberry}, Stephen S. and {Henderson},
Charles P. and {Hayward}, Thomas L. and {Carson}, Joseph C. and {Pirger},
Bruce and {Barry}, Donald J. and {Brandl}, Bernhard R. and {Houck}, James R.
and {Fitzgerald}, Gregory J. and {Stolberg}, T.~M.},

        title = "{A Wide-Field Infrared Camera for the Palomar 200-inch Telescope}",

    booktitle = {Instrument Design and Performance for Optical/Infrared Ground-based
Telescopes},

         year = 2003,

       editor = {{Iye}, Masanori and {Moorwood}, Alan F.~M.},

       series = {Society of Photo-Optical Instrumentation Engineers (SPIE)
Conference Series},

       volume = {4841},

        month = mar,

        pages = {451-458},

          doi = {10.1117/12.460336},

       adsurl = {https://ui.adsabs.harvard.edu/abs/2003SPIE.4841..451W}

}

@ARTICLE{Stefansson2017,

       author = {{Stefansson}, Gudmundur and {Mahadevan}, Suvrath and {Hebb}, Leslie
and {Wisniewski}, John and {Huehnerhoff}, Joseph and {Morris}, Brett and
{Halverson}, Sam and {Zhao}, Ming and {Wright}, Jason and {O'rourke}, Joseph
and {Knutson}, Heather and {Hawley}, Suzanne and {Kanodia}, Shubham and {Li},
Yiting and {Hagen}, Lea M.~Z. and {Liu}, Leo J. and {Beatty}, Thomas and
{Bender}, Chad and {Robertson}, Paul and {Dembicky}, Jack and {Gray}, Candace
and {Ketzeback}, William and {McMillan}, Russet and {Rudyk}, Theodore},

        title = "{Toward Space-like Photometric Precision from the Ground with
Beam-shaping Diffusers}",

      journal = {\apj},

         year = 2017,

        month = oct,

       volume = {848},

       number = {1},

          eid = {9},

        pages = {9},

          doi = {10.3847/1538-4357/aa88aa},

archivePrefix = {arXiv},

       eprint = {1710.01790},

 primaryClass = {astro-ph.IM},

       adsurl = {https://ui.adsabs.harvard.edu/abs/2017ApJ...848....9S}

}

@ARTICLE{Vissapragada2020,

       author = {{Vissapragada}, Shreyas and {Jontof-Hutter}, Daniel and {Shporer},
Avi and {Knutson}, Heather A. and {Liu}, Leo and {Thorngren}, Daniel and
{Lee}, Eve J. and {Chachan}, Yayaati and {Mawet}, Dimitri and
{Millar-Blanchaer}, Maxwell A. and {Nilsson}, Ricky and {Tinyanont}, Samaporn
and {Vasisht}, Gautam and {Wright}, Jason T.},

        title = "{Diffuser-assisted Infrared Transit Photometry for Four Dynamically
Interacting Kepler Systems}",

      journal = {\aj},

         year = 2020,

        month = mar,

       volume = {159},

       number = {3},

          eid = {108},

        pages = {108},

          doi = {10.3847/1538-3881/ab65c8},

archivePrefix = {arXiv},

       eprint = {1907.04445},

 primaryClass = {astro-ph.EP},

       adsurl = {https://ui.adsabs.harvard.edu/abs/2020AJ....159..108V}

}

@ARTICLE{GomezBarrientos2025,

       author = {{Gomez Barrientos}, Jonathan and {Greklek-McKeon}, Michael and
{Knutson}, Heather A. and {Giacalone}, Steven and {Levine}, W. Garrett and
{Saidel}, Morgan and {Vissapragada}, Shreyas and {Ciardi}, David R. and
{Collins}, Karen A. and {Latham}, David W. and {Watkins}, Cristilyn N. and
{Budnikova}, Polina A. and {Cheryasov}, Dmitry V. and {Fukui}, Akihiko and
{Bieryla}, Allyson and {Shporer}, Avi and {Tofflemire}, Benjamin M. and
{Clark}, Catherine A. and {Stockdale}, Chris and {Littlefield}, Colin and
{Gilbert}, Emily and {Palle}, Enric and {Girardin}, Eric and {Murgas}, Felipe
and {Bergsten}, Galen J. and {Osborn}, Hugh P. and {Crossfield}, Ian J.~M.
and {de Leon}, Jerome and {Higuera}, Jesus and {Isogai}, Keisuke and
{Everett}, Mark E. and {Lund}, Michael B. and {Narita}, Norio and {Schwarz},
Richard P. and {Zambelli}, Roberto and {Howell}, Steve B.},

        title = "{Validation of TESS Planet Candidates with Multicolor Transit
Photometry and TRICERATOPS+}",

      journal = {\aj},

         year = 2025,

        month = sep,

       volume = {170},

       number = {3},

          eid = {148},

        pages = {148},

          doi = {10.3847/1538-3881/ade68b},

archivePrefix = {arXiv},

       eprint = {2508.02782},

 primaryClass = {astro-ph.EP},

       adsurl = {https://ui.adsabs.harvard.edu/abs/2025AJ....170..148G}

}

@article{safonov2017speckle,
  title={The speckle polarimeter of the 2.5-m telescope: Design and calibration},
  author={Safonov, BS and Lysenko, PA and Dodin, AV},
  journal={Astronomy Letters},
  volume={43},
  number={5},
  pages={344--364},
  year={2017},
  publisher={Springer}
}

@article{smith2012kepler,
  title={Kepler presearch data conditioning II-a Bayesian approach to systematic error correction},
  author={Smith, Jeffrey C and Stumpe, Martin C and Van Cleve, Jeffrey E and Jenkins, Jon M and Barclay, Thomas S and Fanelli, Michael N and Girouard, Forrest R and Kolodziejczak, Jeffery J and McCauliff, Sean D and Morris, Robert L and others},
  journal={Publications of the Astronomical Society of the Pacific},
  volume={124},
  number={919},
  pages={1000--1014},
  year={2012},
  publisher={University of Chicago Press}
}

@article{strakhov2023speckle,
  title={Speckle interferometry with CMOS detector},
  author={Strakhov, IA and Safonov, BS and Cheryasov, DV},
  journal={Astrophysical Bulletin},
  volume={78},
  number={2},
  pages={234--258},
  year={2023},
  publisher={Springer}
}

@article{conzo2025characterization,
  title={Characterization of the Visual Binary TOI-6883AB and its dynamical implications for the planetary companion TOI-6883Ab},
  author={Conzo, Giuseppe and Campos, Fran and Conti, Francesco and Sharp, Ian},
  journal={Research Notes of the AAS},
  volume={9},
  number={6},
  pages={139},
  year={2025},
  publisher={The American Astronomical Society}
}

@article{Nielsen_Planet_Formation_2023,
  title={Planet formation throughout the Milky Way-Planet populations in the context of Galactic chemical evolution},
  author={Nielsen, Jesper and Gent, Matthew Raymond and Bergemann, Maria and Eitner, Philipp and Johansen, Anders},
  journal={Astronomy \& Astrophysics},
  volume={678},
  pages={A74},
  year={2023},
  publisher={EDP Sciences}
}

@article{Hallatt_thickdisk_planets_2025,
  title={On the Formation of Planets in the Milky Way’s Thick Disk},
  author={Hallatt, Tim and Lee, Eve J},
  journal={The Astrophysical Journal},
  volume={979},
  number={2},
  pages={120},
  year={2025},
  publisher={IOP Publishing}
}

@ARTICLE{2025arXiv251121817F,
       author = {{Frommer}, Arielle C. and {Eastman}, Jason D. and {Ciardi}, David R. and {Howell}, Steve B. and {Zhou}, George and {Bieryla}, Allyson and {Ziegler}, Carl and {Owen}, Kepler},
        title = "{A Paradigm Shift in Exoplanet False Positive Identification with HIP-44302}",
      journal = {arXiv e-prints},
         year = 2025,
        month = nov,
          eid = {arXiv:2511.21817},
        pages = {arXiv:2511.21817},
          doi = {10.48550/arXiv.2511.21817},
archivePrefix = {arXiv},
       eprint = {2511.21817},
 primaryClass = {astro-ph.EP},
       adsurl = {https://ui.adsabs.harvard.edu/abs/2025arXiv251121817F}
}

@article{huang2020photometry,
  title={Photometry of 10 million stars from the first two years of TESS full frame images: part I},
  author={Huang, Chelsea X and Vanderburg, Andrew and P{\'a}l, Andras and Sha, Lizhou and Yu, Liang and Fong, Willie and Fausnaugh, Michael and Shporer, Avi and Guerrero, Natalia and Vanderspek, Roland and others},
  journal={Research Notes of the AAS},
  volume={4},
  number={11},
  pages={204},
  year={2020},
  publisher={The American Astronomical Society}
}

@ARTICLE{2021FrASS...8..138S,
       author = {{Scott}, Nicholas J. and {Howell}, Steve B. and {Gnilka}, Crystal L. and {Stephens}, Andrew W. and {Salinas}, Ricardo and {Matson}, Rachel A. and {Furlan}, Elise and {Horch}, Elliott P. and {Everett}, Mark E. and {Ciardi}, David R. and {Mills}, Dave and {Quigley}, Emmett A.},
        title = "{Twin High-resolution, High-speed Imagers for the Gemini Telescopes: Instrument description and science verification results}",
      journal = {Frontiers in Astronomy and Space Sciences},
         year = 2021,
        month = sep,
       volume = {8},
          eid = {138},
        pages = {138},
          doi = {10.3389/fspas.2021.716560},
       adsurl = {https://ui.adsabs.harvard.edu/abs/2021FrASS...8..138S}
}

@ARTICLE{2011AJ....142...19H,
       author = {{Howell}, Steve B. and {Everett}, Mark E. and {Sherry}, William and {Horch}, Elliott and {Ciardi}, David R.},
        title = "{Speckle Camera Observations for the NASA Kepler Mission Follow-up Program}",
      journal = {\aj},
         year = 2011,
        month = jul,
       volume = {142},
       number = {1},
          eid = {19},
        pages = {19},
          doi = {10.1088/0004-6256/142/1/19},
       adsurl = {https://ui.adsabs.harvard.edu/abs/2011AJ....142...19H}
}

@INPROCEEDINGS{Bonfils2015,
       author = {{Bonfils}, X. and {Almenara}, J.~M. and {Jocou}, L. and {Wunsche}, A. and {Kern}, P. and {Delboulb{\'e}}, A. and {Delfosse}, X. and {Feautrier}, P. and {Forveille}, T. and {Gluck}, L. and {Lafrasse}, S. and {Magnard}, Y. and {Maurel}, D. and {Moulin}, T. and {Murgas}, F. and {Rabou}, P. and {Rochat}, S. and {Roux}, A. and {Stadler}, E.},
        title = "{ExTrA: Exoplanets in transit and their atmospheres}",
    booktitle = {Techniques and Instrumentation for Detection of Exoplanets VII},
         year = 2015,
       editor = {{Shaklan}, Stuart},
       series = {Society of Photo-Optical Instrumentation Engineers (SPIE) Conference Series},
       volume = {9605},
        month = sep,
          eid = {96051L},
        pages = {96051L},
          doi = {10.1117/12.2186999},
archivePrefix = {arXiv},
       eprint = {1508.06601},
 primaryClass = {astro-ph.IM},
       adsurl = {https://ui.adsabs.harvard.edu/abs/2015SPIE.9605E..1LB}
}
\begin{acknowledgements}
We warmly thank the OHP staff for their support on the observations. We received funding from the French Programme National de Physique Stellaire (PNPS) and the Programme National de Planétologie (PNP) of CNRS (INSU). N.H. acknowledges support from a postdoctoral fellowship funded by the Domaine de Recherche et d’Innovation Majeur (DIM), financed by the Île-de-France Region. B.S.S. and I.A.S. acknowledge the support of M.V. Lomonosov Moscow State University
Program of Development.

DD acknowledges support from the TESS Guest Investigator Program grant 80NSSC23K0769. We acknowledge the use of public TESS data from pipelines at the TESS Science Office and at the TESS Science Processing Operations Center. Resources supporting this work were provided by the NASA High-End Computing (HEC) Program through the NASA Advanced Supercomputing (NAS) Division at Ames Research Center for the production of the SPOC data products. This work makes use of observations from the LCOGT network. Part of the
 LCOGT telescope time was granted by NOIRLab through the Mid-Scale
 Innovations Program (MSIP). MSIP is funded by NSF.


 This paper is based on observations made with the Las Cumbres
 Observatory’s education network telescopes that were upgraded through
 generous support from the Gordon and Betty Moore Foundation.


 This research has made use of the Exoplanet Follow-up Observation Program
 (ExoFOP; DOI: 10.26134/ExoFOP5) website, which is operated by the
 California Institute of Technology, under contract with the National
 Aeronautics and Space Administration under the Exoplanet Exploration
 Program.


 Funding for the \textit{TESS} mission is provided by NASA's Science Mission
 Directorate. KAC acknowledges support from the \textit{TESS} mission via subaward
 s3449 from MIT.

Some of the observations in this paper made use of the High-Resolution Imaging instrument ‘Alopeke and were obtained under Gemini Proposal Number: GN-2025A-DD-101. ‘Alopeke was funded by the NASA Exoplanet Exploration Program and built at the NASA Ames Research Center by Steve B. Howell, Nic Scott, Elliott P. Horch, and Emmett Quigley. Alopeke was mounted on the Gemini North telescope of the international Gemini Observatory, a program of NSF’s OIR Lab, which is managed by the Association of Universities for Research in Astronomy (AURA) under a cooperative agreement with the National Science Foundation. on behalf of the Gemini partnership: the National Science Foundation (United States), National Research Council (Canada), Agencia Nacional de Investigación y Desarrollo (Chile), Ministerio de Ciencia, Tecnología e Innovación (Argentina), Ministério da Ciência, Tecnologia, Inovações e Comunicações (Brazil), and Korea Astronomy and Space Science Institute (Republic of Korea). This work is based on observations obtained at the Hale Telescope, Palomar
Observatory, as part of a collaborative agreement between the Caltech Optical Observatories and the Jet Propulsion Laboratory (operated by Caltech for NASA). F. M. acknowledges the financial support from the Agencia Estatal de Investigaci\'{o}n del Ministerio de Ciencia, Innovaci\'{o}n y Universidades (MCIU/AEI) through grant PID2023-152906NA-I00.  E.M. acknowledges funding from FAPEMIG under project number APQ-02493-22 and a research productivity grant number 309829/2022-4 awarded by the CNPq, Brazil. O.D.S.D. acknowledges support from e-CHEOPS (PEA No 4000142255) and from the
Fundação para a Ciência e a Tecnologia (FCT) and Fundo Europeu de Desenvolvimento
Regional (FEDER) via COMPETE2020 through the research grants UIDB/04434/2020,
UIDP/04434/2020, 2022.06962.PTDC. A.J. acknowledges support from Fondecyt project 1251439. R.B. acknowledges support from FONDECYT Project 1241963.
We acknowledge funding from the European Research Council under the ERC Grant Agreement n. 337591-ExTrA.

\end{acknowledgements}
\onecolumn
\begin{appendix}

\section{\textit{TESS} observations summary}
\begin{table*}[h!]
\centering
\small
\caption{\textit{TESS} observations summary}
\resizebox{0.5\columnwidth}{!}{%
\begin{tabular}{ll}
\hline 
System & Sectors (EXP) \\
\hline
TOI-6457 & 44 (120), 45 (120), 60 (120), 71 (120), 72 (120) \\
TOI-6649 & 17 (120), 57 (20), 84 (20) \\
TOI-6842A & 42 (600), 43 (600), 70 (120) \\
TOI-6893A & 42 (120), 43 (20), 70 (120) \\
TOI-7025 & 14 (1800), 15 (1800), 16 (1800), 17 (1800), 18 (1800), 19 (1800), \\
& 20 (1800), 21 (1800), 22 (1800), 23 (1800), 24 (1800), 26 (1800), \\
& 41 (600), 47 (600), 48 (600), 49 (600), 50 (600), 51 (600), \\
& 53 (600), 54 (600), 55 (600), 57 (200), 59 (200), 60 (200), \\
& 73 (200), 74 (200), 76 (200), 77 (200), 79 (200), 84 (120), \\
& 85 (120), 86 (120) \\
TOI-7176 & 56 (200), 83$^{*}$ (200) \\
TOI-5621 & 14 (1800), 15 (1800), 20 (1800), 21 (1800), \\
& 22 (1800), 40 (600), 41 (600), 47 (600), 48 (600), \\
& 53 (600), 60 (120), 74 (20), 75 (20) \\
TOI-5321 & 17 (1800), 42 (600), 43 (600), 57 (120), 84 (120) \\
TOI-3678 & 18 (1800), 42 (120), 43 (120), 44 (120), 58 (120), 85 (120) \\
TOI-5217A & 40$^{*}$ (600), 41$^{*}$ (600), 54$^{*}$ (600), 74 (120) \\
TOI-6883A & 55 (120), 81$^{*}$ (200) \\
TOI-5893 & 55 (600), 82 (120) \\
TOI-5906 & 15 (1800), 55 (120), 56 (120), 75 (120), 76 (120), 82 (120), \\
& 83 (120) \\
TOI-5308 & 4 (1800), 42 (600), 43 (600), 70 (120)\\
\hline
\end{tabular}%
}
\begin{tablenotes}
\small
\item Notes: This table is based on the \textit{TESS}-point Web Tool.\protect\footnote{\url{https://heasarc.gsfc.nasa.gov/wsgi-scripts/TESS/TESS-point_Web_Tool/TESS-point_Web_Tool/wtv_v2.0.py}} 
The exposure time (in seconds) for each \textit{TESS} sector observation is indicated in parentheses. $^{*}$ The SAP data were used for these sectors.
\end{tablenotes}
\label{tess_obs}
\end{table*}

\section{RV observations summary}
\begin{table*}[h!]
\caption{\label{rv_ob} RV observations summary}
\centering
\resizebox{0.8\columnwidth}{!}{%
\begin{tabular}{lllllllllll}
\hline
System &instrument&Mode&N RVs (used)& Start [UT]/end [UT]& RMS [m/s]&$\sigma_{RV}$[m/s] &S/N& EXP [s]\\
\hline
TOI-6457&SOPHIE&HR&15 (13)&November 2024/January 2026&25&6.3&31&1934\\
        &TRES&---&9(9)&October 2023/ February 2024 &30&22.2&38&347\\
TOI-6649&SOPHIE&HR&78(76)&November 2023/January 2026&51&7.7 &33&1009\\
TOI-6842A&SOPHIE&HE&18(18)&July 2023/August 2025&33&11.6&32&1379\\
TOI-6893A&SOPHIE&HR&19(19)&July 2024/October 2025&26&5.0&49&672\\
& FEROS & HR & 7 & August 2024 / October 2025& 15.4 & 7.9 & 120 & 600 \\
& PLATOspec & HR & 25 & July 2025/ Janurary 2026 & 22.6 & 10.7 & 100 & 900 \\
TOI-7025&SOPHIE&HE&28(27)&August 2024/October 2025&50&14.4&36&1537\\
        &TRES&---&12(12)&June 2024/September 2024&75&28.3&36 & 1029\\
TOI-5621&SOPHIE &HE&16(16)&March 2024/ May 2025&3662&21.5&22&1926 \\
TOI-5321&SOPHIE &HR&9(9)& September 2022/ September 2023&6350&26.4 &26&1455\\
        &TRES&---&8(8)&July 2022/September 2024&5575& 50.9&36&428\\
        &FEROS&HR&2(2)&August 2022/ August 2022&7977&31.7&85&900\\
TOI-3678&SOPHIE &HR&11(11)& August 2021/October 2025&8381&47.2&26&1232\\
TOI-5217A&SOPHIE&HR&61(59)& July 2024/ October 2025& 173& 8.8&23&1770\\
         &TRES&---&16(16)&March 2022/November 2024&123.2 &27.1&32&709\\
TOI-6883A&SOPHIE&HR&19(19)& May 2024/December 2024&207& 3.1&37&884\\
         &TRES&---&10(10)&April 2024/June 2024&186&17.1&43&207\\
TOI-5893&SOPHIE&HE&43(42)&December 2022/December 2025& 46& 14.2&29&1175\\
         &FEROS&HR&36(36)&June 2023/ October 2025&41.8 &12.2&40&1200\\

TOI-5308&SOPHIE&HR&11(10)&September 2022/ December 2024&6278 &11.4 &17&1342\\
        &TRES&---&8(8)&January 2019/September 2024&6058&41.4&31&953\\
TOI-5906&SOPHIE&HE&16(16)&December 2022/September 2024&23906&15.6& 31&1101\\
TOI-7176&SOPHIE&HE&18(17)& August 2024/November 2025&34&8.5&40&1521\\
        &TRES&---&4(4)&January 2025/November 2025&47&16.8&27&900\\
\hline
\end{tabular}
}
\tablefoot{\footnotesize
In this table, `HR` and `HE` refer to the high-resolution and high-efficiency modes of spectrographs, respectively. `N RVs` lists the total number of RV measurements obtained for each target, with the number actually used in our analysis given in parentheses. The observing start/end dates indicate the temporal span of the observations, but do not imply continuous monitoring. The `RMS` column reports the root-mean-square scatter of the final RV time series after applying the corrections described in Sect.~\ref{sophie_observation}. $\sigma_{RV}$ corresponds to the median uncertainty of the RV measurements. The S/N for SOPHIE is given per pixel at 550~nm, whereas for FEROS and PLATOspec it is given per resolution element at 515~nm. The values listed for S/N, exposure time (`EXP`), and $\sigma_{RV}$ correspond to their mean values for each target.}
\end{table*}

\section{High resolution image}

\begin{table*}[h!]
\centering
\caption{Summary of high-spatial-resolution imaging observations. The 0.25\arcsec{} and 1.0\arcsec{} limits correspond to the 5$\sigma$ detection contrast ($\Delta$mag) achieved at angular separations of 0.25\arcsec{} and 1.0\arcsec{} from the target star, respectively.}
\begin{tabular}{lccccccc}
\hline
Object & Telescope & Band & Detector & Detection? & Date & 0.25\arcsec{} limit & 1.0\arcsec{} limit \\
\hline
TOI-6649  & SAI (2.5 m) & $I_c$   & CMOS  & no & 2023-09-29 & 4.8 & 6.3 \\
TOI-6842A & SAI (2.5 m) & $I_c$   & CMOS  & no & 2024-08-29 & 4.0 & 4.9 \\
TOI-6893A & SAI (2.5 m) & 625,nm & CMOS  & no & 2024-11-10 & 5.6 & 7.6 \\
TOI-7025  & SAI (2.5 m) & $I_c$   & CMOS  & no & 2024-07-30 & 4.7 & 5.8 \\
TOI-5621  & SAI (2.5 m) & $I_c$   & CMOS  & no & 2023-05-04 & 3.8 & 5.4 \\
TOI-5321  & SAI (2.5 m) & $I_c$   & CMOS  & no & 2022-11-08 & 4.1 & 7.5 \\
TOI-3678  & SAI (2.5 m) & $I_c$   & EMCCD & no & 2021-10-21 & 4.4 & 5.9 \\
TOI-5217A & SAI (2.5 m) & $I_c$   & CMOS  & no & 2023-08-31 & 4.4 & 7.0 \\
TOI-6883A & SAI (2.5 m) & 625,nm & CMOS  & no & 2024-11-12 & 5.2 & 6.0 \\
TOI-5893  & SAI (2.5 m) & $I_c$   & CMOS  & no & 2023-09-23 & 4.9 & 6.9 \\

TOI-5906  & SAI (2.5 m) & $I_c$   & CMOS  & no & 2023-08-28 & 4.6 & 7.2 \\
TOI-5308  & SAI (2.5 m) & $I_c$   & CMOS  & no & 2022-11-20 & 3.4 & 6.4 \\
TOI-7176  & Gemini (8 m) & 562 \& 832\,nm & EMCCD & no & 2025-06-14 & 5.4 & 8.0 \\
TOI-6457  & Gemini (8 m) & 562 \& 832\,nm & EMCCD & no & 2022-02-10 & 5.4 & 8.0 \\ 
\hline
\end{tabular}
\label{high_image}
\end{table*}

\begin{figure*}
\centering
\includegraphics[width=0.8\columnwidth]{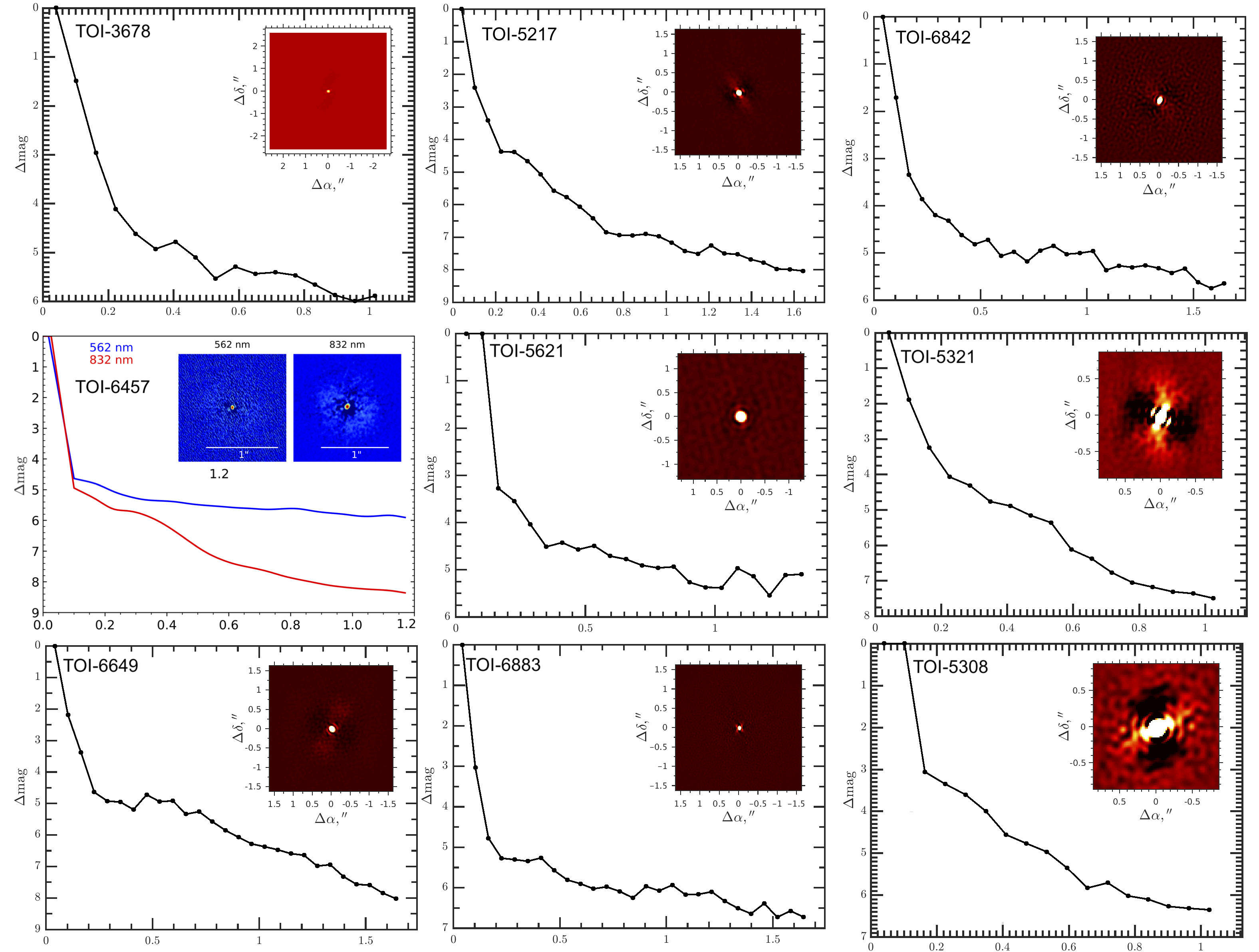}
\includegraphics[width=0.8\columnwidth]{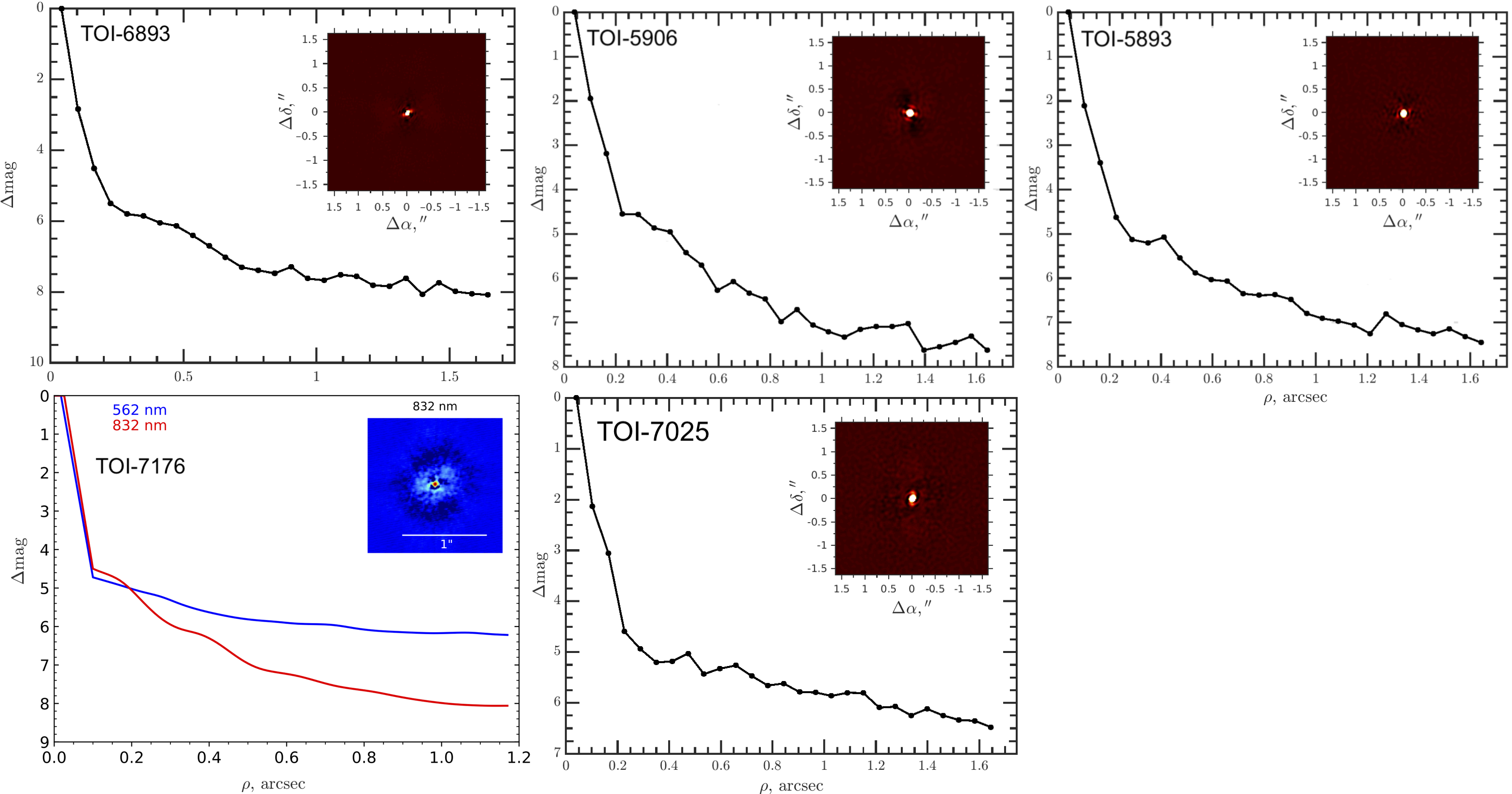}
\caption{Contrast curves from the speckle interferometry observations of these stars are displayed as solid black, blue, and red lines. The star name appears on each plot, which also includes an inset in the upper right corner showing the final reconstructed image. For TOI-7176, only the 832 nm image is shown due to the low S/N of the 562 nm image.}
\label{image_high}
\end{figure*}

\onecolumn
\section{Summary of ground-based light curve observations}

\begin{table*}[h!]
\caption{Summary of ground-based light curve observations}\label{table:groundobservatories}
\begin{tabular}{llcccccc}
\hline
TOI & Observatory & UTC Date & Filter & Ap. Radius  & Detrending &
Coverage & Result \\
    &             &          &        & (arcsec)    &
&          &        \\
\hline
6842A & RFAC-0.3m        & 2024-07-24 & $\rm R_c$  & 5.8 & X  & Ingress &
detection on target\\
6842A & LCOGT-CTIO-0.4m  & 2024-10-30 & Sloan $i'$ & 3.7 & AM & Full    &
detection on target\\
6842A & LCOGT-Hal-0.4m   & 2024-10-30 & Sloan $i'$ & 3.7 & AM, TOTC &
Egress  & detection on target\\
6842A & LCOGT-Hal-0.4m A & 2024-11-15 & Sloan $g'$ & 3.7 & TOTC & Full    &
detection on target\\
6842A & LCOGT-Hal-0.4m B & 2024-11-15 & Sloan $g'$ & 3.7 & FWHM & Ingress &
detection on target\\
5217A & LCOGT-McD-0.4m   & 2025-07-05 & Sloan $i'$ & 2.9 & AM, FWHM &
Full    & detection on target\\
6649 & LCOGT-Teid-1m    & 2023-20-28 & Pan-STARRS $\rm z_s$ & 6.2 & AM,
FWHM & Ingress & detection on target\\
6649 & ULMT-0.6m        & 2024-12-02 & Sloan $i'$ & 5.8 & AM & OOT     &
n/a\\
6883A & LCOGT-CTIO-0.4m  & 2024-07-02 & Sloan $g'$ & 3.7 & TM, TOTC &
Full    & detection on target\\
6883A & LCOGT-CTIO-0.4m  & 2024-07-02 & Sloan $i'$ & 3.7 & TM, TOTC &
Full    & detection on target\\
7025 & WIRC-5.1m& 2026-02-05 & $\rm J$ &4.25&FWHM&
Full& detection on target\\
\hline
\end{tabular}
\begin{list}{}{}
\item {\bf{Notes.}}
See Section \ref{sec:groundtransits} for observatory identifications.
Pan-STARRS $\rm z_s$ band  has $\lambda_{\rm c} = 8700$\,\AA\ and ${\rm
Width} =1040$\,\AA. OOT indicates ``out of transit''. See the text for
detrending codes and methodology.
\end{list}
\end{table*}

\section{Priors on EXOFASTv2 joint analysis}

\begin{table*}[h!]
\caption{Adopted priors used in the joint modeling with \texttt{EXOFASTv2}. $\mathcal{N}(a,,b)$ represents a Gaussian prior with mean $a$ and standard deviation $b$, and $\mathcal{U}(a,b)$ a uniform prior between $a$ and $b$.}
\label{tab:priors}
\begin{tabular}{lcccccc}
\hline
\hline
Target & $T_{\rm eff}$ (K) & [Fe/H] & $A_V$ (mag) & Parallax (mas) & $M_\star$ ($M_\odot$) & $R_\star$ ($R_\odot$) \\
\hline
TOI-5893 & $\mathcal{N}(5963,\,136)$ & $\mathcal{N}(0.149,\,0.094)$ & $\mathcal{U}(0.00,\,0.38)$ & $\mathcal{N}(3.078,\,0.017)$ & --- & ---\\
TOI-5217A & $\mathcal{N}(6142,\,147)$ & $\mathcal{N}(0.299,\,0.080)$ &--- & --- & $\mathcal{N}(1.51,\,0.09)$ & $\mathcal{N}(2.31,\,0.07)$ \\
TOI-5308 & $\mathcal{N}(5820,\,194)$ & $\mathcal{N}(0.188,\,0.140)$ & $\mathcal{U}(0.00,\,0.12)$ & $\mathcal{N}(2.947,\,0.057)$ & ---&--- \\
TOI-5621 & $\mathcal{N}(5989,\,144)$ & $\mathcal{N}(0.014,\,0.080)$ & $\mathcal{U}(0.00,\,0.10)$ & $\mathcal{N}(1.423,\,0.014)$ &--- & ---\\
TOI-6649 & $\mathcal{N}(6397,\,138)$ & $\mathcal{N}(0.181,\,0.094)$ & $\mathcal{U}(0.00,\,0.10)$ & $\mathcal{N}(4.981,\,0.019)$ & --- & --- \\
TOI-6883A & $\mathcal{N}(5831,\,140)$ & $\mathcal{N}(0.275,\,0.080)$ & $\mathcal{U}(0.00,\,0.32)$ & $\mathcal{N}(10.644,\,0.018)$ & ---& --- \\
TOI-6457 & $\mathcal{N}(6230,134)$ & $\mathcal{N}(0.23,0.05)$ & $\mathcal{U}(0.00,0.19)$ & $\mathcal{N}(3.8127,0.0237)$ & --- & ---\\
TOI-6842A & $\mathcal{N}(5786,128)$ & $\mathcal{N}(-0.007,0.09)$ & $\mathcal{U}(0.0000,0.1496)$ & $\mathcal{N}(2.5449,0.0242)$ & --- & ---\\
TOI-6893A & $\mathcal{N}(6239, 138)$ & $\mathcal{N}(0.25,0.1)$ & $\mathcal{U}(0.00,0.09)$ & $\mathcal{N}(6.008,0.02)$ &---& ---\\
TOI-5621 & $\mathcal{N}(5989,144)$ & $\mathcal{N}(0.014,0.08)$ & $\mathcal{U}(0.00,0.0979)$ & $\mathcal{N}(1.42333,0.01428)$ & ---& ---\\
TOI-5321 & $\mathcal{N}(6530,140)$ & $\mathcal{N}(-0.22,0.13)$ & --- & --- & $\mathcal{N}(1.39,0.08)$ & $\mathcal{N}(1.972,0.095)$ \\
TOI-3678 & $\mathcal{N}(6777,205)$ & $\mathcal{N}(-0.03,0.22)$ 
& ---& --- & $\mathcal{N}(1.60,0.10)$ & $\mathcal{N}(2.35,0.14)$ \\
TOI-7176& $\mathcal{N}(4484,171)$& $\mathcal{N}(-0.08,0.09)$
& $\mathcal{U}(0.00,0.479)$& $\mathcal{N}(14.5007,0.02353)$& --- & --- \\
TOI-5906 & --- & $\mathcal{N}(-0.19, 0.05)$ & $\mathcal{U}(0.00, 9.46)$ & $\mathcal{N}(3.455, 0.018)$ & ---& --- \\
TOI-7025 & {$\mathcal{N}(6164,67)$} & {$\mathcal{N}(0.08, 0.05)$} & --- & --- & {$\mathcal{N}(1.27, 0.08)$} & {$\mathcal{N}(1.53, 0.03)$}\\
\hline
\end{tabular}

\vspace{0.3cm}
\small
Notes. $\mathcal{N}(\mu,\,\sigma)$ denotes a normal prior with mean $\mu$ and standard deviation $\sigma$, and $\mathcal{U}(a,\,b)$ denotes a uniform prior between $a$ and $b$. For TOI-6649, additional normal priors were applied to the orbital period ($P = 5.987 \pm 0.500$~d), the time of conjunction ($T_{\rm c} = 2460587.9234 \pm 0.500$ BJD$_{TDB}$), and the planet-to-star radius ratio ($p = 0.034 \pm 0.100$). For TOI-5321, a uniform prior between 0 and 185 was applied to the velocity offsets of both the SOPHIE and FEROS data sets. For TOI-5906, we imposed a uniform prior on the time of secondary eclipse, $\mathcal{U}(2459831.24,,2459831.29)$ BJD$_{\mathrm{TDB}}$. In addition, we applied a lower bound of $\cos i > 0.03$, corresponding to $i < 88.3^\circ$, consistent with the non-detection of the primary eclipse. Finally, for TOI-7025, a normal prior was imposed on the time of conjunction ($T_{\rm c} = 2460340.03 \pm 0.01$ BJD$_{\rm TDB}$), together with uniform bounds corresponding approximately to $0.6 < e < 1$ and $57^\circ < \omega_* < 108^\circ$. The SOPHIE and TRES RV jitter terms were assigned uniform priors over the intervals $[0,\,70]$ and $[0,\,141]$~m\,s$^{-1}$, respectively.
\end{table*}

\section{Less probable solutions from bimodal posterior}
\label{bimodal}

\begin{table*}
\centering
\caption{Stellar and planetary parameters corresponding to the secondary, less probable modes of the posterior probability distributions for TOI-5308.}
\begin{tabular}{l l}
\hline
Parameter & TOI-5308 (Prob. = 15\%) \\

\hline
$M_*$ (\msun) & $1.069^{+0.048}_{-0.054}$ \\
$R_*$ (\rsun) & $1.553^{+0.037}_{-0.036}$ \\
$L_*$ (\lsun) & $2.20^{+0.11}_{-0.10}$ \\
$\rho_*$ (cgs) & $0.401^{+0.028}_{-0.025}$ \\
$\log g$ (cgs) & $4.083\pm0.021$ \\
$T_{\rm eff}$ (K) & $5643^{+74}_{-73}$ \\
$[{\rm Fe/H}]$ (dex) & $0.19\pm0.12$ \\
Age (Gyr) & $9.0^{+1.9}_{-1.5}$ \\
EEP & $450.2^{+4.7}_{-5.3}$ \\
$A_V$ (mag) & $0.078^{+0.027}_{-0.043}$ \\
$d$ (pc) & $338.0^{+6.1}_{-6.0}$ \\
\hline
\\
& TOI-5308\,B \\
\\
$P$ (days) & $8.3403294\pm0.0000067$ \\
$R_B$ ($R_J$)& $1.405^{+0.047}_{-0.045}$ \\
$M_B$ ($M_J$) & $104.3^{+3.0}_{-3.5}$ \\
$T_C$ (BJD$_{\rm TDB}$) & $2459485.68488^{+0.00055}_{-0.00054}$ \\
$a$ (AU) & $0.0848^{+0.0012}_{-0.0014}$ \\
$i$ (deg) & $85.68\pm0.12$ \\
$e$ & 0 (fixed) \\
$\omega_*$ (deg) & 90 (fixed) \\
$T_{\rm eq}$ (K) & $1165^{+15}_{-14}$ \\
$\tau_{\rm circ}$ (Gyr) & $720^{+120}_{-100}$ \\
$K$ (m/s) & $9390^{+51}_{-42}$ \\
$R_B/R_*$ & $0.0930\pm0.0013$ \\
$a/R_*$ & $11.72^{+0.26}_{-0.25}$ \\
$\delta_{\rm TESS}$ & $0.00746\pm0.00016$ \\
$\tau$ (days) & $0.0498^{+0.0037}_{-0.0033}$ \\
$T_{14}$ (days) & $0.1462\pm0.0022$ \\
$b$ & $0.8835^{+0.0057}_{-0.0062}$ \\
$\rho_P$ (cgs) & $46.5^{+4.7}_{-4.1}$ \\
$M_B\sin i$ ($M_J$) & $104.0^{+3.0}_{-3.5}$ \\
$T_P$ (BJD$_{\rm TDB}$) & $2459485.68443^{+0.00055}_{-0.00054}$ \\
\hline
\end{tabular}
\label{tab:toi_bimodal}
\end{table*}

\section{Summary of the spectroscopic, photometric, and SED analyses}
\label{all_plot}

\begin{figure*}
\centering
\includegraphics[width=0.8\columnwidth]{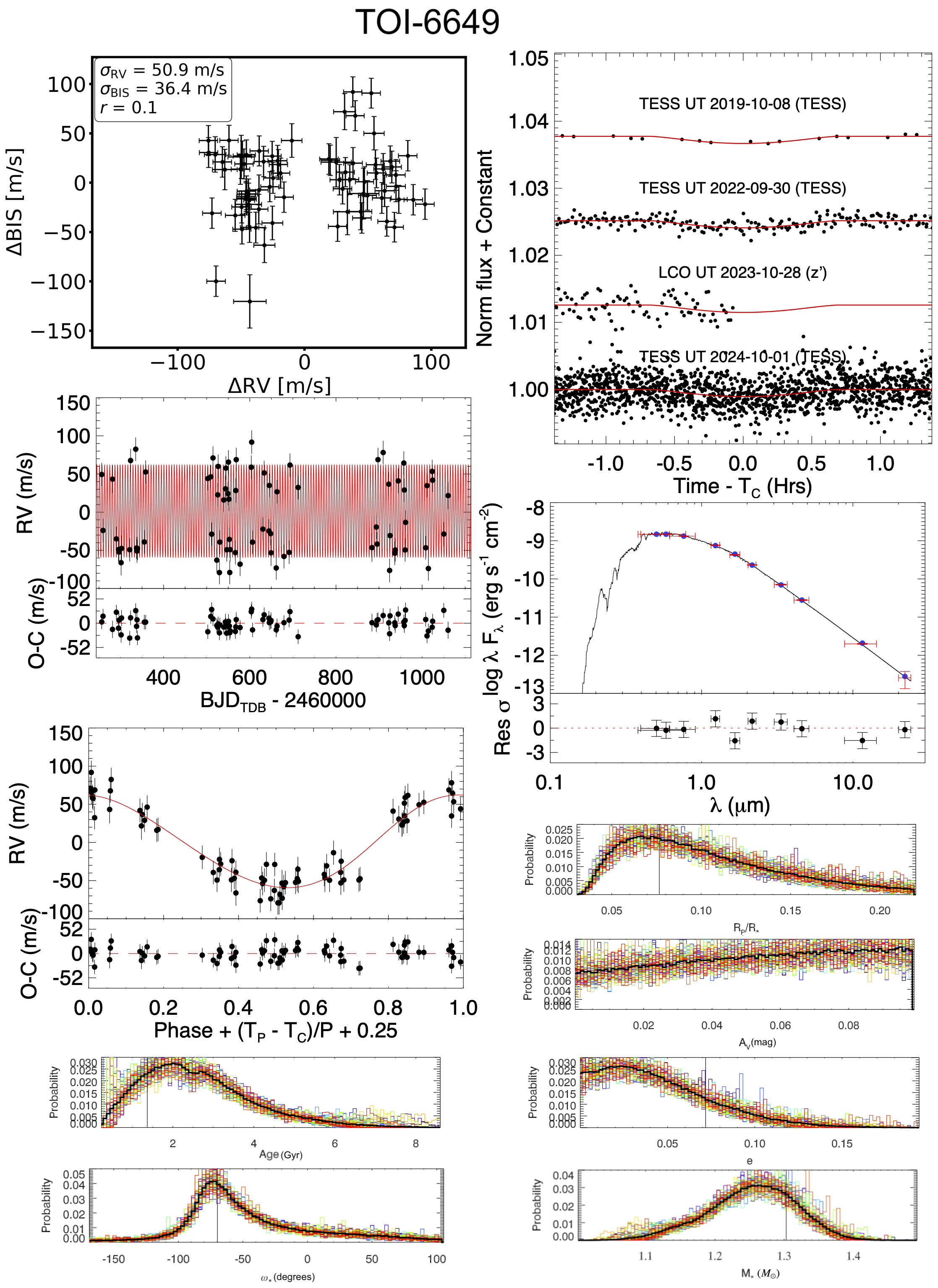}
\caption{Summary of the spectroscopic, photometric, and SED analyses for TOI-6649. \emph{Top-left}:  Bisector span versus RV. \emph{Second-left}: RV time series with the best-fit Keplerian model overplotted in red; residuals are shown below. \emph{Third-left}: Phase-folded RV curve for TOI-6649b with the best-fit model (red line) and residuals. \emph{Top-right:} \textit{TESS} and ground-based light curves, phase-folded on the best-fit period and time of conjunction, with their respective transit models overplotted in red. \emph{Second-right:} Broadband photometry is shown in red, with blue points indicating the best-fit model flux derived from the stellar properties and the MIST bolometric correction grid. For illustrative purposes only, a Kurucz (1993) atmospheric model corresponding to the best-fit stellar parameters is plotted in black; note that the fit is performed directly using the MIST grid. Residuals are displayed below. \emph{Bottom-right} and \emph{Bottom-left:} Posterior probability distributions of non-Gaussian parameters obtained from the EXOFASTv2 fit. Colored curves represent individual MCMC chains, the thick black curve shows their average, and the thin black vertical line marks the median value. TOI-6649\,b is a grazing giant planet with an orbital period of $P = 5.9877995^{+0.000010}_{-0.0000098}$~d, a mass of $M_{\rm p}=0.645^{+0.032}_{-0.033}\,M_{\rm J}$, and a radius of $R_{\rm p}=1.25^{+0.74}_{-0.45}\,R_{\rm J}$.}
\label{image_6649}
\end{figure*}

\begin{figure*}
\centering
\includegraphics[width=0.8\columnwidth]{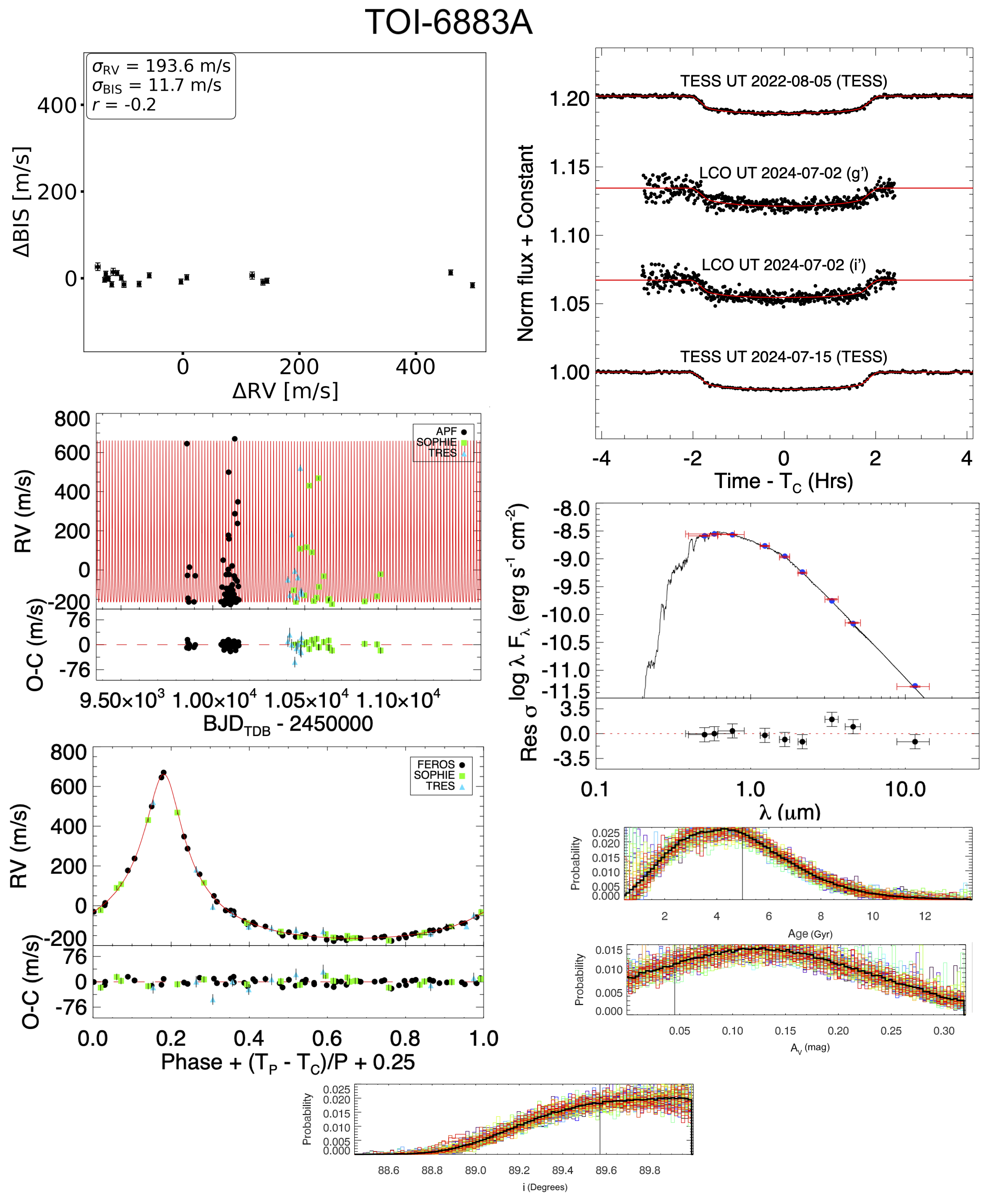}
\caption{Summary of the spectroscopic, photometric, and SED analyses for TOI-6883A. The panels are similar to those shown in Fig.~\ref{image_6649}, except that the posterior probability distributions shown here correspond to age, inclination, and $A_V$. TOI-6883A\,b is a giant planet with an orbital period of $P = 16.2499265\pm0.0000077$~d, a mass of $M_{\rm p}=4.26\pm0.14\,M_{\rm J}$, and a radius of $R_{\rm p}=1.073^{+0.021}_{-0.020}\,R_{\rm J}$.}
\label{image_6883}
\end{figure*}

\begin{figure*}
\centering
\includegraphics[width=0.8\columnwidth]{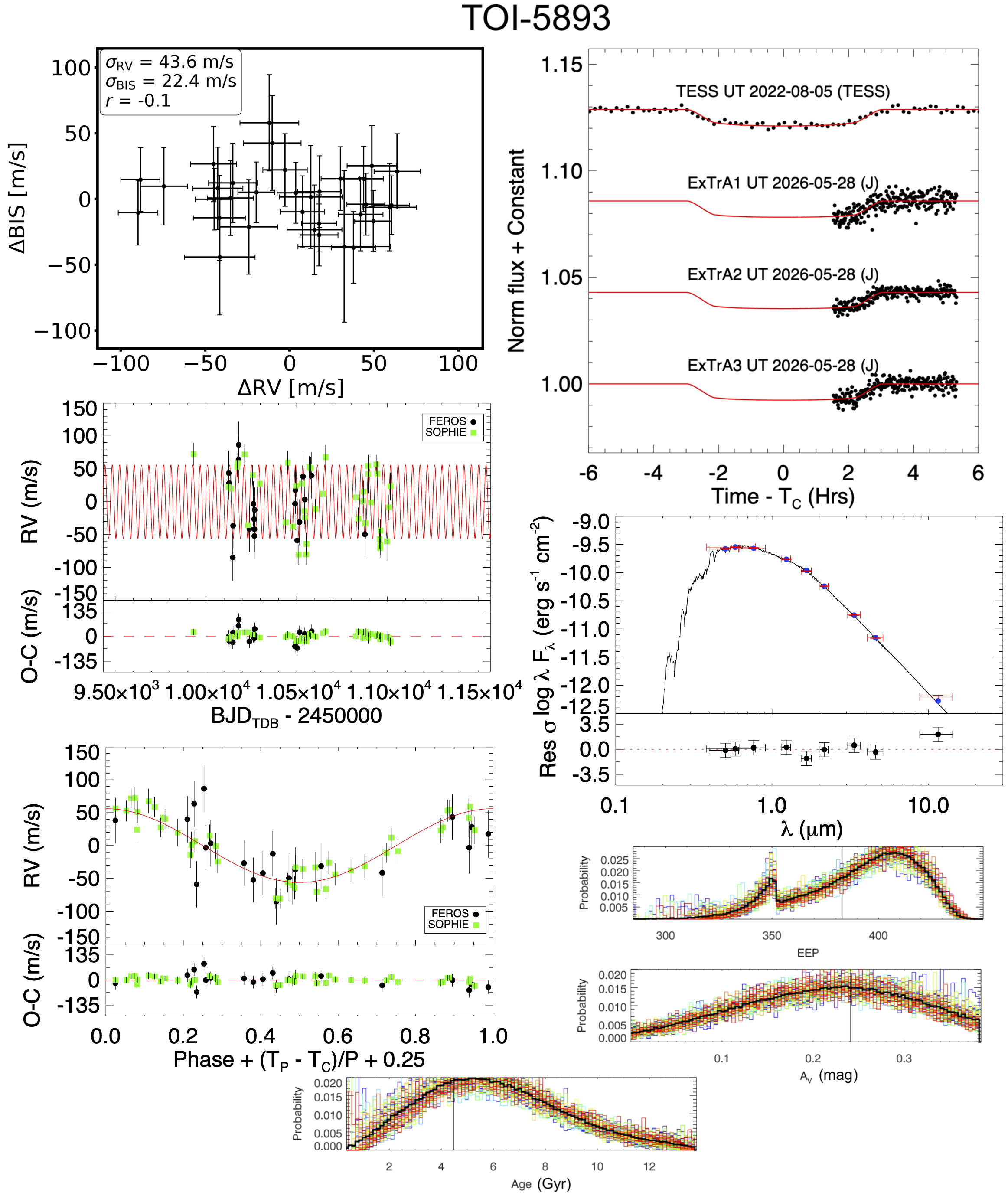}
\caption{Summary of the spectroscopic, photometric, and SED analyses for TOI-5893. The panels are similar to those shown in Fig.~\ref{image_6649}. The posterior probability distributions shown here correspond to the EEP, age, $A_V$, and orbital period. We note that the discontinuities in the EEP probability distribution function are expected due to the correction from a uniform EEP prior to a uniform age prior \citep{eastman2019}. TOI-5893\,b is a warm Jupiter with an orbital period of $P = 40.369376 \pm 0.000100$~d, a mass of $M_{\rm p}=1.012^{+0.073}_{-0.072}\,M_{\rm J}$, and a radius of $R_{\rm p}=0.973^{+0.036}_{-0.035}\,R_{\rm J}$.}
\label{image_5893}
\end{figure*}

\begin{figure*}
\centering
\includegraphics[width=0.8\columnwidth]{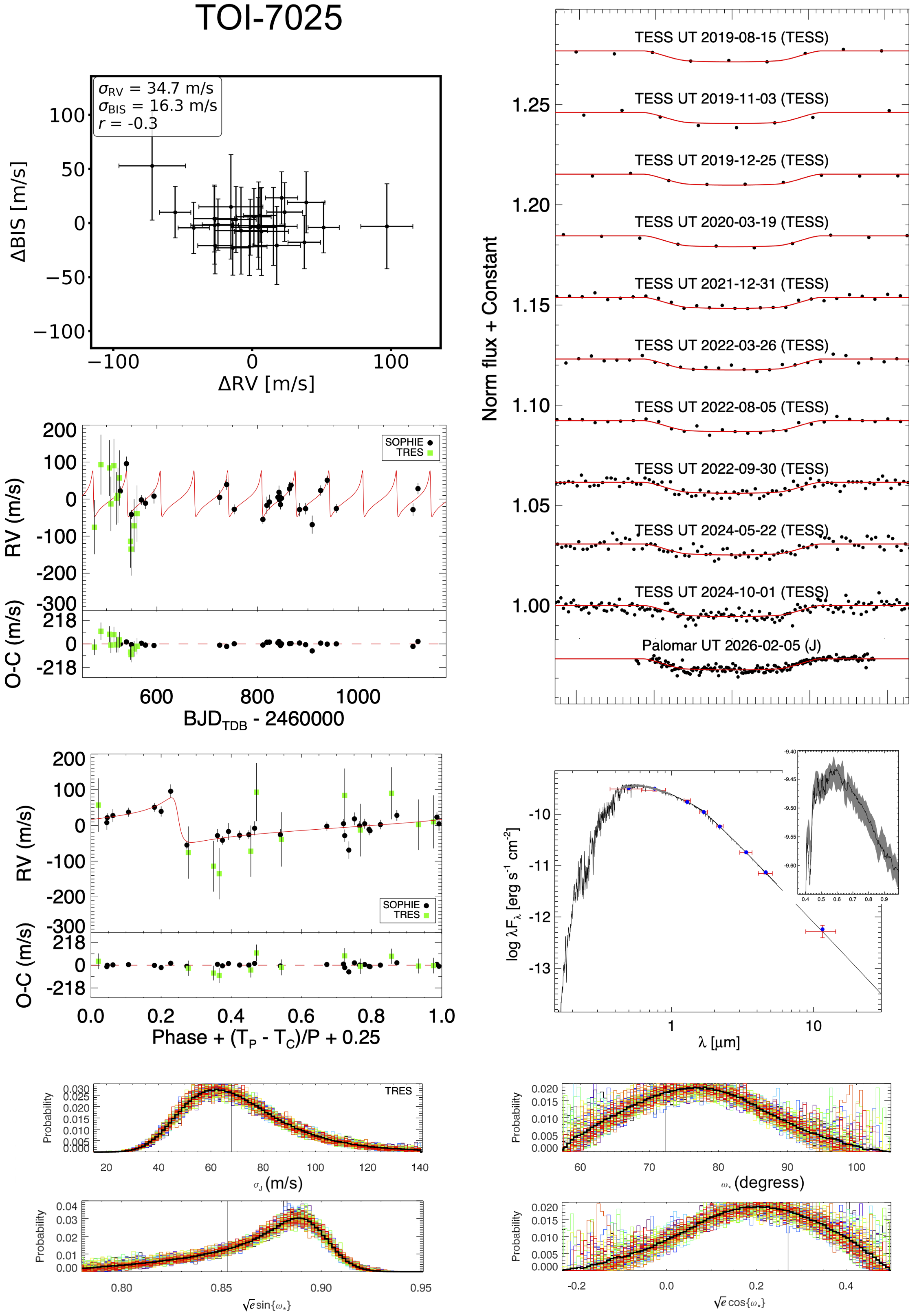}
\caption{Summary of the spectroscopic and photometric analyses for TOI-7025. The panels are similar to those shown in Fig.~\ref{image_6649}. The posterior probability distributions shown correspond to TRES stellar jitter and argument of periastron. TOI-7025\,b is a long-period giant planet with an orbital period of $P = 66.995436 \pm 0.000034$~d, a mass of $M_{\rm p} = 0.873^{+0.099}_{-0.095}\,M_{\rm J}$, and a radius of $R_{\rm p} = 1.152 \pm 0.029\,R_{\rm J}$.}
\label{image_7025}
\end{figure*}

\begin{figure*}
\centering
\includegraphics[width=0.8\columnwidth]{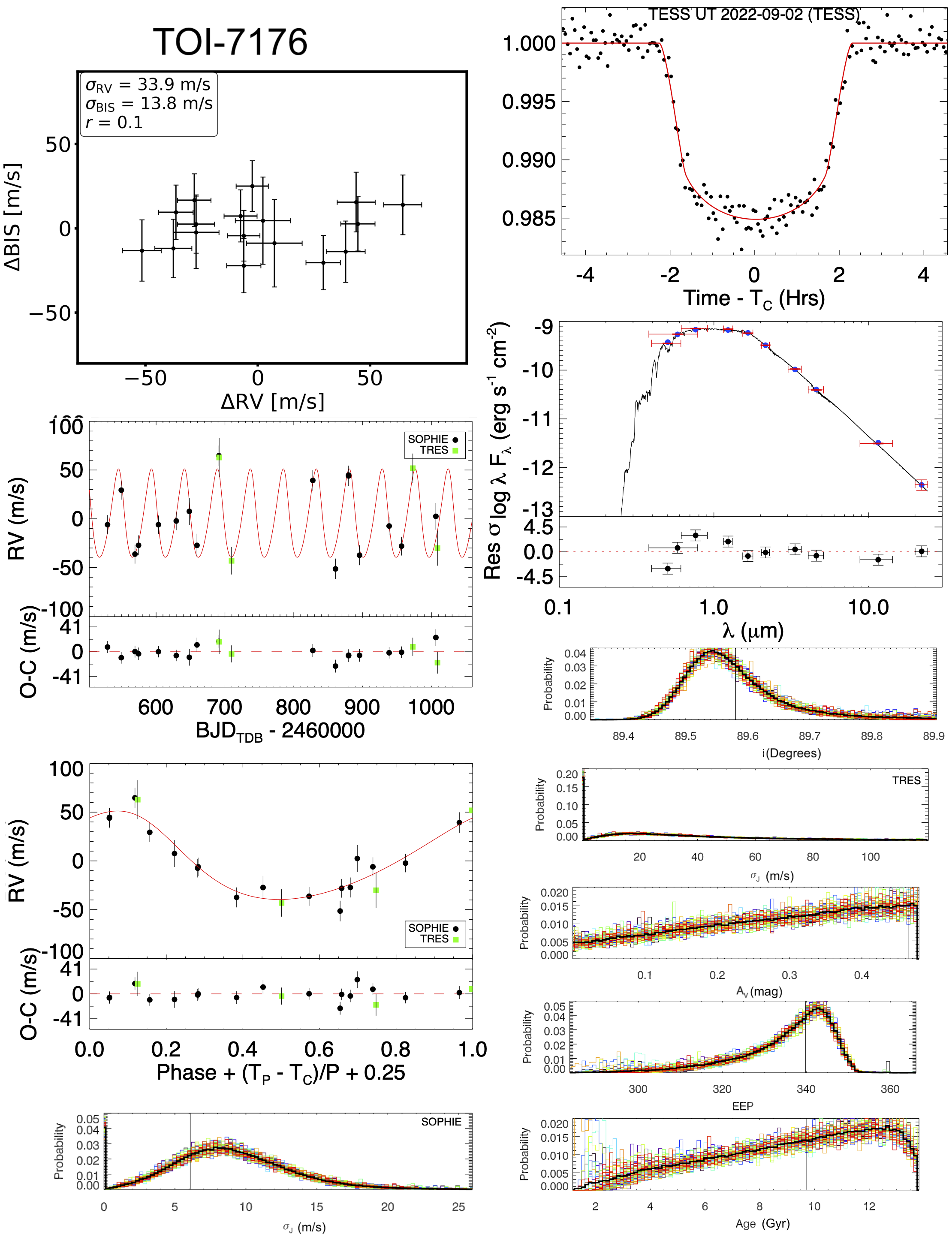}
\caption{Summary of the spectroscopic and photometric analyses for TOI-7176. Same as the caption of Fig. \ref{image_6649}. The panels are similar to those shown in Fig.~\ref{image_6649}. The posterior probability distributions shown correspond to the orbital inclination, radial velocity jitter, $A_V$, EEP, and age. TOI-7176\,b is a warm Jupiter with an orbital period of $P = 47.880^{+0.059}_{-0.065}$~d, a mass of $M_{\rm p}=0.632\pm0.063\,M_{\rm J}$, and a radius of $R_{\rm p}=0.772\pm0.026\,R_{\rm J}$.}
\label{image_7176}
\end{figure*}

\begin{figure*}
\centering
\includegraphics[width=0.8\columnwidth]{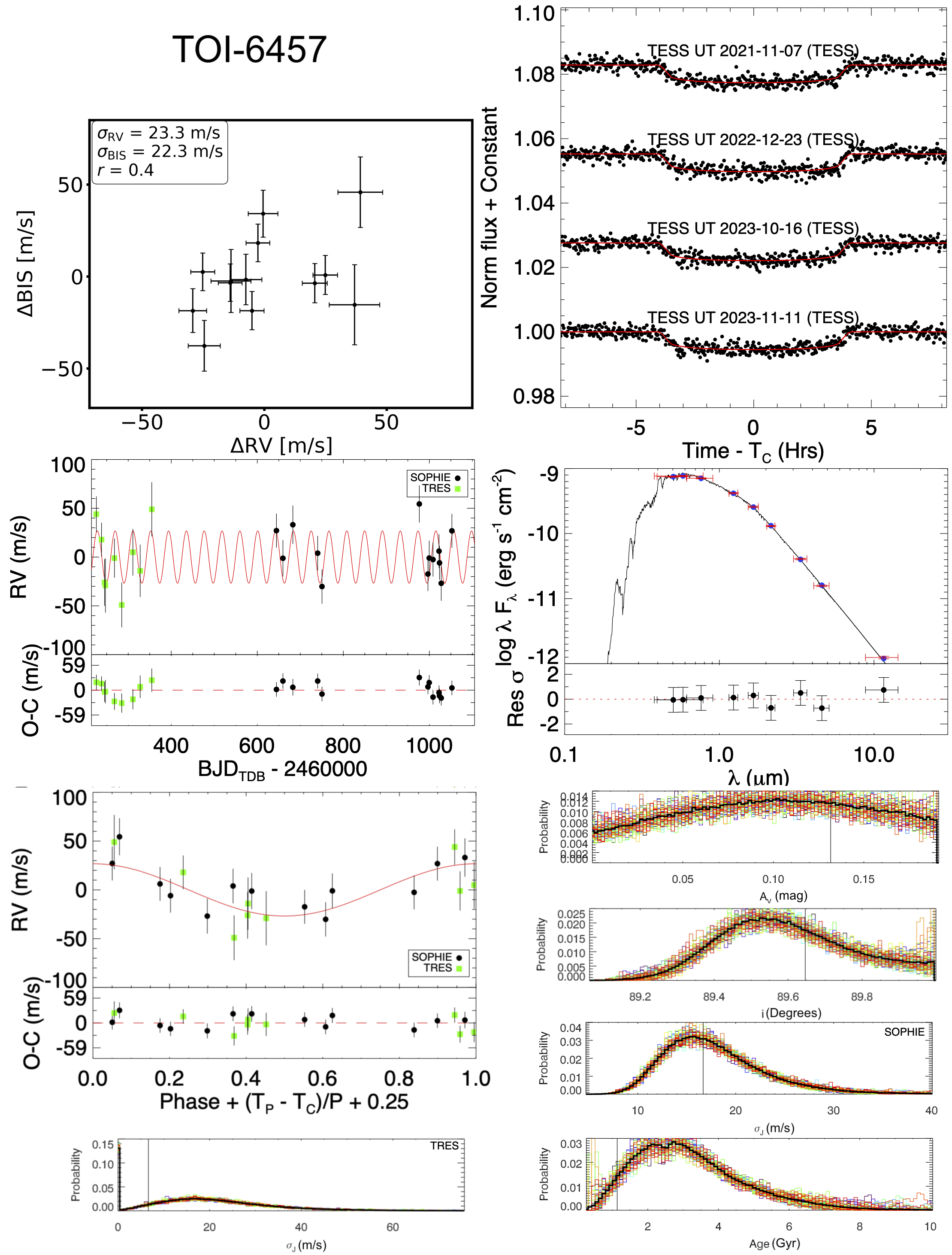}
\caption{Summary of the spectroscopic and photometric analyses for TOI-6457. The panels are similar to those shown in Fig.~\ref{image_6649}. The posterior probability distributions shown correspond to the $A_V$, orbital inclination, radial velocity jitter, and age. TOI-6457\,b is a warm Jupiter with an orbital period of $P = 41.41833 \pm 0.00010$~d, a mass of $M_{\rm p}=0.49\pm0.12\,M_{\rm J}$, and a radius of $R_{\rm p}=0.915^{+0.034}_{-0.029}\,R_{\rm J}$.}
\label{image_6457}
\end{figure*}

\begin{figure*}
\centering
\includegraphics[width=0.8\columnwidth]{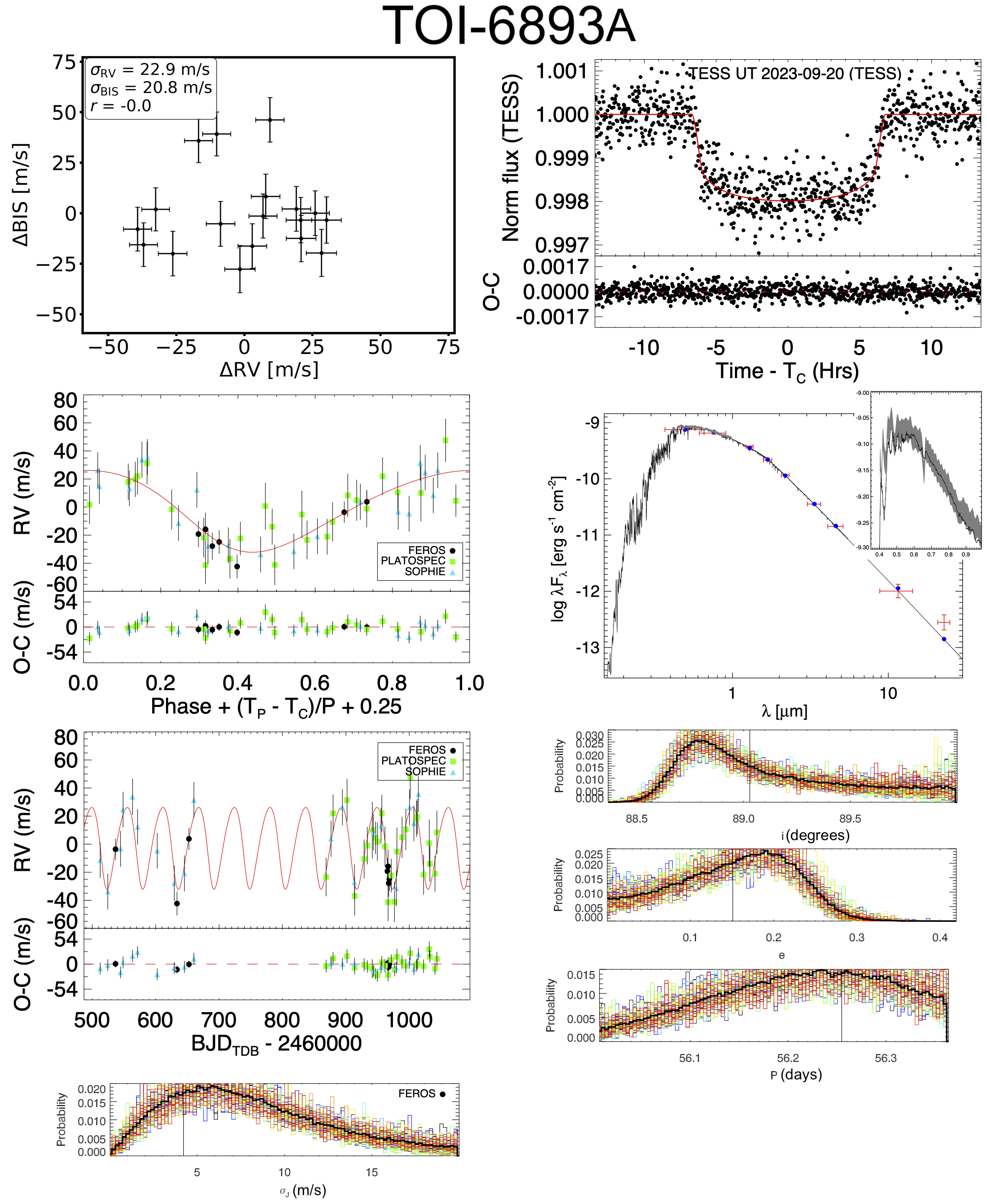}
\caption{Summary of the spectroscopic and photometric analyses for TOI-6893A. The spectroscopic and photometric panels are similar to those shown in Fig.~\ref{image_6649}, and the SED panel is similar to that shown in Fig.~\ref{image_5217}. The posterior probability distributions shown correspond to the FEROS RV jitter, inclination, eccentricity, and orbital period, eccentricity. TOI-6893A\,b is a warm Jupiter with an orbital period of $56.222^{+0.087}_{-0.10}$~d, a mass of $M_{\rm p}=0.627^{+0.078}_{-0.076}\,M_{\rm J}$, and a radius of $R_{\rm p}=1.036^{+0.038}_{-0.035}\,R_{\rm J}$.}
\label{image_6893}
\end{figure*}

\begin{figure*}
\centering
\includegraphics[width=0.8\columnwidth]{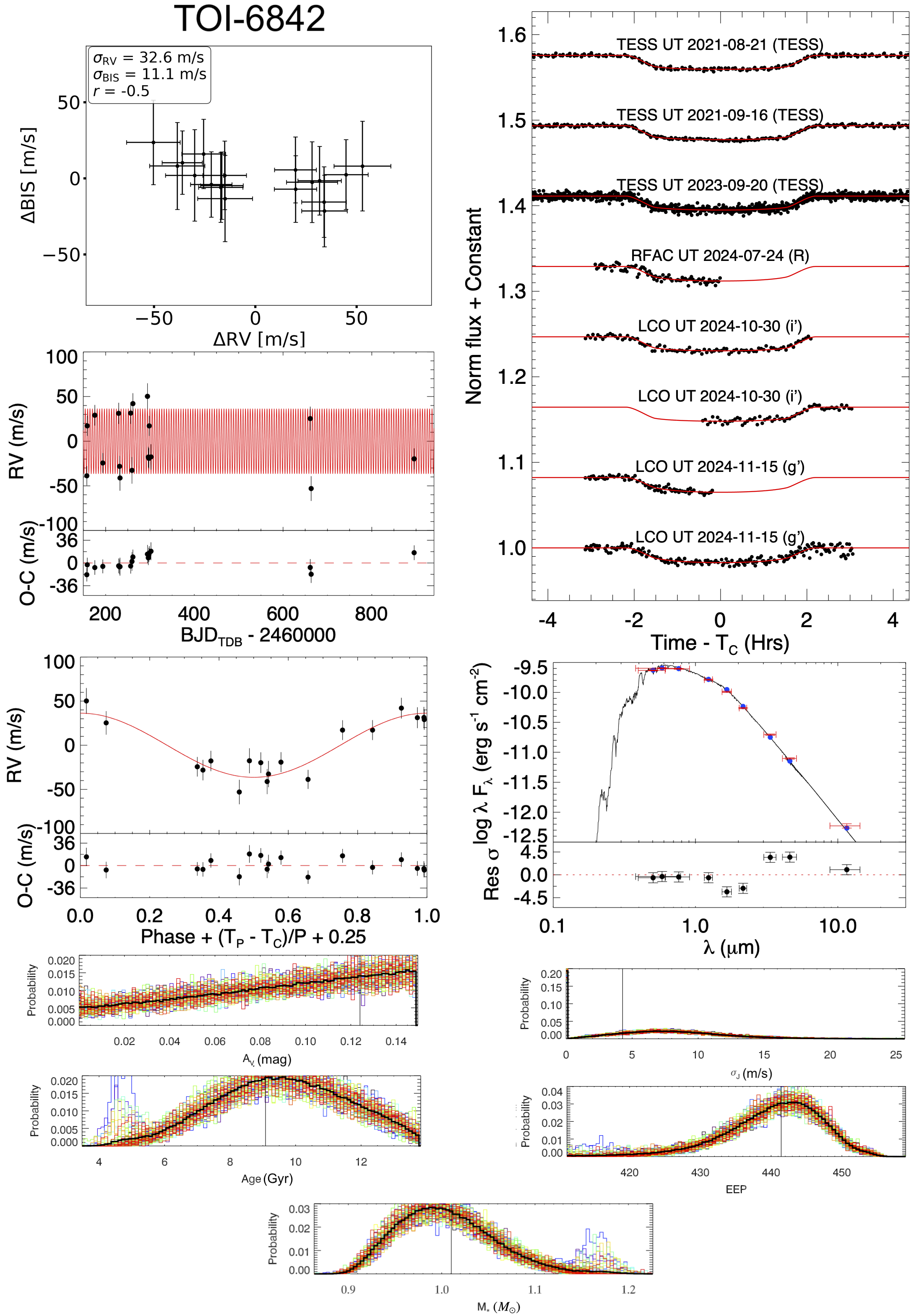}
\caption{Summary of the spectroscopic, photometric, and SED analyses for TOI-6842A. The panels are similar to those shown in Fig.~\ref{image_6649}. The posterior probability distributions shown correspond to the $A_V$, radial velocity jitter, age, EEP, and $M_\star$. TOI-6842\,b is a highly inflated giant planet with an orbital period of $P = 5.3917436 \pm 0.0000025$~d, a mass of $M_{\rm p}=0.302\pm0.035\,M_{\rm J}$, and a radius of $R_{\rm p}=1.647\pm0.039\,R_{\rm J}$.}
\label{image_6842}
\end{figure*}

\begin{figure*}
\centering
\includegraphics[width=0.8\columnwidth]{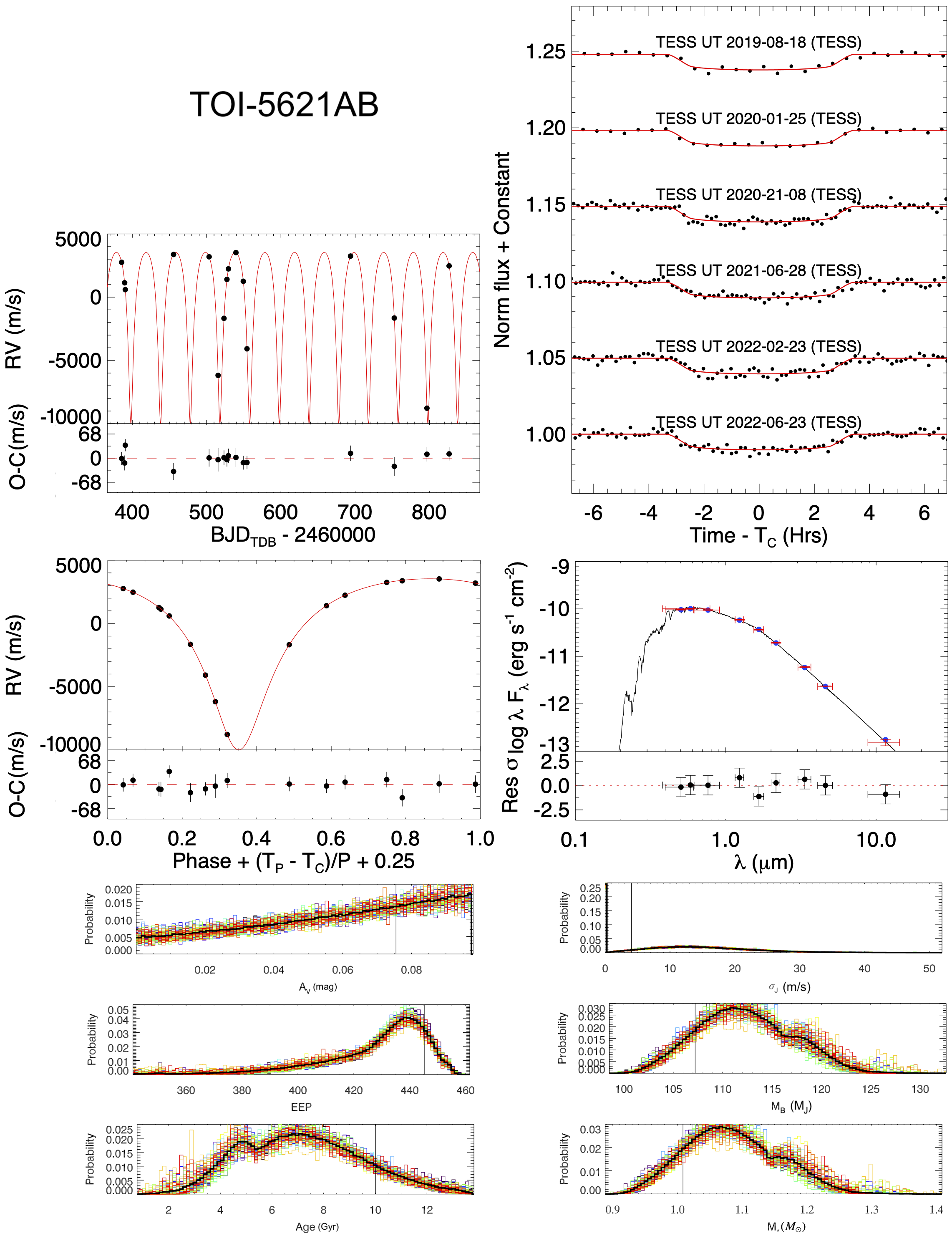}
\caption{Summary of the spectroscopic, photometric, and SED analyses for TOI-5621. The panels are similar to those shown in Fig.~\ref{image_6649}. The RV--bisector correlation panel is omitted because TOI-5621 is a stellar binary with a large RV semi-amplitude. Additionally, the posterior probability distributions shown are those for $A_V$, the radial velocity jitter, EEP, the mass of star B ($M_B$), age, and the mass of star A ($M_A$). The companion TOI-5621B has an orbital period of $P = 39.97547 \pm 0.00012$~d, a mass of $M = 112.4^{+5.5}_{-4.6}\,M_{\rm J}$ ($0.1071\,M_\odot$), and a radius of $R = 1.341^{+0.053}_{-0.052}\,R_{\rm J}$ ($0.1378\,R_\odot$).}
\label{image_5621}
\end{figure*}

\begin{figure*}
\centering
\includegraphics[width=0.8\columnwidth]{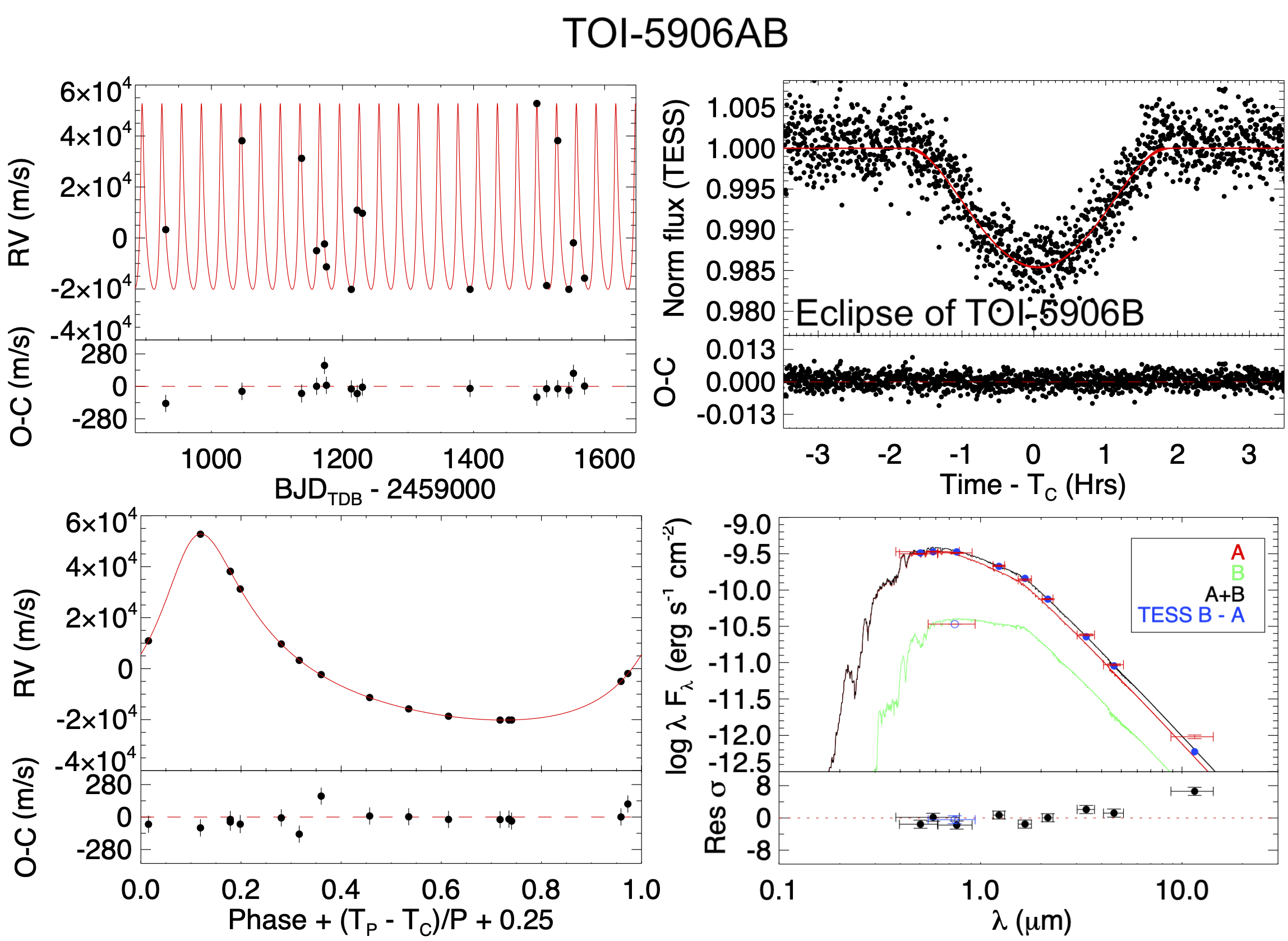}
\caption{Summary of the spectroscopic, photometric, and SED analyses for TOI-5906. The panels are similar to those shown in Fig.~\ref{image_6649}. The RV--bisector correlation panel is omitted because TOI-5906 is a stellar binary with a large RV semi-amplitude. Additionally, all fitted parameters have approximately Gaussian posterior distributions. The eclipsing companion TOI-5906B has an orbital period of $P = 30.116874\pm0.000022$~d, a mass of $M = 0.647\pm0.021\,M_\odot$, and a radius of $R = 0.625\pm0.024\,R_\odot$.}
\label{image_5906}
\end{figure*}

\begin{figure*}
\centering
\includegraphics[width=0.8\columnwidth]{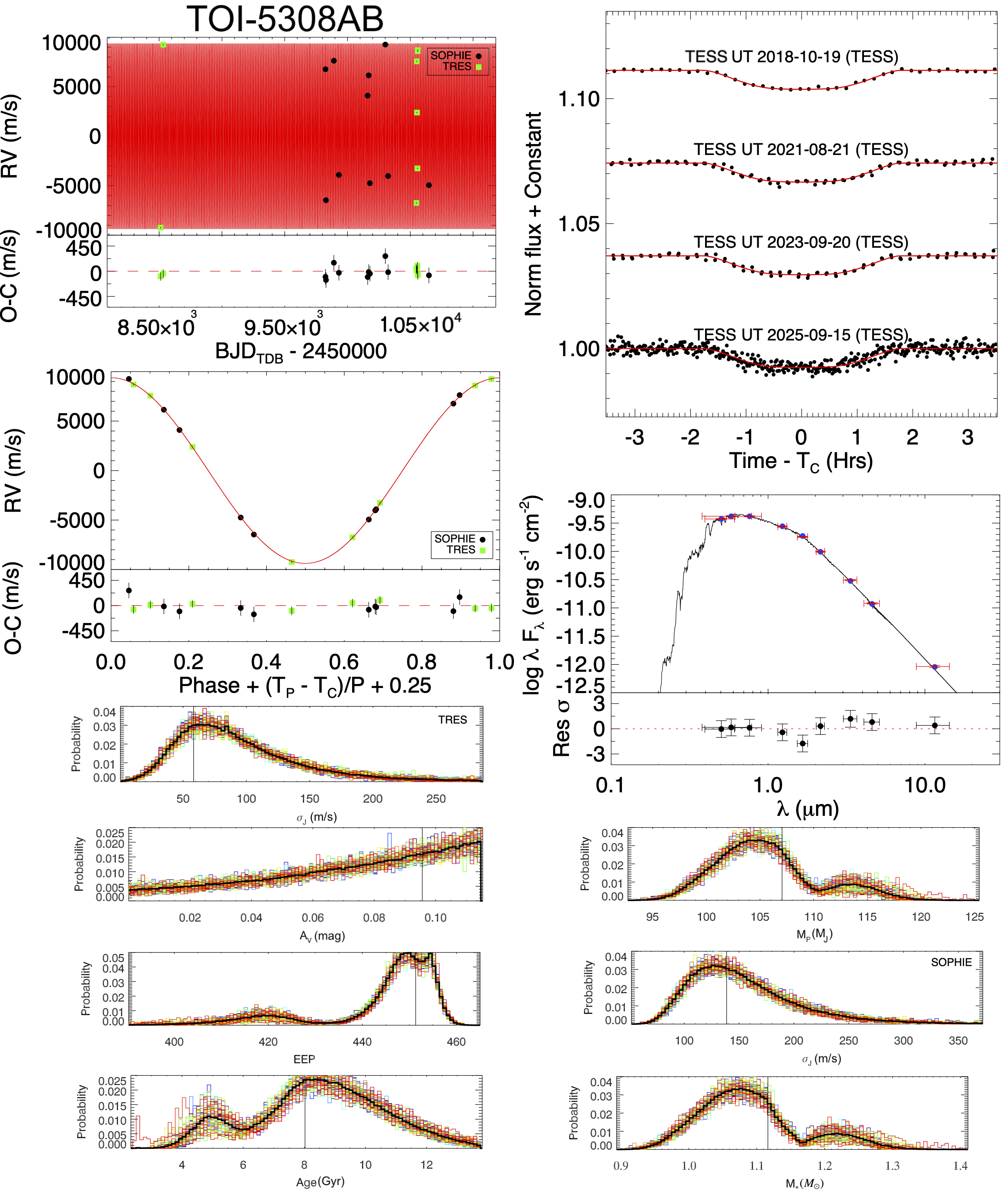}
\caption{Summary of the spectroscopic, photometric, and SED analyses for TOI-5308. The panels are similar to those shown in Fig.~\ref{image_6649}. The RV--bisector correlation panel is omitted because TOI-5308AB is a stellar binary with a large RV semi-amplitude. Additionally, the posterior probability distributions shown are those for the TRES radial velocity jitter, $A_V$, $M_B$, EEP, the SOPHIE radial velocity jitter, age, and $M_A$. The companion TOI-5308B has an orbital period of $P = 8.3403287^{+0.0000066}_{-0.0000067}$~d, a mass of $M = 105.8^{+3.2}_{-3.6}\,M_{\rm J}$ ($0.1088\,M_\odot$), and a radius of $R = 1.405^{+0.046}_{-0.045}\,R_{\rm J}$ ($0.1470\,R_\odot$).}
\label{image_5308}
\end{figure*}

\begin{figure*}
\centering
\includegraphics[width=0.8\columnwidth]{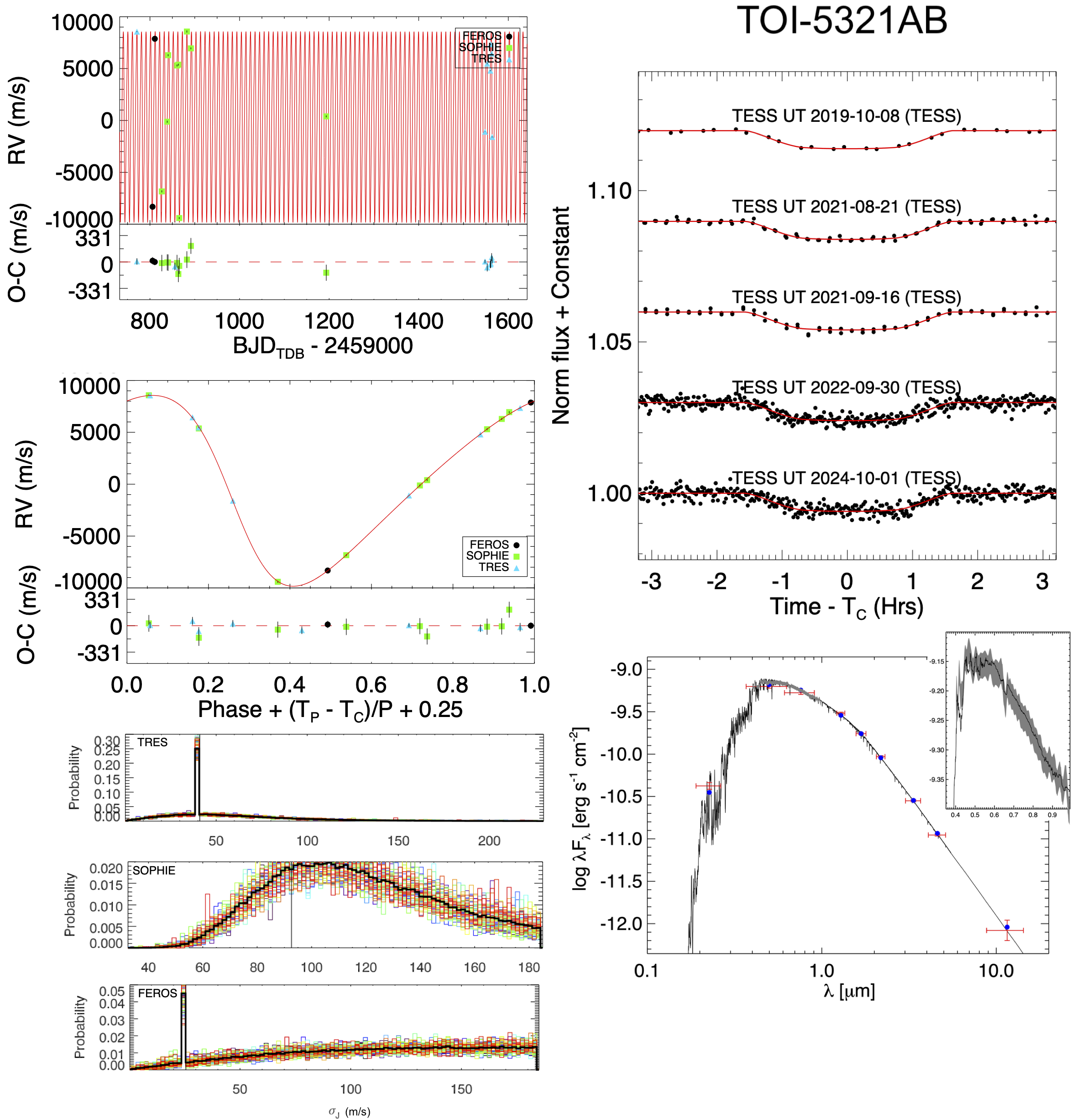}
\caption{Summary of the spectroscopic, photometric, and SED analyses for TOI-5321. The spectroscopic and photometric panels are similar to those shown in Fig.~\ref{image_6649}, and the SED panel is similar to that shown in Fig.~\ref{image_5217}. The RV--bisector correlation panel is omitted because TOI-5321 is a stellar binary with a large RV semi-amplitude. Additionally, the posterior probability distributions shown are those for the TRES, SOPHIE, and FEROS radial velocity jitters. The companion TOI-5321B has an orbital period of $P = 10.1392043^{+0.0000078}_{-0.0000076}$~d, a mass of $M = 127.2\pm4.4\,M_{\rm J}$ ($0.1214\,M_\odot$), and a radius of $R = 1.466^{+0.052}_{-0.053}\,R_{\rm J}$ ($0.1507\,R_\odot$).}
\label{image_5321}
\end{figure*}

\begin{figure*}
\centering
\includegraphics[width=0.8\columnwidth]{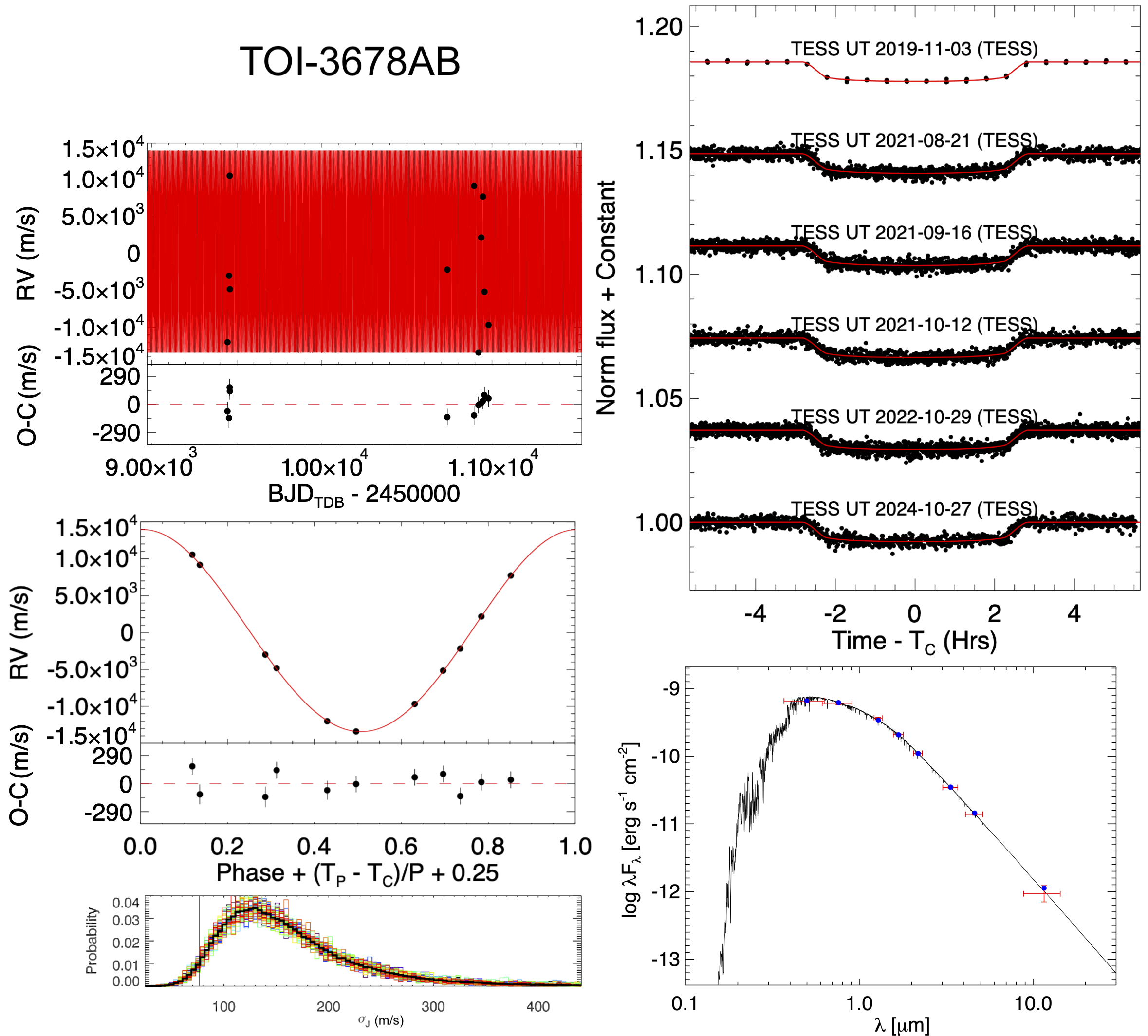}
\caption{Summary of the spectroscopic, photometric, and SED analyses for TOI-3678. The spectroscopic and photometric panels are similar to those shown in Fig.~\ref{image_6649}, while the SED panel is similar to that shown in Fig.~\ref{image_5217}. The RV--bisector correlation panel is omitted because TOI-3678 is a stellar binary with a large RV semi-amplitude. The posterior probability distribution shown corresponds to the SOPHIE radial velocity jitter. The companion TOI-3678B has an orbital period of $P = 4.8542845\pm0.0000024$~d, a mass of $M = 167.9^{+6.5}_{-6.6}\,M_{\rm J}$ ($0.1602\,M_\odot$), and a radius of $R = 1.884^{+0.053}_{-0.055}\,R_{\rm J}$ ($0.1935\,R_\odot$).}
\label{image_3678}
\end{figure*}

\end{appendix}

\end{document}